\documentclass[aps,prl,10pt,twocolumn,superscriptaddress,preprintnumbers,floatfix]{revtex4-2}

\usepackage{amsmath}
\usepackage{amssymb}
\usepackage{graphicx}
\usepackage{braket}
\usepackage{silence}
\usepackage{url}
\usepackage{appendix}
\usepackage{lineno}
\usepackage{float}
\usepackage{physics}

\usepackage{titlesec}

\usepackage{tabularx}
\usepackage[table,x11names,dvipsnames]{xcolor}
\usepackage{longtable}
\usepackage{array}
\usepackage{multirow}
\usepackage{extarrows}

\usepackage{color}
\usepackage{colortbl}
\usepackage{setspace}

\newcolumntype{L}[1]{>{\raggedright\let\newline\\\arraybackslash\hspace{0pt}}m{#1}}
\newcolumntype{C}[1]{>{\centering\let\newline\\\arraybackslash\hspace{0pt}}m{#1}}
\newcolumntype{R}[1]{>{\raggedleft\let\newline\\\arraybackslash\hspace{0pt}}m{#1}}

\usepackage[caption=false,position=top, singlelinecheck=off, justification=raggedright, labelformat=simple, labelfont=bf]{subfig}

\usepackage[unicode]{hyperref}
\usepackage{float}
\hypersetup{unicode=true,colorlinks=true,linkcolor=blue, citecolor=blue,urlcolor=blue}

\makeatletter
\renewcommand\paragraph[1]{\textit{#1}---\!}
\makeatother

\def\be{\begin{equation}}
\def\ee{\end{equation}}
\def\bea{\begin{eqnarray}}
\def\eea{\end{eqnarray}}
\def\X{{\rm X}}
\def\Y{{\rm Y}}

    \def\<{\langle}         \def\>{\rangle}

\def\F{{\mathcal F}}

\newcommand{\nocontentsline}[3]{}
\let\origcontentsline\addcontentsline
\newcommand\stoptoc{\let\addcontentsline\nocontentsline}
\newcommand\resumetoc{\let\addcontentsline\origcontentsline}

\begin{document}
\stoptoc

\title{Squeezing atomic $p$-orbital condensates for detecting gravitational waves} 

\author{Xinyang Yu}
\affiliation{State Key Laboratory of Surface Physics, Institute of Nanoelectronics and Quantum Computing, and Department of Physics, Fudan University, Shanghai 200433, China}
\affiliation{Shanghai Qi Zhi Institute, AI Tower, Xuhui District, Shanghai 200232, China} 

\author{W. Vincent Liu}
\affiliation{Department of Physics and Astronomy, University of
  Pittsburgh, Pittsburgh, PA 15260, USA}

\author{Xiaopeng Li}\email{xiaopeng\_li@fudan.edu.cn}
\affiliation{State Key Laboratory of Surface Physics, Institute of Nanoelectronics and Quantum Computing, and Department of Physics, Fudan University, Shanghai 200433, China}
\affiliation{Shanghai Qi Zhi Institute, AI Tower, Xuhui District, Shanghai 200232, China} 
\affiliation{Hefei National Laboratory, Hefei 230088, China}
\affiliation{Shanghai Artificial Intelligence Laboratory, Shanghai 200232, China}
\affiliation{Shanghai Research Center for Quantum Sciences, Shanghai 201315, China}

\begin{abstract}
   Detecting the faint signal of continuous gravitational waves (CWs) stands as a major frontier in gravitational-wave astronomy, pushing the need for detectors whose sensitivity exceeds the standard quantum limit (SQL).
   Here, we propose an orbital optomechanical (OOM) sensor that exploits the sensitive coupling of an orbitally squeezed $p$-orbital Bose-Einstein condensate to spacetime distortions, enabling the detection of interferometer phase shifts induced by CWs. 
   This sensor achieves a theoretical quantum-noise-limited sensitivity 16 dB below the SQL while reducing the required laser power by five orders of magnitude. The performance arises from a novel noise trade-off: a counter-propagating readout scheme suppresses photonic shot noise, while orbital squeezing minimizes the remaining atomic projection noise.
   By leveraging quantum control over atomic orbital degrees of freedom, this approach establishes a new framework for interferometric sensing with direct applications to searches for CWs and ultralight dark matter.
\end{abstract}

\date{\today}
\maketitle

\paragraph{Introduction}
Rapid progress in gravitational wave (GW) astronomy has opened a new observational window onto previously inaccessible phenomena and fundamental physics~\cite{LIGO2016_GW150914}. While observations of cataclysmic events like binary mergers have been revolutionary~\cite{LIGO2017_GW170817,Abbott_2021}, the next frontier targets signals that are even more elusive. Two of the most compelling targets are persistent continuous gravitational waves (CWs)~\cite{CW2017LRR,bailes2021gravitational} and signals from ultralight dark matter~\cite{ferreira2021ultra}. Continuous waves, expected from sources like rapidly spinning, non-axisymmetric neutron stars, could provide unprecedented insight into the equation of state of dense matter~\cite{Yunes-Miller-Yagi:22rev}, although they have not yet been definitively detected. Similarly, certain models of ultralight dark matter predict a CW-like signal that could reveal its nature and coupling to ordinary matter~\cite{Mosbech2023Gwave-DM,Bertone-Tait:18rev,Bertone:24:NobelSym,Vermeulen:21:ScalarDM,ligo2022PRD,LIGO_vectordarkmatter:24arx}. To access these frontiers, a new generation of detectors is required with sensitivity beyond current limits, particularly in the challenging low-frequency regime~\cite{bailes2021gravitational}.

In atomic quantum technology, spin squeezing has emerged as a powerful technique for enhancing the sensitivity of magnetic field detection by suppressing quantum shot noise in the sensing signal—specifically, the spin Larmor precession angle—from the standard quantum limit (SQL) to the Heisenberg limit~\cite{2018_Treutlein_RMP}. However, due to the equivalence principle, atomic spins are fundamentally decoupled from gravitational forces.  Meanwhile, the motional degrees of freedom of atoms, which naturally couple to gravitational forces, have been successfully quantized through optical lattice confinement in cold atoms, resulting in discrete energy-separated orbitals~\cite{wirth2011evidence,2011_Lewenstein_NP,li2016physics}. Over the past several years, significant progress has been made in the quantum control of $s$, $p$, and $d$ orbitals~\cite{2020_Rey_OrbitalGate,2021_Zhou_PRL,2021_Zhou_OrbitalGate_PRA,2021_Xu_Nature,2023_Thywissen_Nature}. This has led to the realization of exotic multi-orbital superfluids with vestigial orders~\cite{2021_Zhou_PRL,2021_Xu_Nature,2023_Thywissen_Nature} and the development of universal quantum orbital gates with fidelities reaching above 95\%~\cite{2020_Rey_OrbitalGate,2021_Zhou_OrbitalGate_PRA}.

\begin{figure}[htbp]
\centering
\includegraphics[width=\linewidth]{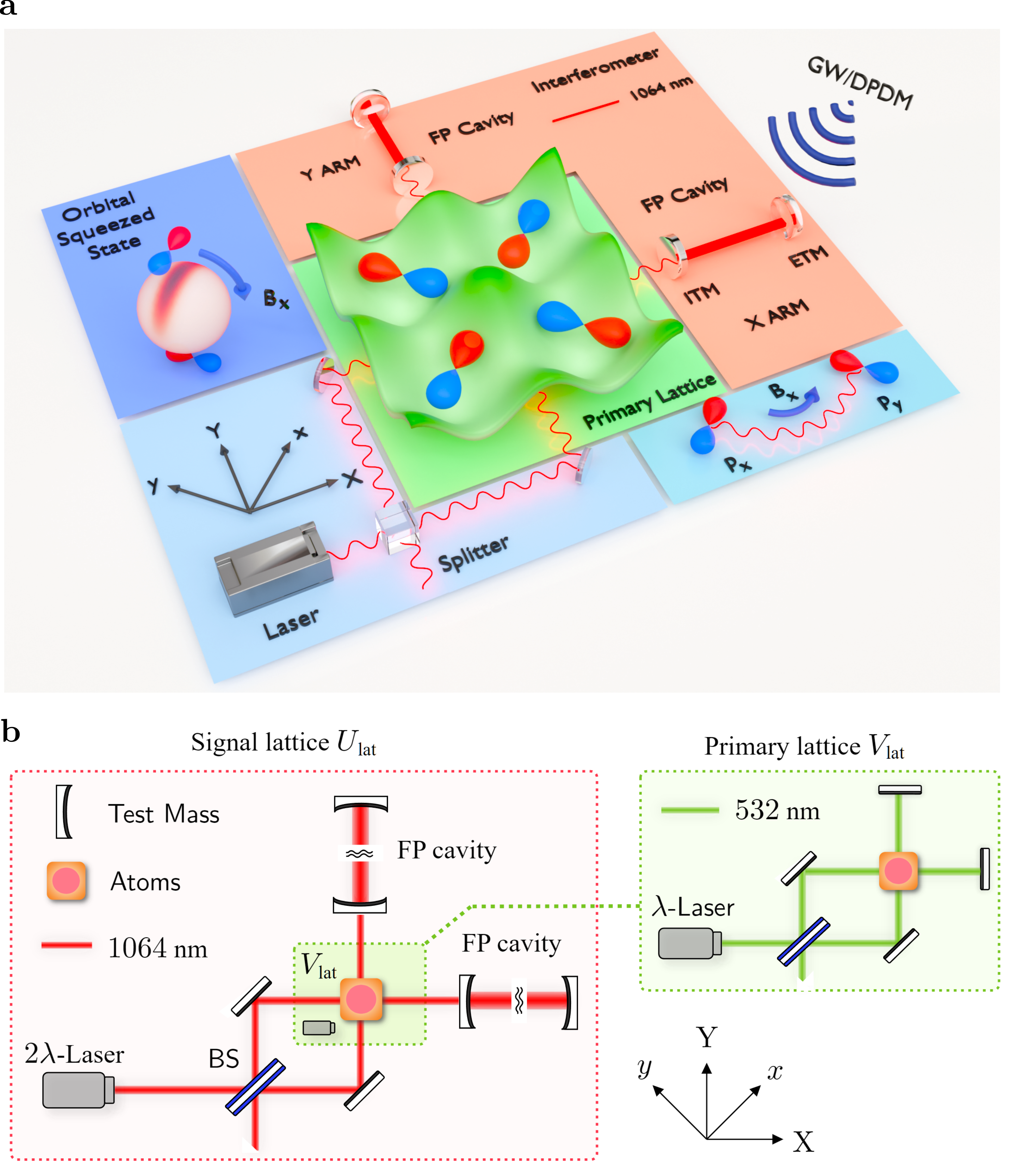}
\caption{\textbf{Schematic illustration of the OOM sensor.} \textbf{a}, An orbital squeezed state is prepared by loading atoms into the $p$-orbital band of the primary optical lattice (shown in green).
The incoming gravitational wave (GWs) or dark photon dark matter (DPDM) modifies the apparent distance between the Input Test Mass (ITM) and End Test Mass (ETM), causing signal lattice deformation and generating a pseudo-magnetic field ($B_x$) that couples the degenerate $p_x$ and $p_y$ orbitals. The resulting orbital rotation can be detected through a time-of-flight experiment that measures the population difference between the two $p$-orbitals. \textbf{b}, Optical setup of the OOM sensor. The primary lattice $V_{\mathrm{lat}}$ confining the atoms is generated by the $\lambda$-laser (532 nm), while the signal lattice $U_{\mathrm{lat}}$ for signal detection is generated by a doubled-wavelength $2\lambda$-laser (1064 nm).  The doubled wavelength of the signal laser provides the spatial symmetry needed to couple the degenerate $p_x$ and $p_y$ states (\cite{supp}, Sec. S-3.1).
}
\label{fig:illustration}
\end{figure}

\begin{figure*}[htbp]
    \centering
    \hspace{-10pt}
    \includegraphics[width=0.9\linewidth]{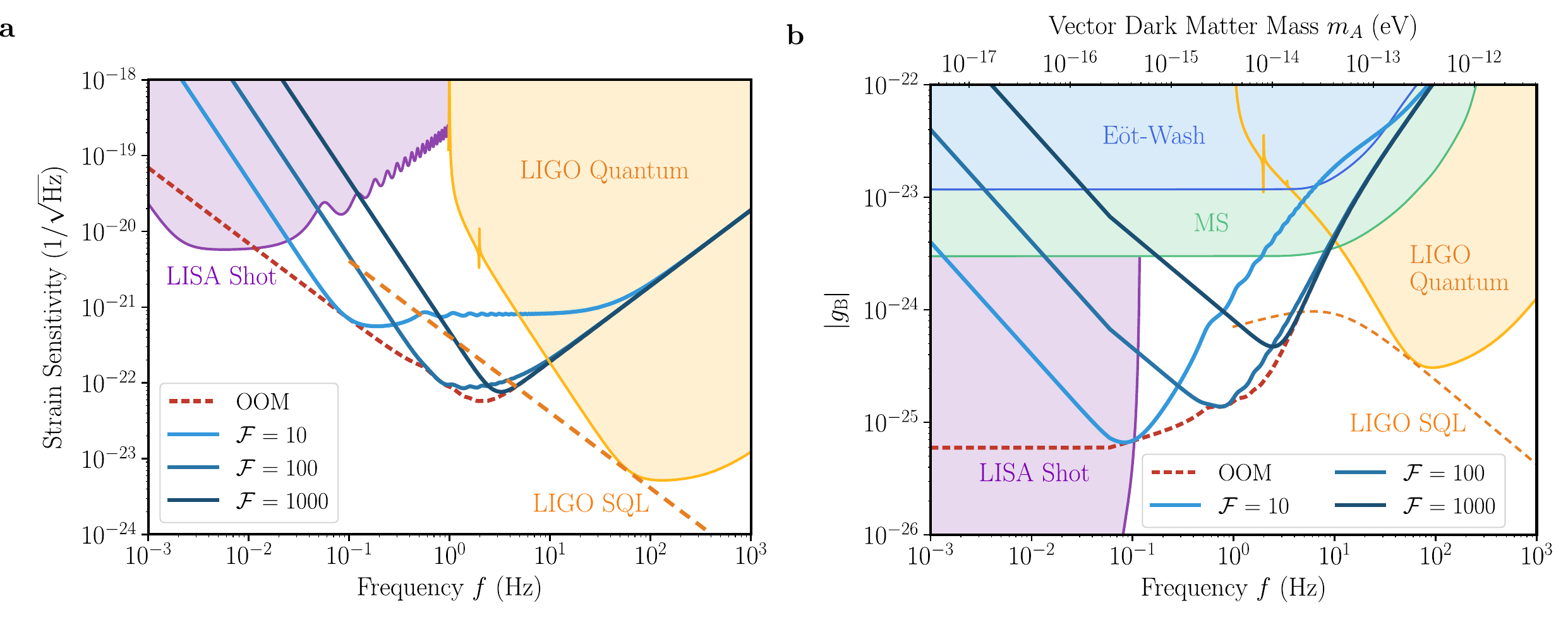}
    \caption{\textbf{Quantum-noise-limited sensitivity curve for detecting GW signals and constraints on DPDM coupling.} 
    \textbf{a}, Sky- and polarization-averaged strain sensitivity curves of the OOM sensor, compared with LIGO~\cite{ligo2015cqg} and LISA detection schemes~\cite{2019LISA}. The blue-solid lines represent the OOM sensitivity for different cavity finesses ($\F$), with the red-dashed line showing the optimized envelope. The `LIGO Quantum' curve shows the total quantum noise floor (shot and radiation pressure noise) for a conventional LIGO-like detector, while the `LIGO SQL' curve represents the fundamental limit arising from their trade-off. Our sensor a $16$ dB (factor of 6.3) improvement over the conventional SQL for LIGO's parameters ($M=40\mathrm{kg}$, $L=4\mathrm{km}$) at frequencies below a critical value of $f_c \simeq 1.87\,\mathrm{Hz}$.
    \textbf{b}, 1$\sigma$ upper limit on the dark photon-baryon coupling $|g_{\mathrm{B}}|$, compared with limits from LIGO~\cite{ligo2022PRD}, LISA~\cite{2019LISA}, Eöt-Wash~\cite{EotWash1,EotWash2}, and MICROSCOPE (MS) experiments~\cite{MICROSCOPE1,MICROSCOPE2,MICROSCOPE3,MICROSCOPE4}. For constraining $|g_{\mathrm{B}}|$, a total observation time of two years is assumed for comparison. The sensor shows an $34$ dB (factor of 50) improvement over the MICROSCOPE experiment in detecting dark photons with masses around $5 \times 10^{-16}\,\mathrm{eV/c}^2$ and $4 \times 10^{-15}\,\mathrm{eV/c}^2$.}
    \label{fig:sensitivity}
\end{figure*}

Here, we theoretically propose and analyze an orbital optomechanical (OOM) sensor by squeezing the orbital degrees of freedom of a Bose-Einstein condensate (BEC) in the degenerate $p_{x,y}$-bands of an optical lattice (Fig.~\ref{fig:illustration}). The two-axis-counter-twisting (TACT) mechanism, essential for generating optimal squeezing, is notoriously difficult to engineer in conventional spin systems~\cite{ma2011quantum,2022PrlTACT,luo2024hamiltonianengineeringcollectivexyz}.
In contrast, this orbital atomic system allows for the realization of the TACT Hamiltonian by tuning the intrinsic atomic interactions.
The sensor's design maps GW-induced phase shifts from a Fabry-Perot-Michelson interferometer onto the quantum state rotation of the BEC. 
This architecture forgoes the complex dual-recycling techniques of current observatories~\cite{ligo2015cqg} and operates with laser power five orders of magnitude lower. The sensor's ability to surpass the SQL stems from a novel noise trade-off.
A readout scheme using counter-propagating light fields intrinsically cancels a significant portion of the photonic shot noise. While this process introduces atomic projection noise, this new dominant noise source is in turn suppressed by TACT-generated orbital squeezing. Consequently, our simulations show that this mechanism enables a quantum-noise-limited sensitivity 16 dB (a factor of 6.3) below the conventional SQL (Fig.~\ref{fig:sensitivity}), opening a new path for the search for CWs and ultralight dark matter. 

\begin{figure}[htb]
    \centering
    \includegraphics[width=\linewidth]{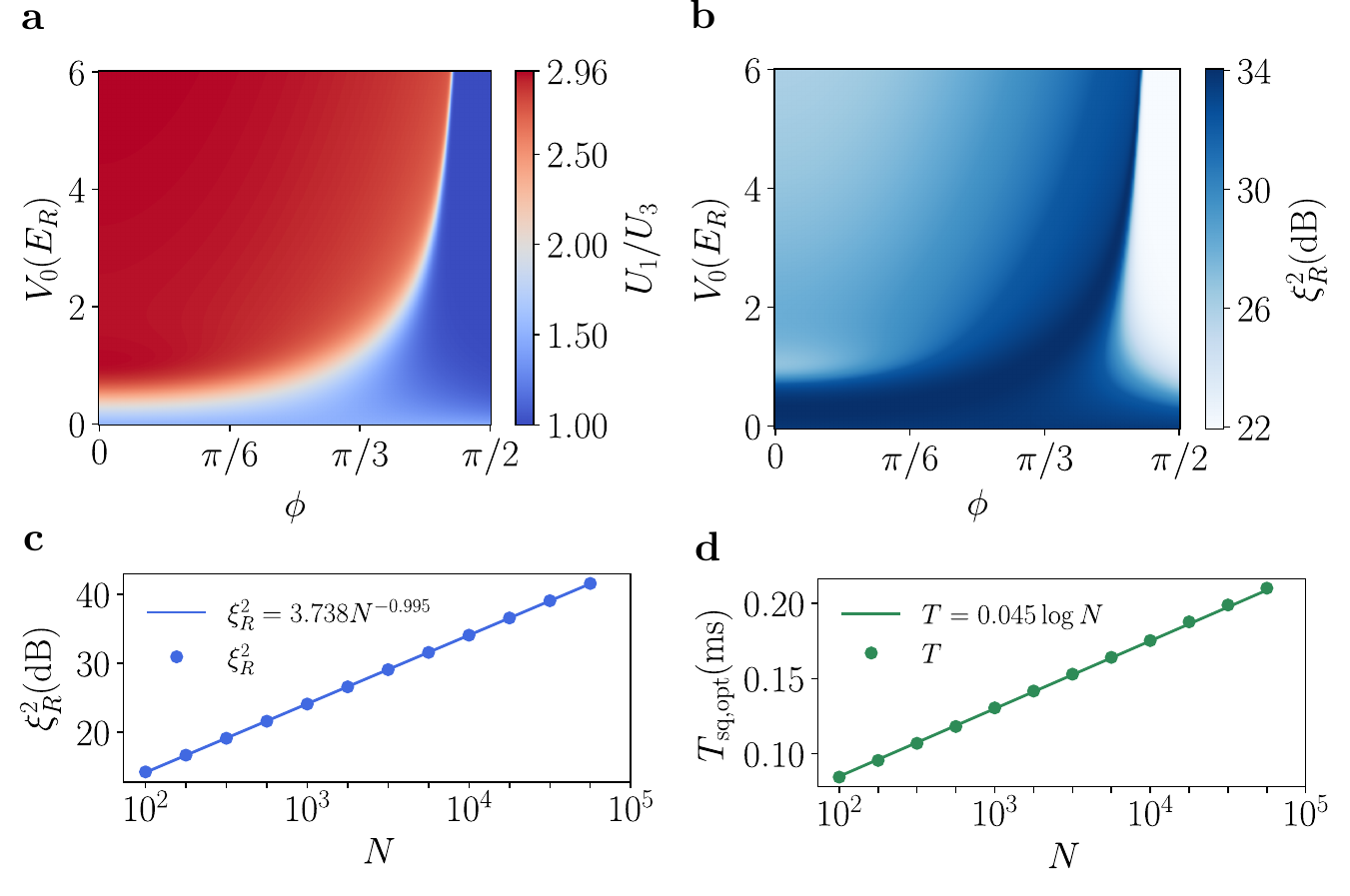}
    \caption{\textbf{Orbital squeezing of the OOM sensor}. 
    \textbf{a}, Ratio of interaction parameters $U_1/U_3$ as a function of the primary lattice depth $V_0$ and the phase difference $\phi$ between the standing waves in the $\mathrm{X}$ and $\mathrm{Y}$ arms (\cite{supp}, Sec.~S-1.1). The interaction parameters in Eq.~\eqref{eqn:SpinHamiltonian} exhibit significant tunability. The white region at $U_1/U_3 = 2$ indicates TACT dynamics, while the red and blue regions approaching $U_1/U_3 \rightarrow 3$ and $U_1/U_3 \rightarrow 1$ reflect OAT dynamics.
    \textbf{b}, Optimal squeezing parameter $\xi^2_R$ for $N = 10^4$ atoms. TACT dynamics achieve an orbital squeezed state with $\xi^2_R \sim N^{-1}$, reaching the Heisenberg limit, while OAT dynamics yield $\xi^2_R \sim N^{-2/3}$.
    \textbf{c}, Scaling behavior of the optimal squeezing time $T_{\text{sq,opt}}$ and the optimal squeezing parameter $\xi^2_{R,\text{opt}}$ as a function of atom number $N$ in the TACT regime. The optical lattice parameters used here are $V_0 = 6E_R$ and $\phi = 0.44875\pi$.
    }
    \label{fig:details}
\end{figure}

\paragraph{Squeezing atomic $p$-orbital states}
We consider a large number ($N$) of bosonic atoms loaded into the $p$-orbital bands of a bipartite optical lattice, which is formed by the primary light potential $V_{\text{lat}}$ with depth $V_0$ (Fig.~\ref{fig:illustration})~\cite{wirth2011evidence,kock2016orbital}. 
The $p$-orbital bands exhibit two inequivalent degenerate minima in quasi-momentum space denoted as $|p_x\rangle$ ($\equiv |\hspace{-0.5ex}\uparrow\rangle$) and $|p_y\rangle$ ($\equiv |\hspace{-0.5ex}\downarrow\rangle$)~\cite{li2016physics}.
The two minima form a two-level (pseudo-spin-1/2) system, with its Hilbert space represented by a Bloch sphere (Fig.~\ref{fig:illustration}\textbf{a}). 
Under the two-mode approximation, the quantum kinematics of the $p$-orbital many-body system is described by pseudo-spin-$N/2$ operators, $\hat{\bf J} \equiv ( \hat{J}_x,  \hat{J}_y, \hat{J}_z)$, with $\hat{J}_x = (\hat{p}_y^\dagger \hat{p}_x + \text{h.c.})/2$, $\hat{J}_y = i(\hat{p}_y^\dagger \hat{p}_x - \text{h.c.})/2$, and $\hat{J}_z=(\hat{p}_x^{\dagger}\hat{p}_x-\hat{p}_y^{\dagger}\hat{p}_y)/2$. Here, both $\hat{J}_{x}$ and $\hat{J}_{y}$ transform $|p_{x}\rangle$ and $|p_{y}\rangle$ into each other, while $\hat{J}_z$ represents the condensed atom number difference between the two $p$-orbitals~\cite{Hall1998a,Hall1998b,Shin2004DoubleWell,schumm2005matter,leggett2001bose,dalton2012two}.

In the experiments, a $p$-orbital Bose-Einstein condensate with bosonic atoms condensing into one single minimum has been prepared~\cite{wirth2011evidence,2021_Zhou_PRL}.
Without loss of generality, we assume the state is $|\Psi_0\rangle$, with $\hat{J}_z |\Psi_0\rangle = \frac{N}{2}|\Psi_0\rangle$. This corresponds to a  coherent spin state $|\theta=0,\varphi\rangle$ in the pseudo-spin representation, with $\theta$ and $\varphi$ the polar and azimuthal angles on the Bloch sphere. 
When subject to a pseudo-magnetic field $\Delta \hat{H} = B_x \hat{J}_x$, the pseudo-spin vector ${\bf J}$ rotates around the $x$-axis, accumulating an angle $\Delta \theta \propto B_x$. Averaging over $N$ independent atoms, the resolution for sensing the pseudo-magnetic field constrained by quantum shot noise corresponds to a standard quantum limit $\Delta \theta_{\rm SQL} = N^{-1/2}$~\cite{Cappellaro2017rev}. 

When the atomic interactions are introduced, the dynamics of the $p$-orbital system are governed by (\cite{supp}, Sec.~S-2.1):
\begin{equation}\label{eqn:SpinHamiltonian}
    \hat{H} = (3U_3 - U_1) \hat{J}_x^2 - (U_1- U_3) \hat{J}_y^2.
\end{equation}  

\noindent This interacting Hamiltonian $\hat{H}$ is a combination of one-axis-twisting (OAT) $\hat{J}_x^2$ (or $\hat{J}_y^2$) and TACT $\hat{J}_x^2-\hat{J}_y^2$ interactions. It reduces to the OAT Hamiltonian when $U_1=U_3$ or $U_1=3U_3$, and to the TACT Hamiltonian when $U_1=2U_3$. 
In this $p$-orbital system, the ratio of $U_1/U_3$ is largely tunable by adjusting the interfering laser beams forming the optical lattice (Fig.~\ref{fig:details}\textbf{a}). This tunability makes the orbital-TACT Hamiltonian accessible, in sharp contrast to spin systems where engineering the TACT Hamiltonian has fundamental challenges owing to the spin SU(2) symmetry~\cite{ma2011quantum,2022PrlTACT,luo2024hamiltonianengineeringcollectivexyz}. 

In quench dynamics, we demonstrate that the interaction effects transform the $p$-band condensate, $|\Psi_0\rangle$, into an orbital squeezed state with its quantum shot-noise ($\Delta \theta_{\rm proj} $) suppressed. 
The extent of this suppression is quantified by the metrological squeezing parameter, $\xi_R \equiv \Delta \theta_{\mathrm{proj}}/\Delta \theta_{\mathrm{SQL}}$, which is calculated as $\xi_R^2 = N( \Delta J_{\varphi})^2/|\langle J_z \rangle|^2$~\cite{Wineland1992,Wineland1994,ma2011quantum}. Here, $J_\varphi = \cos\varphi J_x + \sin\varphi J_y$, and the squeezing angle $\varphi$ is chosen to minimize the variance $(\Delta J_{\varphi})^2$ perpendicular to the mean spin direction. 
Our simulation results  with $N = 10^4$ (Fig.~\ref{fig:details}\textbf{b}) reveal that optimal squeezing is achieved near the TACT regime. The time required to achieve this optimal squeezing ($T_{\text{sq,opt}}$), is approximately $0.2$~$\mathrm{ms}$ (Fig.~\ref{fig:details}\textbf{c}) under standard experimental conditions~\cite{wirth2011evidence}. 
The TACT dynamics produce an orbital squeezed state with $\xi^2_R \sim N^{-0.995}$ (Fig.~\ref{fig:details}\textbf{d}), indicating quantum enhancement in detecting the pseudo-magnetic field.  
This enhancement could suppress $\Delta \theta_{\mathrm{proj}}$ from the standard quantum limit to $N^{-0.998}$, approaching the Heisenberg limit ($N^{-1}$)---a fundamental lower bound set by the non-commutative nature of quantum mechanics~\cite{ma2011quantum}.   

\begin{figure}[htb]
    \centering
    \includegraphics[width=\linewidth]{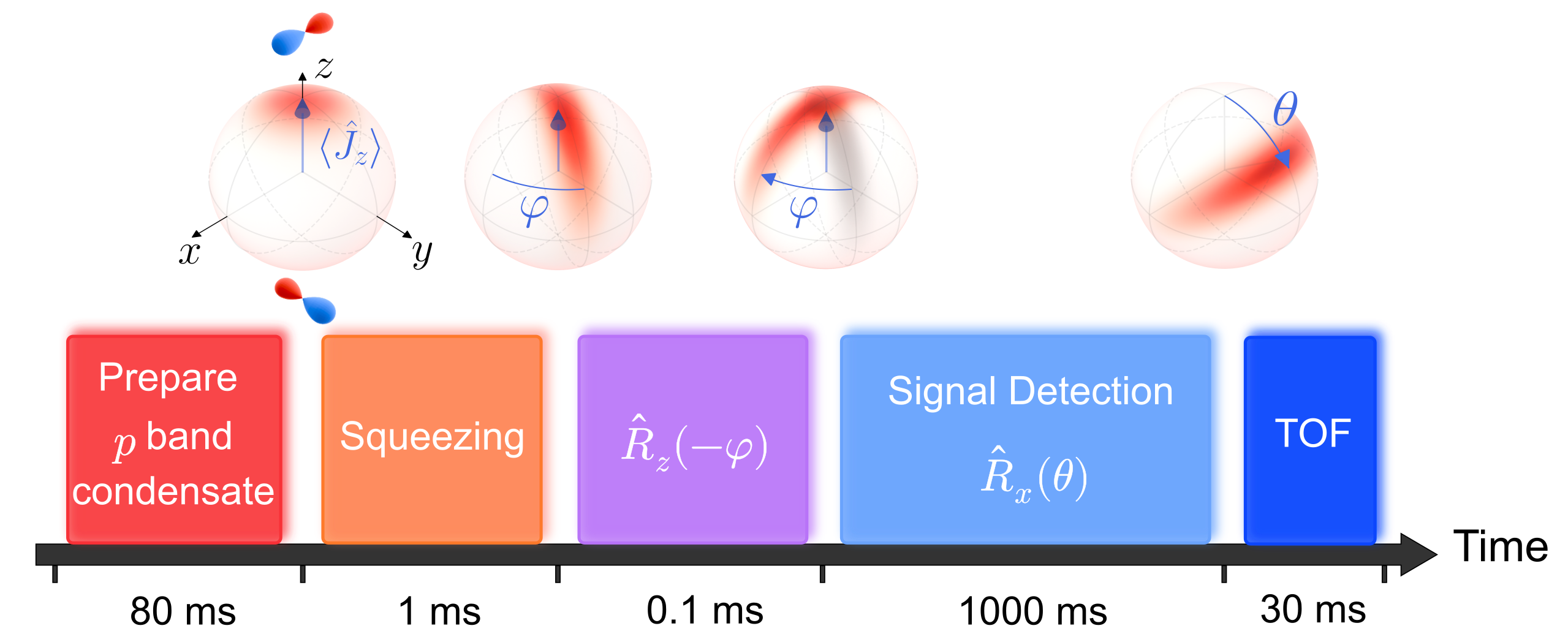}
    \caption{\textbf{Time sequence of high-precision detection with the OOM sensor.} Each measurement cycle consists of five steps: 
    (1) Preparation of bosonic atoms in the condensate state of the $p$-orbital band, which takes approximately $T_{\text{prep}} = 80\,\mathrm{ms}$~\cite{wirth2011evidence,xin2025fast}; 
    (2) Interaction quench via Feshbach resonance techniques to achieve atom squeezing within $T_{\text{sq}} = 1\,\mathrm{ms}$; 
    (3) Adjustment of the optical lattice potential $V_{\text{lat}}$ to align the most sensitive axis of the uncertainty ellipse along the $x$-axis within $T_{\text{rot}} = 0.1\,\mathrm{ms}$ (\cite{supp}, Sec.~S-2.3); 
    (4) Detection of the GW/DPDM by evolving the orbital squeezed state under the GW/DPDM-induced pseudo-magnetic field for a duration of $T_{\text{life}} = 1000\,\mathrm{ms}$, within the lifetime of $p$-orbital atoms~\cite{paul2013formation,kock2016orbital} (See also Fig.~S14 in~\cite{supp}); 
    (5) Execution of a Time-of-Flight (TOF) experiment to measure the rotation angle $\theta$, or equivalently the orbital polarization $\langle J_y\rangle $~\cite{wirth2011evidence} (\cite{supp}, Sec.~S5.4). 
    The total duration of a measurement cycle is $T = T_{\text{prep}} + T_{\text{sq}} + T_{\text{rot}} + T_{\text{life}} + T_{\text{TOF}}\simeq1.11$s.}

    \label{fig:protocol}
\end{figure}

\paragraph{Atomic quantum sensing of gravitational waves}
When a GW passes through a Michelson interferometer, it curves the surrounding spacetime, altering the apparent length of the interferometer arms. The alteration of the apparent arm lengths introduces round-trip phase shifts of light,  $\Delta \Phi_{\X}$, and $\Delta \Phi_{\Y}$ in the $\mathrm{X}$ and $\mathrm{Y}$ arms, respectively. 
The resulting differential phase $\Delta \Phi = \Delta \Phi_{\mathrm{X}} - \Delta \Phi_{\mathrm{Y}}$ is the final observable in conventional interferometers~\cite{ligo2015cqg}.
For our OOM sensor, however, this phase shift serves as an intermediate signal. 
The differential phase $\Delta \Phi$ causes a real-space deformation of the signal optical lattice potential, $U_{\text{lat}}$. 
For a CW, this lattice deformation acts on the $p$-orbital atoms as a coherent, oscillating pseudo-magnetic field, whose strength is given by $B_x = \eta U_0 \Delta \Phi$. 
Here, $\eta\in (0,1)$ is a dimensionless efficiency factor dependent on the primary lattice $V_{\text{lat}}$~(\cite{supp}, Sec.~S-3.2). 
The depth $U_0$ of the signal lattice is proportional to the signal laser power on the order of mW, which is several orders of magnitude smaller than the probe laser power used by a conventional interferometer. This pseudo-magnetic field drives a precession of the collective atomic pseudo-spin, inducing a rotation angle $\theta$ that constitutes the final measurable signal.

The rigorous theoretical description of the interaction between GWs and our OOM sensor is captured by the effective action (in natural units $\hbar = c = 1$)~\cite{Sabin_2014_BEC_GW,Robbins_2022_BEC_GW}:
\begin{equation}\label{eqn:action}
 S = \int \mathrm{d}^4x \sqrt{-g} \Big[g^{\mu \nu} \partial_\mu \Psi^\dagger \partial_\nu \Psi - V({\vec{x}}, g^{\mu\nu})  |\Psi|^2 \Big] \, .
 \end{equation}
Here $g^{\mu \nu}$ is the curved spacetime metric,  $\Psi$ is the atomic field, and $V = m^2 + 2m\left(V_{\text{lat}} + U_{\text{lat}}(\Phi_{\X,\Y})\right)$ is the total potential including the atomic mass terms, the primary lattice potential $ V_{\text{lat}}$ for trapping atoms, and the signal lattice $ U_{\text{lat}}$ carrying the information of GW. 
Crucially, the lattice-mediated coupling described by the second term in Eq.~\eqref{eqn:action} is the dominant interaction; the first direct coupling term between the GW and the atoms via the spacetime metric is approximately ten orders of magnitude weaker and thus negligible~(\cite{supp}, Sec.~S-3.3).

After considering the quantum effects of the light, the effective quantum Hamiltonian description for the $p$-orbital atoms interacting with GW-induced lattice distortion is: 
\begin{equation}
    \hat{H} = \hat{B}_x \hat{J}_x \,,
\end{equation}
where $\hat{B}_x$ incorporates both the classical GW-induced field $B_x$, and the quantum fluctuations of the light (\cite{supp}, Sec.~S-4.1). 
In our protocol, the orbital polarization $\langle J_y \rangle$ is measured after a certain duration ($T$) of evolution under this Hamiltonian, starting from the prepared orbital squeezed state  (Fig.~\ref{fig:protocol}). 
The orbital rotation angle induced by light is $\hat{\theta} = \int_0^T dt y(t) \hat{B}_x(t)/\hbar$ with $y(t)$ a window function associated with the applied pulse sequences (\cite{supp}, Sec.~S-3.5). The final measurement uncertainty of the readout observable $\langle J_y \rangle$ is limited by two fundamental quantum noise sources:
\begin{equation}
    \langle J_y^2 \rangle = |\langle J_z \rangle|^2 \left((\Delta \theta)^2_{\text{proj}} + (\Delta \theta)^2_{\text{photon}}\right)\,,\label{eqn:Jy2Noise}
\end{equation}
where $(\Delta \theta)^2_{\text{proj}} = \xi^2_R/N$ accounts for the orbital projection noise in the initial atomic state, and $(\Delta \theta)^2_{\text{photon}} = \langle \hat{\theta}^2 \rangle$ represents the photonic quantum noise from the dark port input state---composed of photonic shot noise and radiation pressure noise acting on the test masses (\cite{supp}, Sec.~S-4.1). In a conventional interferometer, it is the fundamental trade-off between these two photonic noise that defines the SQL.

 A key innovation of our sensor is how it circumvents this standard noise trade-off. The ability to surpass the SQL stems from two features. 
First, photonic shot noise is intrinsically suppressed because the atomic ensemble probes the phase of the optical lattice formed by counter-propagating fields in the interferometer arms. In this configuration, quantum phase fluctuations from the outgoing and the reflected arm fields partially cancel due to the common-mode rejection, in contrast to a conventional dark-port measurement~\cite{pitkin2011gravitational,danilishin2012quantum}. This suppression breaks the conventional requirement for high laser power and thus minimizes radiation pressure noise.
Second, with the optical noise floor lowered, the atomic ensemble effectively acts as a quantum amplifier. It transduces the weak optical signal into a large, collective atomic rotation, which is then measured with high signal-to-noise ratio (SNR) enhanced by orbital squeezing. While this process introduces atomic projection noise, our full analysis (\cite{supp}, Sec.~S-4.2) shows that this noise trade-off favors a significant net improvement in sensitivity.

The quantum-noise-limited performance of our OOM sensor, presented in Fig.~\ref{fig:sensitivity}\textbf{a}, is most advantageous at low frequencies ($f < f_c \simeq 1.87\mathrm{Hz}$), where the sensitivity is described by:
\begin{equation}\label{eqn:sensitivity}
      S_{\text{OOM}}(f)  = S_{\text{SQL}}(f) C_{\text{atom}}(f,T) \frac{ (\Delta \theta)_{\text{proj}}}{\eta \sqrt{U_0 T}}  \,.
\end{equation} 
Here, $C_{\text{atom}}(f,T)$ is a coefficient dependent solely on the detecting protocol details, and  $S_{\text{SQL}}(f) = 8\hbar/[ML^2(2\pi f)^2]$ represents the SQL of a conventional interferometric GW detector with test mass $M$ and arm length $L$. Our sensor surpasses the SQL because orbital squeezing 
suppresses the atomic projection noise, $(\Delta \theta)_{\text{proj}}$, which ensures the prefactor in Eq.~\eqref{eqn:sensitivity} is less than one. 
As shown in the Supplementary Materials (Fig. S13), without orbital squeezing, the sensor would perform worse than the SQL.

For $f > f_c$, the sensitivity scales worse than a conventional detector due to the limited amount of atomic integration time (\cite{supp}, Sec.~S4.2).

\paragraph{Ultralight dark matter detection}
Like other interferometric gravitational-wave detectors~\cite{zhao2018PRL,ligo2020PRD,ligo2022PRD,2021PRDImproved}, our orbital optomechanical sensor is also a powerful probe for such searches. The ultralight dark photon dark matter $A^\mu$ is expected to couple to normal matter current $J^\mu_\mathrm{D
}$, where the charge $D$ can be the baryon number ($B$) or baryon-minus-lepton number ($B-L$) (\cite{supp}, Sec. S-6.1). 
The Lagrangian describing this coupling is given by (in natural units $\hbar = c = \epsilon_0 = 1$):
\begin{equation}\label{eqn:DM_Lagrangian}
    \mathcal{L} = -\frac{1}{4}F^{\mu \nu} F_{\mu \nu} + \frac{1}{2}m_A^2 A^{\mu} A_{\mu} - eg_\mathrm{D} J^\mu_\mathrm{D} A_\mu \,,
\end{equation}

\noindent\ with $F_{\mu \nu} = \partial_\mu A_\nu - \partial_\nu A_\mu$ the field strength tensor of dark photon, $m_A$ the dark photon mass, and $g_\mathrm{D}$ the coupling constant normalized to the electromagnetic coupling $e$. This coupling also causes oscillations of the apparent length of the arms, altering the differential phase $\Delta \Phi$, which can be detected by our sensor (\cite{supp}, Sec. S-6.2). The upper limit on $g_\mathrm{D}$ can therefore be calculated from the GW strain sensitivity, which is shown in Fig.~\ref{fig:sensitivity}\textbf{b}. Our sensor predicts 34 dB of magnitude (a factor of 50) improvement on the detectable upper limits of the $U(1)_\mathrm{D}$ coupling near $m_A \sim 5 \times 10^{-16} \mathrm{eV/c^2}$, surpassing the current best constraints achieved by the MICROSCOPE experiment~\cite{MICROSCOPE1,MICROSCOPE2,MICROSCOPE3,MICROSCOPE4}.

\paragraph{Discussion} 
We have proposed an OOM sensor that leverages a novel noise trade-off—suppressing photonic shot noise with a counter-propagating readout while minimizing atomic projection noise via orbital squeezing—to achieve sensitivity below the SQL and reduce the required laser power. This scheme not only provides a new path for detecting elusive gravitational waves and ultralight dark matter, but could also advance other interferometry-based quantum technologies such as optical gyroscopes in inertia navigation and Fourier-transform spectroscopy in chemical analysis. 

The photonic quantum noise $(\Delta \theta)^2_{\text{photon}}$, which represents the probe light's back-action, is fully incorporated in our model and poses the fundamental sensitivity limit. While our scheme already suppresses this noise to sub-SQL levels, it could be further mitigated by injecting photonic squeezed states~\cite{LIGOsqueezing1,ma2017proposal,LIGOsqueezing2,LIGOsqueezing3} or through advanced quantum control techniques.
Realizing this quantum-limited sensitivity requires overcoming significant classical noise, a challenge addressed in our full analysis (\cite{supp}, Sec.~5). The analysis confirms that technical requirements on laser stability and atomic readout are met by state-of-the-art systems. For a ground-based scheme, environmental noise targets align with goals for next-generation observatories ~\cite{ET_design,ET_ScienceCase,ETandCE,CE_horizon}. Crucially, for a space-based scheme, the stability demands are less stringent than performance already demonstrated by the LISA Pathfinder mission~\cite{2016LISA_pathfinder}.
Additionally, although our sensor is designed for low power to preserve the stability of the atomic sample, future architectures could  employ arrays of multiple atomic ensembles. Such an approach would enable the use of higher total laser power, potentially enhancing sensitivity by a factor of $1/\sqrt{M}$ with $M$ the number of ensembles, though this may introduce new sources of technical noise. 

\paragraph{Acknowledgments}The authors wish to acknowledge helpful discussions with 
Rainer Blatt, Peter Zoller, Dan Stamper-Kurn, Bo Yan, Gordon Baym, Daniel Vanmaekelbergh and Cristiane Morais Smith.
This work is supported by 
National Key Research and Development Program of China (Grant No.~2021YFA1400900), 
Innovation Program for Quantum Science and Technology of China (Grant No.~2024ZD0300100), 
National Natural Science Foundation of China (Grant No.~11934002), 
and Shanghai Municipal Science and Technology (Grant No.~2019SHZDZX01, 24DP2600100) 
(X.Y., X.L.),  
and by AFOSR Grant No.~FA9550-23-1–0598, MURI-ARO Grant No.~W911NF-17-1-0323 through UC Santa Barbara and PQI Community Collaboration Award (W.V.L.). 
W.V.L. expresses gratitude for the visit fellowship at the IQOQI Innsbruck, Austrian Academy of Sciences, where this manuscript was completed.    

\paragraph{Data availability}The data that support the findings of this article are openly available~\cite{Yu_OOM-Sensor_Code_2025}.

\bibliographystyle{apsrev4-2}
\bibliography{main}

\end{document}


\baselineskip 24pt
\title{Supplementary Materials for \\ \textbf{\Large ``Squeezing atomic $p$-orbital condensates for detecting gravitational waves"}}
\author{Xinyang Yu, W. Vincent Liu, Xiaopeng Li$^\ast$\and
\small$^\ast$Corresponding author. Email: xiaopeng\_li@fudan.edu.cn}
\date{}
\maketitle

\tableofcontents
\newpage

\nocite{LIGO2016_GW150914,LIGO2017_GW170817,Abbott_2021,CW2017LRR,bailes2021gravitational,ferreira2021ultra,Yunes-Miller-Yagi:22rev,Mosbech2023Gwave-DM,Bertone-Tait:18rev,Bertone:24:NobelSym,Vermeulen:21:ScalarDM,ligo2022PRD,LIGO_vectordarkmatter:24arx,2018_Treutlein_RMP,wirth2011evidence,2011_Lewenstein_NP,li2016physics,2020_Rey_OrbitalGate,2021_Zhou_PRL,2021_Zhou_OrbitalGate_PRA,2021_Xu_Nature,2023_Thywissen_Nature,supp,ma2011quantum,2022PrlTACT,luo2024hamiltonianengineeringcollectivexyz,ligo2015cqg,2019LISA,EotWash1,EotWash2,MICROSCOPE1,MICROSCOPE2,MICROSCOPE3,MICROSCOPE4,kock2016orbital,Hall1998a,Hall1998b,Shin2004DoubleWell,schumm2005matter,leggett2001bose,dalton2012two,Cappellaro2017rev,Wineland1992,Wineland1994,xin2025fast,paul2013formation,Sabin_2014_BEC_GW,Robbins_2022_BEC_GW,pitkin2011gravitational,danilishin2012quantum,zhao2018PRL,ligo2020PRD,2021PRDImproved,pitkin2011gravitational,LIGOsqueezing1,ma2017proposal,LIGOsqueezing2,LIGOsqueezing3,ET_design,ET_ScienceCase,ETandCE,CE_horizon,2016LISA_pathfinder,Yu_OOM-Sensor_Code_2025}
\section{ Multi-orbital BEC in Bipartite Square Lattice}\label{Ssec:pbec}

Controlling Bose-Einstein condensates (BECs) in the $p$-orbital bands of optical lattices is a key element of our proposed orbital optomechanical (OOM) sensor. This approach builds upon nearly two decades of theoretical and experimental advances that have established high-fidelity control over such systems~\cite{2011_Lewenstein_NP,li2016physics,kock2016orbital}. Multi-orbital BECs have been successfully realized in various lattice geometries, from quasi-1D lattices~\cite{2007_Bloch_PRL,2018_Xiaoji_PRL} to the 2D checkerboard~\cite{wirth2011evidence,2021_Hachmann_PRL,2021_Vargas_PRL} and hexagonal lattices~\cite{2021_Zhou_PRL,2021_Xu_Nature}. These experimental advances followed early theoretical proposals ~\cite{2005_Isacsson_PRA,2006_Kuklov_PRL,2006_Liu_PRA}.

\begin{figure}[htp]
    \centering
    \includegraphics[width=\linewidth]{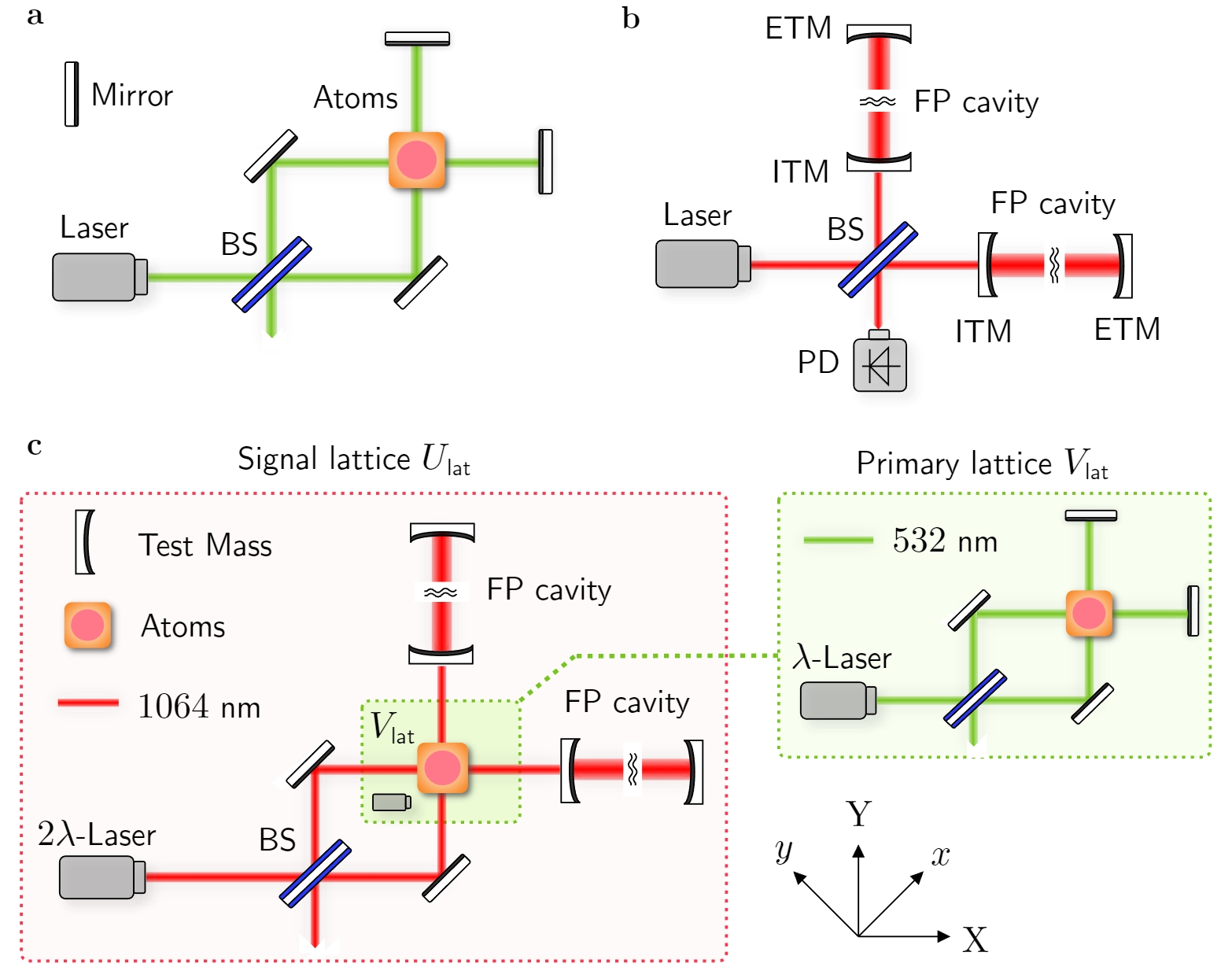}
    \caption{\textbf{Comparison of optical setups.} \textbf{a}, Sketch of the optical lattice setup~\cite{wirth2011evidence} where two optical standing waves are superimposed. \textbf{b}, A standard gravitational-wave observatory such as LIGO~\cite{ligo2011gravitational}, based on a Fabry-Perot-Michelson interferometer. Advanced observatories incorporate additional techniques like power and signal recycling to enhance sensitivity. The setup proposed in this work uses the core Fabry-Perot-Michelson configuration without advanced optical techniques..
    \textbf{c}, The proposed orbital optomechanical sensor, , which integrates a cold-atom system with a large-scale interferometer. The primary lattice $V_{\mathrm{lat}}$ confining the atoms is generated by the $\lambda$-laser (532 nm), while the signal lattice $U_{\mathrm{lat}}$ for signal detection is generated by $2\lambda$-laser (1064 nm). The acronyms used are BS = beam splitter, ITM = input test mass, ETM = end test mass, PD = photon detector.}
    \label{Sfig:MI}
\end{figure}

The methods employed in our work are directly adapted from the well-established Hamburg checkerboard optical lattice experiments~\cite{wirth2011evidence,kock2016orbital}. Consequently, the multi-orbital BEC at the heart of our proposed sensor is feasible with current technology and requires minimal modification to existing setups. This section reviews the fundamental properties of $p$-orbital BECs in this lattice configuration, which provides the foundation for our squeezing and detection protocols. The proposed experimental setup of our orbital optomechanical sensor is shown in Fig.~\textcolor{magenta}{1} in the main text and Fig.~\ref{Sfig:MI}. 

\subsection{ Optical Lattice Setup}\label{Ssubsec:setup}
The primary lattice potential \( V_{\mathrm{lat}} \), used for trapping \(p\)-orbital bosons, is generated by superimposing counter-propagating laser beams with a wavelength of \( \lambda = 532\mathrm{nm} \) along the \( \mathrm{X,Y} \) directions (Fig.~\ref{Sfig:MI}). The potential is expressed as:
\begin{equation}\label{Seqn:Vlat in flat}
    V_{\mathrm{lat}}(\mathbf{r}) =-\frac{V_{0}}{4} \abs{\left[(\mathbf{e}_{z}\cos \alpha +\mathbf{e}_{\Y}\sin \alpha ) e^{i\K\X} +r \mathbf{e}_{z} e^{-i\K\X}\right] + t  e^{i\phi }\mathbf{e}_{z}\left( e^{i\K\Y} +r e^{-i\K\Y}\right) }^{2}\,.
\end{equation}
Here, $\K=2\pi/\lambda$ is the wave-vector of 
the laser with wavelength $\lambda$; this laser is hereafter referred to as the $\lambda$-laser in the following. The parameters $t<1$ and $r<1$ account for imperfect optical transmission and reflection efficiencies,respectively. The angle $\alpha$ is used to adjust the anisotropy between $p_{x,y}$ orbitals by rotating the polarizations of the light. When $r = t =\cos \alpha =1$, we return to the case of perfect optics and the optical lattice has an exact $D_4$ point group symmetry:
\begin{align}
V_{\mathrm{lat}}(\X,\Y) & =-V_{0}\left[\cos^{2} \K\X+\cos^{2} \K\Y +2\cos \phi \cos \K\X\cos \K\Y\right]\,. \label{Seqn:V} 
\end{align}

In terms of the rotated coordinate $x=\frac{1}{\sqrt{2}}(\X+\Y)$ and $y=\frac{1}{\sqrt{2}} (\Y-\X)$, Eq.~\eqref{Seqn:V}  takes the form of:
\begin{equation}\label{Seqn:Vlat}
    V_{\mathrm{lat}}(x,y) = -V_{0} \left[ 1+ \cos kx \cos ky + \cos\phi (\cos kx + \cos ky)\right]\,,
\end{equation}

where $k=\sqrt{2}\K$ is the magnitude of reciprocal lattice vectors. (The photon recoil energy is still defined as $E_R = \hbar^2\K^2/2m$ instead of $\hbar^2k^2/2m$, where $m$ is the atomic mass.) The lattice vectors of $V_{\mathrm{lat}}$ in $(x,y)$ coordinate are $\mathbf{a}_1=a(1,0)$ and $\mathbf{a}_2=a(0,1)$ with $a=2\pi/k$ (Fig.~\ref{Sfig:band structure}). Its two local minima are $V_A = -2V_0(1+\cos \phi)$ at $\mathbf{r}_A=(0,0)$ and $V_B = -2V_0(1-\cos \phi)$ at $\mathbf{r}_B=(a/2,a/2)$. Hereafter, we refer to $V_{\mathrm{lat}}$ as the primary lattice. A much weaker signal lattice, $U_{\mathrm{lat}}$, will be superimposed for signal detection (Sec.~\ref{Ssec:GW}).

\subsection{ Band Structure}\label{Ssubsec:band}
The band structure of $V_{\mathrm{lat}}$ is obtained by diagonalizing the non-interacting Hamiltonian $\hat{H_0}$:
\begin{equation}\label{Seqn:H0}
    \hat{H_0} = \int \d^3\mathbf{r} \ \hat{\Psi}_{3\mathrm{d}}^{\dagger}(\mathbf{r}) \left(-\frac{\hbar^2}{2M}\nabla^2 + V_{\mathrm{lat}}(\mathbf{r}) \right) \hat{\Psi}_{3\mathrm{d}}(\mathbf{r})\,,
\end{equation}
\begin{figure}[htp]
    \centering
    \includegraphics[width=\linewidth]{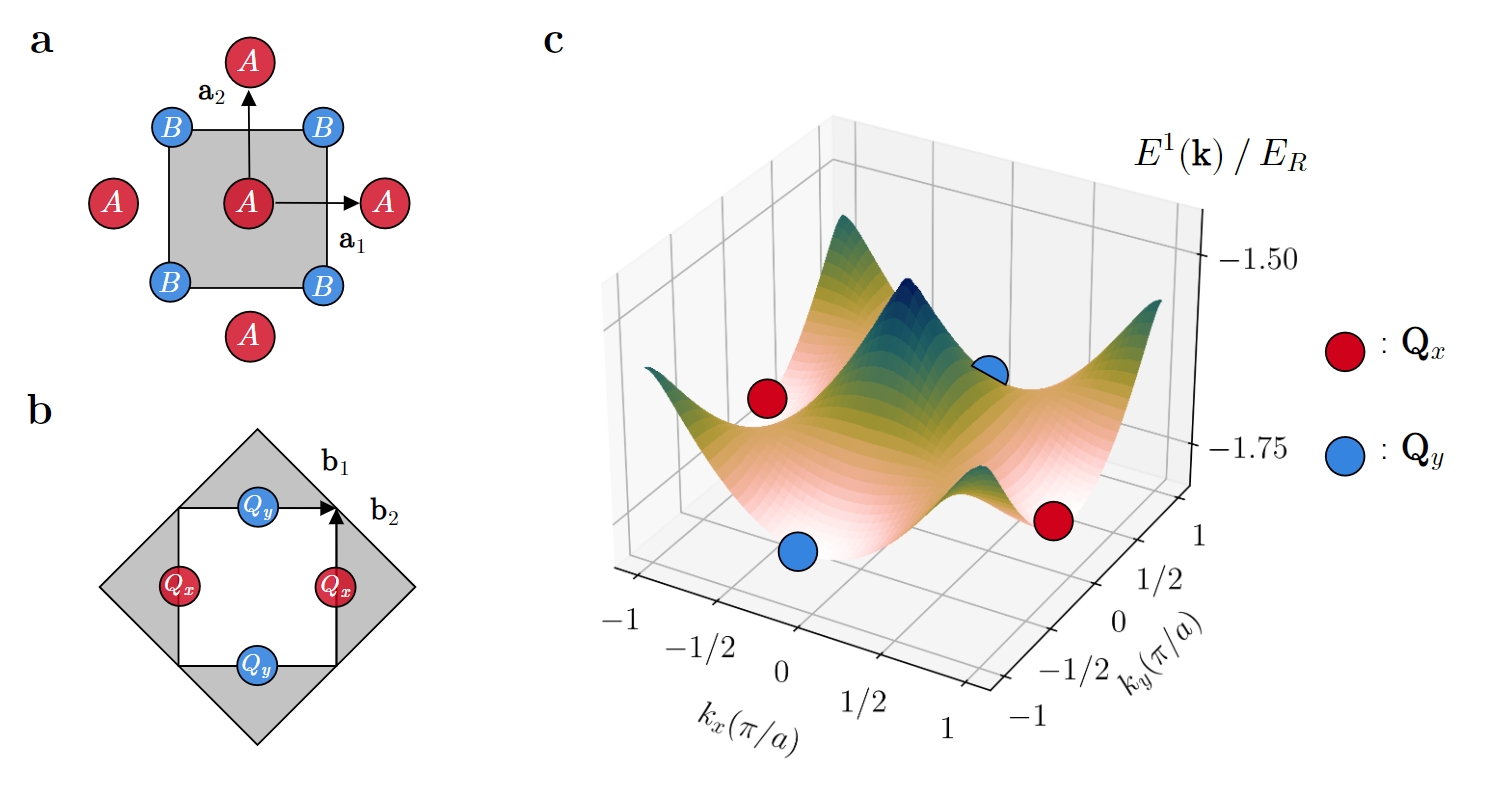}
    \caption{\fontsize{12pt}{14pt}\selectfont \textbf{a}, Real-space configuration of the lattice potential $V_{\mathrm{lat}}$ in the $(x,y)$ coordinate, with local minima at $A(0,0)$ and $B(a/2,a/2)$. \textbf{b}, The first Brillouin Zone of $V_{\mathrm{lat}}$. $\mathbf{Q}_{x,y}$ represent the  two degenerate band minima in the first excited band. \textbf{c}, The first excited band spectrum $E^{1}(\mathbf{k})$ obtained by numerically diagonalizing the Hamiltonian in Eq.~\eqref{Seqn:H0}. The parameters are set as $V_0 = 2E_R$ and $\phi = 0.4\pi$. } 
    \label{Sfig:band structure}
\end{figure}
where $\hat{\Psi}_{\mathrm{3d}}(\mathbf{r}) = \hat{\Psi}(\mathbf{r})\phi_0(z)$ is the bosonic field operators, and $\phi_0(z)$ is the ground state of the harmonic oscillator with circular frequency $\omega_z$ along the $z$ direction. 
The operator $\hat{\Psi}(\mathbf{r})$ is expanded in the Bloch states basis as:
\begin{equation}\label{Seqn:Field}
    \hat{\Psi}(\mathbf{r}) = \sum_{\mathbf{k}\in 1\mathrm{BZ}} \sum_{n} \psi_{n,\mathbf{k}}(\mathbf{r})\hat{b}_{n}(\mathbf{k}) \,.
\end{equation}
The dispersion relation of the first excited band is shown in Fig.~\ref{Sfig:band structure}. The first excited band possesses two degenerate energy minima with time-reversal (TR) invariant quasi-momentum $\mathbf{Q}_x = (\pi,0)/a$ and $\mathbf{Q}_y = (0,\pi)/a$~\cite{li2016physics}. A $p$-orbital BEC refers to a condensate formed in one or both of these p-band minima.

The ground state manifold of $N$ non-interacting $p$-orbital bosons are spanned by:
\begin{equation}
    |N_x,N_y\rangle = \frac{[\hat{b}_x^{\dagger}(\mathbf{Q}_x)]^{N_x} [\hat{b}_y^{\dagger}(\mathbf{Q}_y)]^{N_y}}{\sqrt{N_x!N_y!}}|0\rangle \ \mathrm{with} \ N_x+N_y = N \,.
\end{equation}
In experiments, we are able to prepare normal $p$-BEC states $|p_x\rangle^{\otimes N} = |N,0\rangle$ and $|p_y\rangle^{\otimes N} = |0,N\rangle$ as well as chirality-breaking $p_{\pm}$ states $(|p_x\rangle \pm i|p_y\rangle)^{\otimes N}$~\cite{wirth2011evidence,kock2016orbital}
by tuning the interaction strength between atoms through Feshbach resonance techniques.

\subsection{ Two-Mode Approximation}
Inspired by previous studies on interference of fragmented BEC of two internal states~\cite{Hall1998a,Hall1998b} and in double wells~\cite{Shin2004DoubleWell,schumm2005matter}, we propose the two-mode approximation~\cite{leggett2001bose,dalton2012two} of $p$-orbital bosons. This model describes the evolution of bosons from the initial $p$-BEC state as:
\begin{equation}\label{Seqn:TMA}
    \hat{\Psi}(\mathbf{r}) = \psi_{x}(\mathbf{r})\hat{p}_x + \psi_{y}(\mathbf{r})\hat{p}_y \,.
\end{equation}
Here, we use the shorthand notation $\psi_{\alpha}(\mathbf{r})$ for the single particle wavefunctions $\psi_{p_{\alpha},\mathbf{Q}_{\alpha}}(\mathbf{r})$ and $\hat{p}_{\alpha}$ for the annihilation operators $\hat{b}_{p_{\alpha}}(\mathbf{Q}_{\alpha})$ from Eq.~\eqref{Seqn:Field}. These operators satisfy the commutation relation $[\hat{p}_{\alpha},\hat{p}_{\beta}^{\dagger}]=\delta_{\alpha,\beta}$ for $\alpha,\beta \in \{x,y\}$.

In the two-mode approximation, the single-body Hamiltonian is given by:
\begin{equation}
    \hat{H}_0 = \epsilon(\hat{n}_x + \hat{n}_y) \,,
\end{equation}
with $\epsilon = E^{1}(\mathbf{Q}_{x,y})$ (Fig.~\ref{Sfig:band structure}) and $\hat{n}_{\alpha}$ the number operator of its corresponding mode. 

The contact interaction $V_\mathrm{int}(\mathbf{r}-\mathbf{r'})=g_{\mathrm{3d}}\delta(\mathbf{r}-\mathbf{r'})$ between bosons  is described by:
\begin{equation}\label{Seqn:Hint}
        \hat{H}_{\mathrm{int}} = \frac{g_{\mathrm{3d}}}{2}\int \d^3 \mathbf{r} \hat{\Psi}_{\mathrm{3d}}^{\dagger}(\mathbf{r}) \hat{\Psi}_{\mathrm{3d}}^{\dagger}(\mathbf{r}) \hat{\Psi}_{\mathrm{3d}}(\mathbf{r}) \hat{\Psi}_{\mathrm{3d}}(\mathbf{r}) = \sum_{\alpha_{1 \sim 4} = x,y}U_{\alpha_1\alpha_2\alpha_3\alpha_4} \hat{p}_{\alpha_1}^{\dagger} \hat{p}_{\alpha_2}^{\dagger} \hat{p}_{\alpha_3} \hat{p}_{\alpha_4} \,,
\end{equation}
where $g_{3d}$ is given by the 3d scattering length $a_{\mathrm{3d}}$ as $g_{\mathrm{3d}}=4\pi\hbar^2 a_{\mathrm{3d}}/M$. The interaction coefficients $U_{\alpha_1\alpha_2\alpha_3\alpha_4}$ are given by:
\begin{equation}
    U_{\alpha_1\alpha_2\alpha_3\alpha_4} = \frac{g_{\mathrm{2d}}}{2}\int \d^2 \mathbf{r} \psi_{\alpha_1}^{*}(\mathbf{r}) \psi_{\alpha_2}^{*}(\mathbf{r}) \psi_{\alpha_3}(\mathbf{r}) \psi_{\alpha_4}(\mathbf{r}) \,,
\end{equation}
with $g_{\mathrm{2d}} = g_{\mathrm{3d}} \int |\phi_0(z)|^4 dz$.
The $D_4$ lattice symmetry restricts that the only non-vanishing (up to permutation of indices) coefficients are:
\begin{align}
    U_1 = U_{xxxx} = U_{yyyy} &= \frac{g_{\mathrm{2d}}}{2}\int \d^2 \mathbf{r} |\psi_{x}(\mathbf{r})|^4 \label{Seqn:U1} \,,\\
    U_2 = U_{xyyx} &= \frac{g_{\mathrm{2d}}}{2}\int \d^2 \mathbf{r} |\psi_{x}(\mathbf{r})|^2|\psi_{y}(\mathbf{r})|^2 \,,\\
    U_3 = U_{xxyy} = U_{yyxx}^{*} &= \frac{g_{\mathrm{2d}}}{2}\int \d^2 \mathbf{r} \left(\psi_{x}^{*}(\mathbf{r})\right)^2 \left(\psi_{y}(\mathbf{r})\right)^2 \label{Seqn:U3} \,.
\end{align}
In the harmonic limit $V_0/E_R \gg 1$, $U_1 = 3U_2 = 3U_3$, while in a general case the ratio $U_1/U_3$ varies with the parameters $(V_0,\phi)$, see Fig.~\textcolor{magenta}{3a} in main text. It is worth noting that $U_2 \equiv U_3$ exactly for our checkerboard lattice because $\psi_{x,y}(\mathbf{r})$ are TR-invariant and thus real-valued. This crucial identity simplifies the interaction Hamiltonian, paving the way for the desired spin squeezing dynamics. Furthermore, $U_1\geq U_2$ because $U_1 - U_2 \propto \int \d^2 \mathbf{r} (|\psi_{x}(\mathbf{r})|^2 - |\psi_{y}(\mathbf{r})|^2)^2 \geq 0$.

The interaction Hamiltonian thus simplifies to three terms:
\begin{equation}
    \hat{H}_{\mathrm{int}} = \frac{U_1}{2} \left(\hat{n}_x(\hat{n}_x-1) + \hat{n}_y(\hat{n}_y-1)\right) + 2U_2 \hat{n}_x \hat{n}_y + \frac{U_3}{2} \left(\hat{p}_{x}^{\dagger} \hat{p}_{x}^{\dagger} \hat{p}_{y} \hat{p}_{y} + \mathrm{h.c.} \right) \,.
    \label{Seq:HOrbital}
\end{equation} 
The presence of the pair-hopping term $\hat{p}_{x}^{\dagger} \hat{p}_{x}^{\dagger} \hat{p}_{y} \hat{p}_{y} + \mathrm{h.c.}$, is the essential physical mechanism that allows for the engineering of a TACT Hamiltonian for optimal squeezing, as we detail in Sec.~\ref{Ssec:squeezing}.

\medskip 
\section{ Orbital Squeezing of \texorpdfstring{$p$}{}-orbital Bosons}\label{Ssec:squeezing}
This section details how the intrinsic interactions within the $p$-orbital condensate can be harnessed to generate highly squeezed quantum states. We model the system as a pseudo-spin ensemble and analyze its dynamics under the interaction Hamiltonian $\hat{H}_{\mathrm{int}}$.
\begin{figure}[htp]
    \centering
    \includegraphics[width=\linewidth]{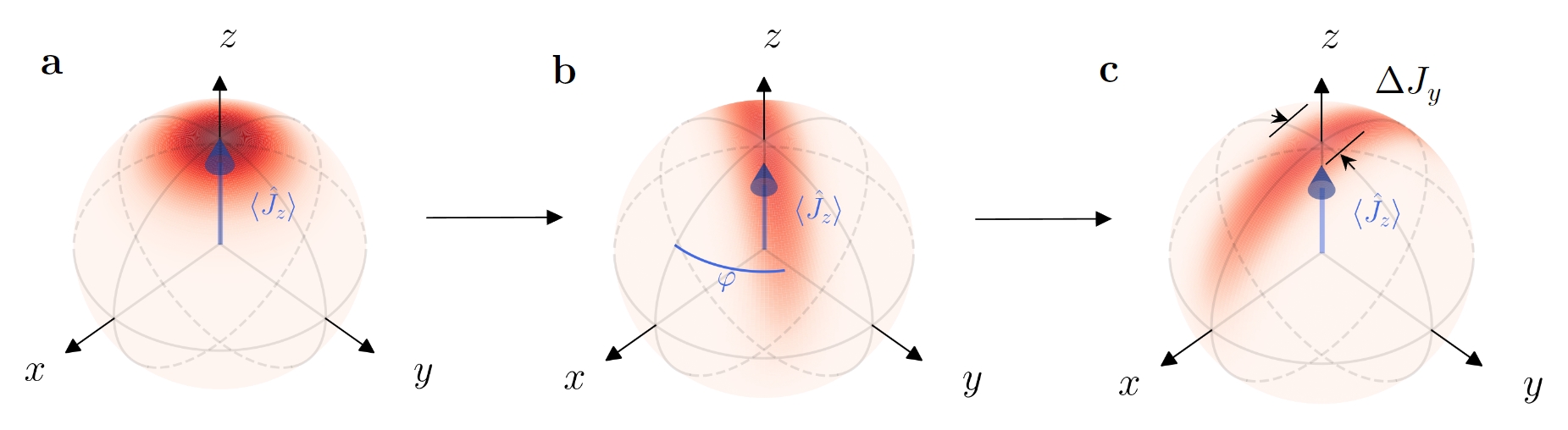}
    \caption{ Husimi-Q function representation of pseudo-spin states. \textbf{a}, The initial pseudo-spin coherent state, corresponding to $|J,J_z=J\rangle$. \textbf{b}, An orbital-squeezed state resulting from the evolution of $|J,J\rangle$ under the Hamiltonian Eq.~\eqref{Seqn:SpinHamiltonian}, exhibiting a squeezing angle $\varphi$ relative to the $z$-axis. \textbf{c}, The orbital-squeezed state following a rotation around the $z$-axis by $\varphi$, enhancing sensitivity to rotations about the $x$-axis. The parameters are $U_1=2.5U_2=2.5U_3$ 
     (Eq.~\eqref{Seq:HOrbital}) 
    and $N=20$.}
    \label{Sfig:HusimiQ}
\end{figure}
\subsection{ Schwinger Representation of Angular Momentum}\label{Ssubsec:Schwinger Rep}
Using the Schwinger representation~\cite{ma2011quantum}, the two orbital modes ($p_x, p_y$) can be mapped onto a large pseudo-spin system with total angular momentum $J=N/2$:
\begin{align}
    \hat{J}_z &= \frac{1}{2}\left(\hat{n}_{x}-\hat{n}_y \right) \,, \\
    \hat{J}_{+} &= \hat{J}_x + i\hat{J}_y = \hat{p}^\dagger_{x} \hat{p}_y \,,  \ \hat{J}_{-}=\hat{p}^\dagger_{y} \hat{p}_x \,,
\end{align}
Here, the $p_x$($p_y$) represents the pseudo-spin up (down) state. The operator $\hat{J}_z$ measures the population difference between $p_x$ and $p_y$ modes. 
The initial p-BEC state is a pseudo-spin coherent state $|\theta,\varphi\rangle$ (Fig. S3). For instance, a condensate fully in the $p_x$ mode corresponds to the state $|J=\frac{N}{2},J_z=\frac{N}{2}\rangle$, which is equivalent to the coherent state $|\theta=0,\varphi\rangle$. A pure $p_y$ condensate corresponds to $|\theta=\pi,\varphi\rangle$, and a pure $p_+$ condensate corresponds to $|\theta=\pi/2,\varphi=\pi/2\rangle$.
Without loss of generality, we will always assume the initial state is prepared in the $p_x$ condensate $|p_x\rangle^{\otimes N}$.

The full pseudo-spin (ps) Hamiltonian, denoted as $\hat{H}_{\mathrm{ps}} = \hat{H}_0 + \hat{H}_{\mathrm{int}}$, is rewritten in terms of angular momentum operators as: 
\begin{align}
    \hat{H}_{\mathrm{ps}} &= \epsilon \hat{N} + \frac{U_1}{2} \left(\hat{n}_x^2 + \hat{n}_y^2 -\hat{N}\right) + 2U_2 \hat{n}_x \hat{n}_y + \frac{U_3}{2} \left(\hat{p}_{x}^{\dagger} \hat{p}_{x}^{\dagger} \hat{p}_{y} \hat{p}_{y} + \mathrm{h.c.} \right) \,, \notag \\
    &= (\epsilon-\frac{U_1}{2})\hat{N} + \frac{1}{4}(U_1+2U_2)\hat{N}^2 + (U_1-2U_2)\hat{J}_z^2 + U_3 (\hat{J}_x^2 -\hat{J}_y^2) \,.
\end{align}
Since the total particle number $\hat{N} = \hat{n}_{x}+\hat{n}_y$ is a conserved quantity, $\hat{J}^2 = \hat{J}_x^2 + \hat{J}_y^2 + \hat{J}_z^2 = \frac{\hat{N}}{2}(\frac{\hat{N}}{2}+1)$ is also conserved, i.e., the state always evolve in the $J=N/2$ subspace. The full Hamiltonian (up to a constant energy shift involving only $\hat{N}$) then takes the form:
\begin{equation}\label{Seqn:SpinHamiltonian}
    \hat{H}_{\mathrm{ps}} = (3U_3 - U_1) \hat{J}_x^2 - (U_1- U_3) \hat{J}_y^2 \,. 
\end{equation}
which reproduces Eq.~\textcolor{magenta}{(2)} of the main text.

This is a form of the Lipkin-Meshkov-Glick (LMG) model~\cite{Ma2009PRA,LMG1}, which is known to generate spin squeezing. Remarkably, due to the time-reversal symmetry of the checkerboard lattice ($U_2 = U_3$), this Hamiltonian contains only $\hat{J}_x^2$ and $\hat{J}_y^2$ terms. By tuning the lattice parameters $(V_0, \phi)$, which control the ratio $U_1/U_3$, the system can be engineered to manifest different squeezing dynamics. Notably, when $U_1=2U_2=2U_3$, the Hamiltonian naturally reduces to the two-axis-counter-twisting (TACT) form, $\hat{H}_{\mathrm{TACT}} \propto (\hat{J}_x^2 - \hat{J}_y^2)$, which is known to be one of the most effective methods for generating entanglement and reaching the Heisenberg limit of squeezing.

\begin{figure}[htp]
    \centering
    \subfloat[\textbf{a}][]{
        \includegraphics[width=0.48\textwidth]{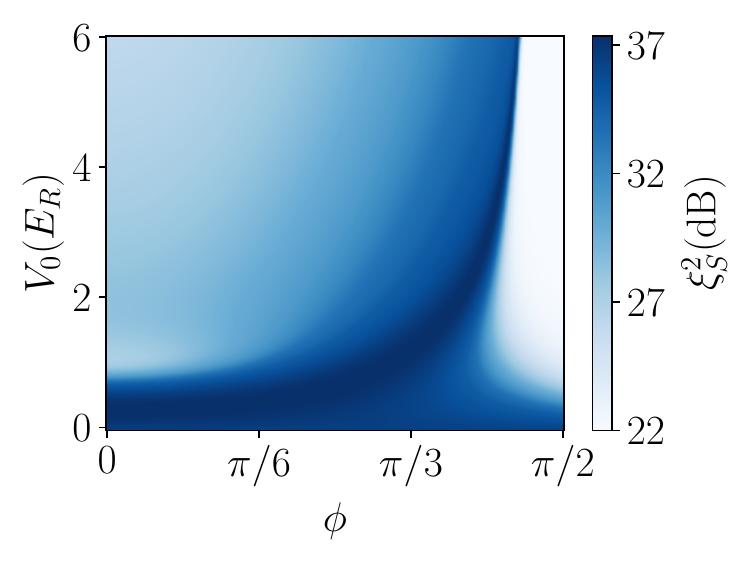}\label{Ssubfig:5a}
    }
    \subfloat[\textbf{b}][]{
        \includegraphics[width=0.47\textwidth]{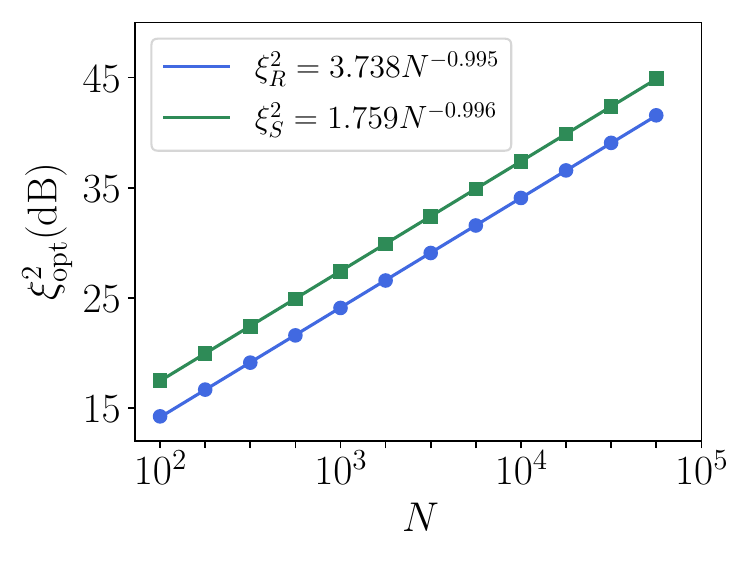}\label{Ssubfig:5b}
    }
    \caption{\fontsize{12pt}{14pt}\selectfont  Squeezing parameter $\xi^2$ of the state $|J,J_z=J\rangle$ under the time evolution of the  Hamiltonian Eq.~\eqref{Seqn:SpinHamiltonian}. \textbf{a}, $\xi_{S}^2$ for a total particle number $N=10^4$. The optimal squeezing parameter is approximately $37$ dB, approaching the Heisenberg limit $\xi^2 \sim 1/N$, which corresponds to a $40$ dB squeezing. \textbf{b}, The scaling of the optimal squeezing parameters with $N$. The parameters used are $V_0 = 6E_R$, $\phi = 0.44857\pi$, and $\omega_z = 20\pi \ \mathrm{rad/s}$. Points represent numerically calculated data, and lines represent fitted curves. }
    \label{Sfig:Xi2}
\end{figure}

\subsection{ Squeezing \texorpdfstring{$p$}{}-orbital Bosons to Heisenberg Limit}
\label{Ssubsec:SS}
We characterize the level of orbital squeezing by using the squeezing parameter $\xi_S$ \cite{Ueda1993SpinSqueezing} or the metrological squeezing parameter $\xi_R$~\cite{Wineland1992,Wineland1994} as:
\begin{align}
    \xi_S^2 &= \frac{(\min \Delta J_{\perp})^2}{N/4} \,, \\
    \xi_R^2 &= \frac{N(\min \Delta J_{\perp})^2}{|\langle \mathbf{J} \rangle|^2} = \left(\frac{N/2}{|\langle \mathbf{J}\rangle|}\right)^2 \xi_S^2 \geq \xi_S^2 \,.
\end{align}
where $(\min \Delta J_{\perp})^2$ is the minimal variance of the spin component perpendicular to the mean spin direction $\langle \mathbf{J}\rangle$. For our system $\langle J_x \rangle = \langle J_y \rangle = 0$, so the mean spin direction is $\mathbf{e}_z$. The correlations $\langle J_x J_y \rangle = \langle J_y J_z \rangle = \langle J_x J_z \rangle$ also vanish. The squeezing effect of different parameters $(V_0,\phi,N)$ is shown in Fig.~\textcolor{magenta}{3} in the main text and Fig.~\ref{Sfig:Xi2}.

\begin{figure}[htp]
    \centering
    \subfloat[\textbf{a}][]{
        \includegraphics[width=0.48\textwidth]{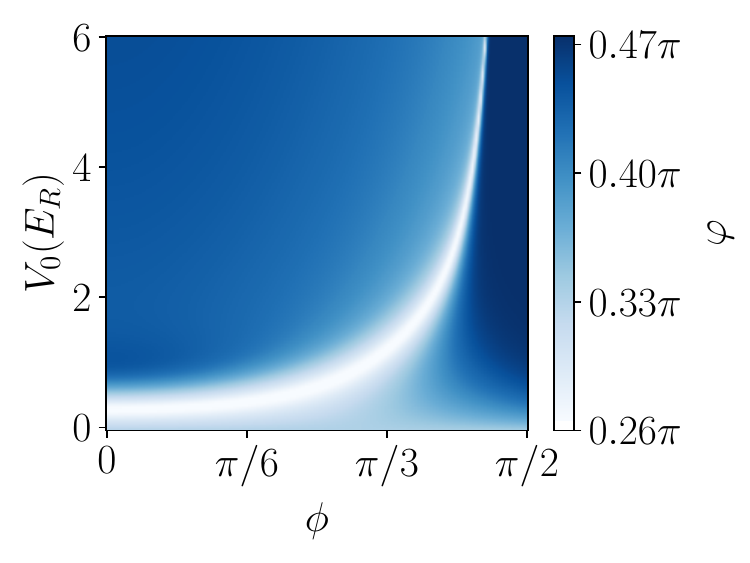}\label{Ssubfig:6a}
    }
    \subfloat[\textbf{b}][]{
        \includegraphics[width=0.47\textwidth]{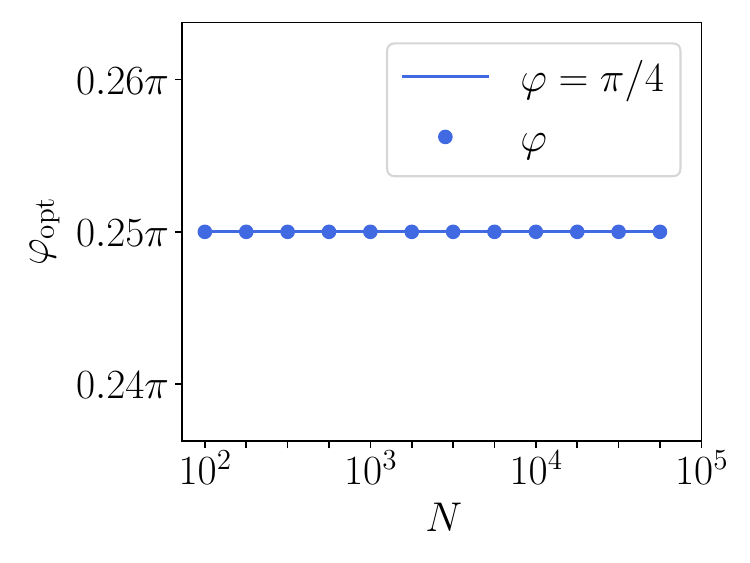}\label{Ssubfig:6b}
    }\\
    \subfloat[\textbf{c}][]{
        \includegraphics[width=0.48\textwidth]{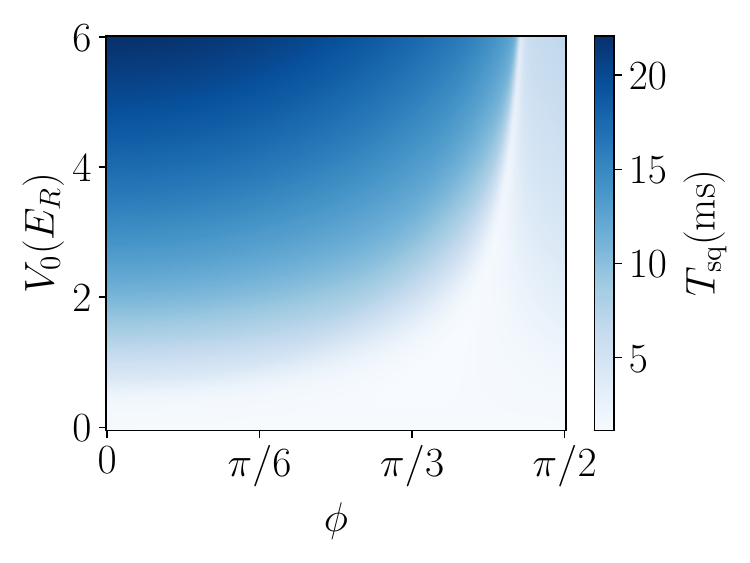}\label{Ssubfig:7a}
    }
    \subfloat[\textbf{d}][]{
        \includegraphics[width=0.47\textwidth]{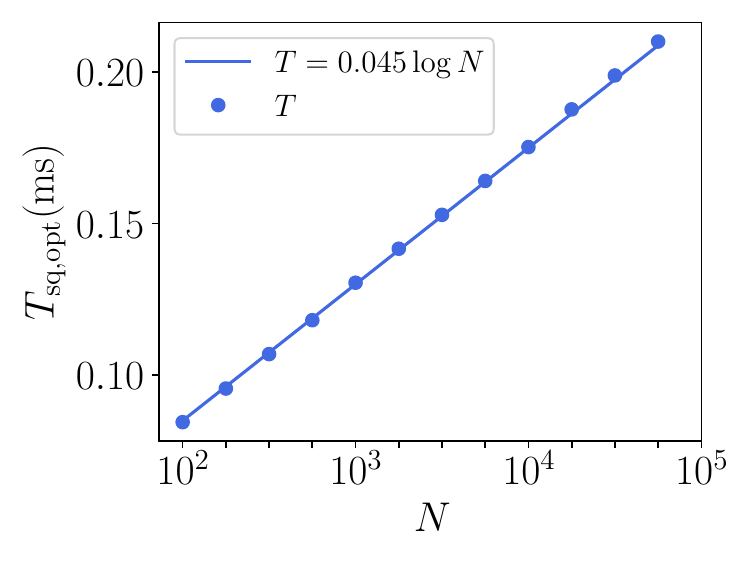}\label{Ssubfig:7b}
    }
    \caption{\fontsize{12pt}{14pt}\selectfont Squeezing angle $\varphi$ and squeezing time $T_{\mathrm{sq}}$ of the spin state $|J,J_z=J\rangle$ under the time evolution of the Hamiltonian in Eq.~\eqref{Seqn:SpinHamiltonian}. \textbf{a}, $\varphi$ as a function of potential parameters with total particle number $N=10^4$. In the OAT limit, $\varphi$ is close to $\pi/2$, while in the TACT limit, $\varphi$ approaches $\pi/4$. \textbf{b}, The scaling of the optimal squeezing angle with $N$. \textbf{c}, $T_{\mathrm{sq}}$ with total particle number $N=10^4$. \textbf{d}, The optimal squeezing time as a function of $N$. The parameters here corresponds to $V_0 = 6E_R, \ \phi=0.44857\pi, \omega_z = 20\pi \ \mathrm{rad/s}.$
   } 
    \label{Sfig:phi}
\end{figure}

The metrological squeezing parameter is associated with phase sensitivity in Ramsey spectroscopy as $\Delta\phi = \xi_R/\sqrt{N}$. 
The state is squeezed for $\xi_R<1$, as its noise beats the standard quantum limit (shot-noise limit) $\Delta \phi < 1/\sqrt{N}$. Conversely, the Heisenberg uncertainty principle imposes a fundamental limit on this squeezing, asserting that $\xi^2_R \geq \xi^2_S \geq 1/N$ and consequently $\Delta \phi \geq 1/N$, establishing the lower bounds for the uncertainty in phase that quantum mechanics allows. The fitted curves in Fig.~\ref{Ssubfig:5b} $\xi_R^2=3.738 N^{-0.995}$ and $\xi_S^2=1.759 N^{-0.996}$ indicate that the squeezing effect generated by the interaction between $p$-orbital bosons achieves the Heisenberg limit.
Comparing Fig.~\textcolor{magenta}{3a} in main text and Fig.~\ref{Ssubfig:5a}, we observe that the optimal squeezing effect is achieved at $U_1=2U_2=2U_3$, where $\hat{H}_{\mathrm{spin}}$ becomes a TACT Hamiltonian. 

Two other quantities crucial for our detecting protocol are the optimal squeezing time $T_{\mathrm{opt}}$ and the optimal squeezing angle $\varphi$. The results are provided in Fig.~\ref{Sfig:phi}.

\begin{figure}[htb]
    \centering
    \subfloat[\textbf{a}][]{
        \includegraphics[width=0.47\textwidth]{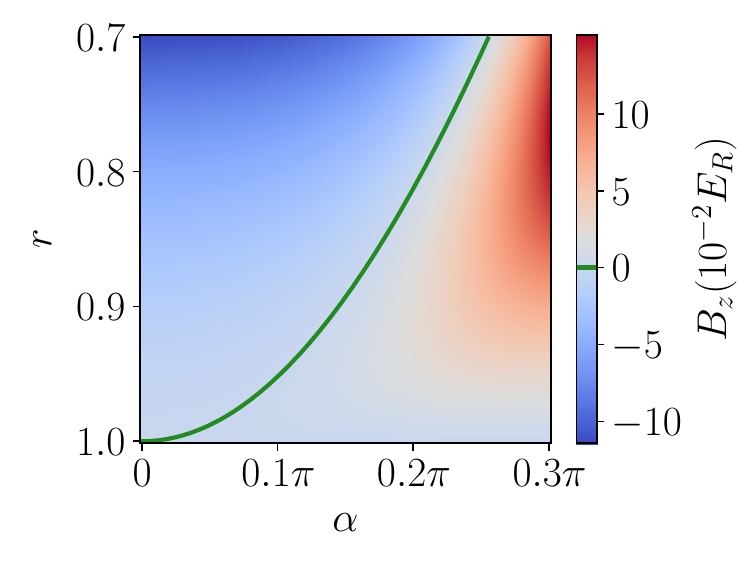}\label{Ssubfig:8a}
    }
    \subfloat[\textbf{b}][]{
        \includegraphics[width=0.47\textwidth]{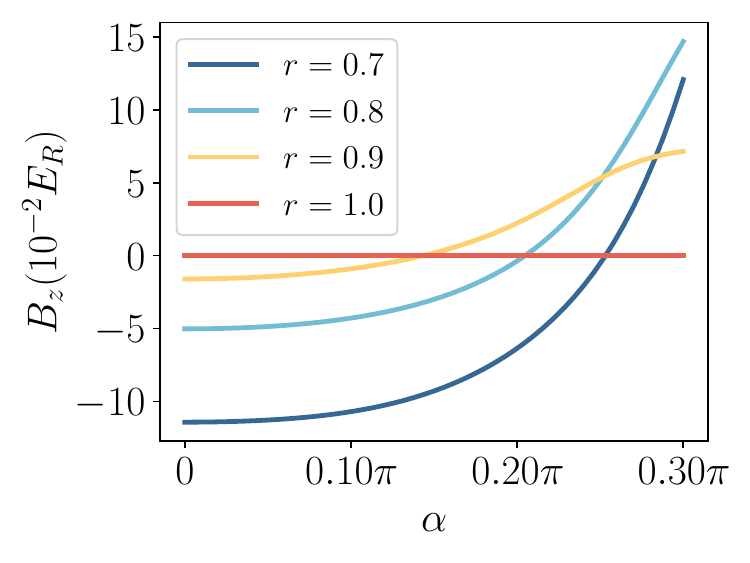}\label{Ssubfig:8b}
    }
    \caption{\fontsize{12pt}{14pt}\selectfont Effective pseudo-magnetic field $B_z$ arising from the energy asymmetry between $p_x$ and $p_y$ orbitals. (\textbf{a}) $B_z$ plotted against the reflection coefficient $r$ and polarization angle $\alpha$. The solid green line marks the condition $r=\cos\alpha$ where the two $p$-orbitals are degenerate and the $D_4$ point group symmetry is partially recovered. (\textbf{b}) $B_z$ as a function of $\alpha$ for specific $r$ values. $B_z$ is of typical magnitude $10^{-2}E_R \sim 10^{-1}E_R$.
   } 
    \label{Sfig:Hz}
\end{figure}

\subsection{ Rotation about \texorpdfstring{$z$}{}-axis}\label{Ssubsec:Rz}
A key step in the detection protocol, detailed in Sec.~\ref{Ssubsec:protocol}, is to rotate the squeezed state so that its most sensitive axis aligns with the direction of the GW-induced signal. As shown in Fig.~\ref{Sfig:HusimiQ}, this requires a rotation about the pseudo-spin $z$-axis.

A straightforward method to achieve this alignment involves a rotation about the $z$-axis, which can be induced by introducing a pseudo-magnetic field $B_z \hat{J}_z = B_z (\hat{n}_x-\hat{n}_y)/2$. 
This term accounts for the energy asymmetry between $p_x$ and $p_y$ orbitals, which can be adjusted by tuning the lattice parameters $r,t,\alpha$ as defined in Eq.~\eqref{Seqn:Vlat in flat}. 
Taking  $r<1$,  we have an energy shift between $p_x$ and $p_y$ orbitals due to the resulting loss of lattice $D_4$ symmetry~\cite{wirth2011evidence,kock2016orbital,2011PRACaizi}. This generates a pseudo-magnetic field along the $z$-axis.

In Fig.~\ref{Sfig:Hz}, we provide the results of $B_z = 2(E^1(\mathbf{Q}_x)-E^1(\mathbf{Q}_y))$. We find that $B_z$ has a typical value about $5\times 10^{-2}E_R$. Therefore, the typical time required to rotate a state squeezed by TACT Hamiltonian about $z$-axis by $\varphi=\pi/4$ is $T_{\mathrm{rot}}(\pi/4)=\pi \hbar/(4\times 5\times  10^{-2}E_R) \simeq 25 \mu\mathrm{s}$.

\medskip 
\section{ Detecting Gravitational Waves}\label{Ssec:GW}
This section describes the core mechanism for sensing gravitational waves (GWs) with our orbital optomechanical sensor. The fundamental idea is that a passing GW induces a minute rotation of the p-orbital pseudo-spin (see Fig.~\textcolor{magenta}{4} in the main text), which can be measured with high precision. This rotation, as we will show, is mediated by the GW's effect on the optical lattices that confine and interact with the atoms.  Throughout this paper, we use the convention $\eta^{\mu \nu} = \mathrm{diag}(1, -1, -1, -1)$ and $x^\mu = (ct, \mathbf{r})$.

\subsection{ Theory}\label{Ssubsec:theory}
Consider a continuous plane gravitational wave traveling along the $\mathbf{e}_z$ direction, which perturbs the spacetime metric as $g_{\mu \nu} = \eta_{\mu \nu} + h_{\mu \nu}$. In the transverse-traceless (TT) gauge, the spacetime metric is given by:
\begin{equation}\label{Seqn:h_metric}
    h_{\mu \nu} = \begin{pmatrix}
0 & 0 & 0 & 0 \\
0 & h_{+} & h_{\times} & 0 \\
0 & h_{\times} & -h_{+} & 0 \\
0 & 0 & 0 & 0
\end{pmatrix}\,,
\end{equation}
where $h_{+}(t,z)$ and $h_{\times}(t,z)$ are the strain fields corresponding to the two independent polarization of the GW,  plus and cross respectively. The spacetime interval is then expressed as:
\begin{equation}
    \d s^2 = g_{\mu \nu} \d x^{\mu} \d x^{\nu} = -c^2 \d t^2 + (1-h_+) \d \X^2 + 2 h_{\times} \d \X \d \Y + (1+h_+) \d \Y^2 + \d z^2 \,.
\end{equation}
The coordinate frame employed here is the TT frame, in which the coordinate position of a free, initially stationary test mass is unaffected by the passing wave~\cite{2007BookMaggiore}.
Note that the polarization described here is aligned with the $\X, \Y$ axes. A complete general-relativistic treatment of $p$-orbital bosons in curved spacetime $g_{\mu \nu}$ involves the interaction between GWs and the bosons, as well as the effect of the GW on the trapping potentials. This is captured by the following action~\cite{mukhanov2007introduction}:
\begin{equation}
    S[\Psi; g^{\mu \nu}, V] = \int \d^4x \sqrt{-g} \left[ g^{\mu \nu} \partial_\mu \Psi^\dagger \partial_\nu \Psi - \kappa |\Psi|^4 - V(\mathbf{r}; g^{\mu \nu}) |\Psi|^2  \right] \,,
\end{equation}
where $g=\det(g_{\mu \nu})$ and $\kappa$ is the atomic interaction coefficient. The first two terms represent the minimal coupling between atoms and GWs, and the third term $V$ is given by:
\begin{equation}
V (\mathbf{r};g^{\mu \nu}) = \frac{m^2c^2}{\hbar^2} + \frac{2m}{\hbar^2} \left(V_{\mathrm{lat}}(\mathbf{r};g^{\mu \nu}) + U_{\mathrm{lat}}(\mathbf{r};g^{\mu \nu}) \right),
\end{equation}
which accounts for the atomic mass, and the GW-induced deformation of the primary lattice $V_{\mathrm{lat}}$ for trapping atoms and signal lattice potential $U_{\mathrm{lat}}$  for signal detection.
The complete atomic Hamiltonian is expressed as:
\begin{equation}
\hat{H} = \hat{H}_{\mathrm{flat}} + \hat{H}_{\mathrm{GW-B}} + \hat{H}_{\mathrm{GW-EM}}.
\end{equation}
Here, $\hat{H}_{\mathrm{flat}}$ corresponds to the Hamiltonian of $p$-orbital atoms in flat spacetime, while $\hat{H}_{\mathrm{GW-B}}$ and $\hat{H}_{\mathrm{GW-EM}}$ describe the direct coupling between GWs and bosons, and the photon-mediated coupling between GWs and atoms, respectively. 
As detailed in~\ref{Ssubsec:bosons}, the direct GW-boson coupling induces a pseudo-magnetic field along the $z$-axis, but with a negligible strength.
Additionally, the impact of GWs on the primary optical lattice potential $V_{\mathrm{lat}}$ is also found to be insignificant, as shown in~\ref{Ssec:lattice}. In contrast, the GW-induced deformation of the signal optical lattice $U_{\mathrm{lat}}$, which has a wavelength twice that of $V_{\mathrm{lat}}$ and half the wave-vector, results in a detectable pseudo-magnetic field $B_x$ that couples $p_x$ and $p_y$ orbital states. For instance, $V_{\mathrm{lat}}$ and $U_{\mathrm{lat}}$ could be generated using lasers with wavelengths of $\lambda = 532$~nm and $2\lambda = 1064$~nm, respectively. The expression for $U_{\mathrm{lat}}$ is given by:
\begin{align}\label{Seqn:Ulat}
    U_{\mathrm{lat}}(\mathbf{r};\phi_U,\Phi_x,\Phi_y) &= -\frac{U_{0}}{4} \abs{e^{i\K\X/2} + e^{-i\K\X/2}e^{-i\Phi_{\X}} +  e^{i(\phi_U+\Phi_{\Y}/2)}\left( e^{i\K\Y} +  e^{-i\K\Y}e^{-i\Phi_{\Y}}\right) }^{2}\,, \notag \\
    &= -U_0 \bigg\{ 1 + \cos (\dfrac{kx}{2}+\Phi_x) \cos (\dfrac{ky}{2}+\Phi_y) \notag \\
    &\ \hspace{9.5ex}  + \cos\phi_U \left[\cos (\dfrac{kx}{2}+\Phi_x) + \cos (\dfrac{ky}{2}+\Phi_y) \right] \bigg\} \,,
\end{align}
where $\Phi_{x,y}$ are tunable phases, and $\phi_U$ denotes the phase difference between the standing waves of the laser  in the $\X,\Y$ arms having wavelength  $2\lambda$ 
 (to be referred to as $2\lambda$-laser in the following). 
The phases are related to the phase accumulation in $\X,\Y$ arms  by $\Phi_{\X}=\Phi_x-\Phi_y$ and $\Phi_{\Y}=\Phi_x+\Phi_y$. We require $U_0 \ll V_0 = O (E_R)$ so that $U_{\mathrm{lat}}$ has a negligible effect on the $p$-band and lattice structure associated with $V_{\mathrm{lat}}$. In this limit, the two-mode approximation Eq.~\eqref{Seqn:TMA} is still valid and the contribution of $U_{\mathrm{lat}}$ to the pseudo-spin Hamiltonian is:
\begin{align}
  \hat{H}_0^U &=  \sum_{\alpha \beta} J^{U}_{\alpha \beta} \hat{p}_{\alpha}^{\dagger} \hat{p}_{\beta} = J^U_{xx} \hat{N} + B_x \hat{J}_x \,, \label{Seqn:H0U}\\
   J^{U}_{\alpha \beta} &= \int \d^2\mathbf{r} \ \psi^{*}_{\alpha}(\mathbf{r})  U_{\mathrm{lat}}(\mathbf{r};\phi_U,\Phi_x,\Phi_y) \psi_{\beta}(\mathbf{r}) \,, \label{Seqn:JU}
\end{align}

\begin{figure}[ht]
    \centering
    \includegraphics[width=0.6\linewidth]{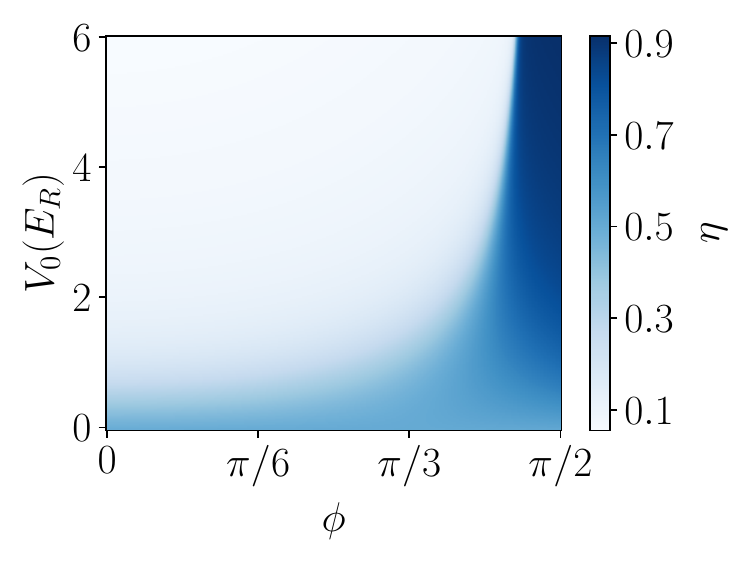}
    \caption{\fontsize{12pt}{14pt}\selectfont Efficiency parameter $\eta$ in Eq.~\eqref{Seqn:eta_def} as a function of potential parameters $V_0$ and $\phi$. The efficiency increases with $V_0$ and exceeds its harmonic limit (Eq.~\eqref{Seqn:eta_har}), $1/e$,  across a broad range of potential parameters.}
    \label{Sfig:eta}
\end{figure}

Therefore, the interaction between GWs and $U_{\mathrm{lat}}$ effectively produces a pseudo-magnetic field $B_x$ along the $\mathbf{e}_x$ direction, given by:
\begin{align}\label{Seqn:Bx}
B_x(\Phi_x,\Phi_y;V_0,\phi) &= 2J^U_{xy}=-2U_0 \sin \Phi_x \sin \Phi_y  \eta(V_0,\phi),
\end{align}
with the dimensionless ``efficiency" parameter $\eta$ depicted in Fig.~\ref{Sfig:eta}. The derivation of this expression and the definition of $\eta$ are detailed in~\ref{Ssubsec:Ulattice}.

For simplicity,  we first consider a plus-polarized GW propagating along the $\mathbf{e}_z$ direction (the polarization and angular patterns are studied in Sec.~\ref{Ssubsubsec:average}). In the presence of such a GW, $U_{\mathrm{lat}}$ gets a real space shift as $U_{\mathrm{lat}}(\mathbf{r};\phi_U,\Phi_x,\Phi_y) \rightarrow U_{\mathrm{lat}}(\mathbf{r};\phi_U + \Delta \phi_U,\Phi_x,\Phi_y-\Delta \Phi/2) = U_{\mathrm{lat}}(\mathbf{r}+\mathbf{l}_U;\phi_U + \Delta \phi_U,\Phi_x,\Phi_y)$. Here, $\mathbf{l}_U = (0,-\Delta \Phi/k)$ (shown in Eq.~\eqref{Seqn:Ulat in curved}). Similarly, the pseudo-magnetic field $B_x$ shift as $B_x(\Phi_x,\Phi_y) \rightarrow B_x(\Phi_x,\Phi_y - \Delta \Phi/2)$, which is most sensitive to $h_+$ when $\Phi_x = \pi/2$ and $\Phi_y = 0$ in the flat spacetime. As a result, $B_x = 0$ in the absence of GW, 
and it gains a finite value when a GW propagates through the detector:
\begin{equation}\label{Seqn:Hx}
    B_x^{\mathrm{GW}}(h_{+};V_0,\phi,\F) = \eta(V_0,\phi) U_0 \Delta \Phi \,.
\end{equation} 
The measured GW signal also depends on the frequency of GWs and measurement protocol. The rotation angle $\theta$ acquired for a measurement with duration $T$ is:
\begin{align}\label{Seqn:theta_def}
    \theta = \int _0^T \d t \ y(t) B_x^{\mathrm{GW}}(t)/\hbar \,,
\end{align}
where $y(t)$ is the modulation function (window function) of the pulse sequence applied in the measurement protocol. The rotated angle is detected at the end of each detection cycle by measuring the orbital polarization $\langle J_y \rangle$ using a time-of-flight experiment (see Fig.~\textcolor{magenta}{4} in the main text and Fig.~\ref{Sfig:tof_schematic}).

For a plus-polarized CW $h_+(t)=h_{0,+} \cos(2\pi f t+\alpha)$, the rotation angle is:
\begin{equation}\label{Seqn:T_theta}
    \theta = \T_{\theta}(f)h_{0,+} = \eta \dfrac{U_0 T}{\hbar} W(f,\alpha,T) \T_\Phi(f) h_{0,+} \,,
\end{equation}
where the protocol-dependent weight function $W(f,\alpha,T)$ is defined as the Fourier transformation of $y(t)$:
\begin{equation}\label{Seqn:Wdef}
    W(f,\alpha,T) = \dfrac{1}{T}\int _0^T \d t \ y(t) \cos(2\pi f t+\alpha) \,.
\end{equation}
The weight function $W$ is analyzed in detail in Sec.~\ref{Ssubsec:protocol}. The transfer function of the Fabry-Perot Michelson Interferometer  is~\cite{2007BookMaggiore}:
\begin{equation} \label{Seqn:T_Phi}
    \T_{\Phi}(f) = \dfrac{\Delta \Phi}{h_{0,+}} = 
        \dfrac{1}{2}\times 2\K L  \, \sinc\left(2\pi f \dfrac{L}{c}\right)  
        \dfrac{2 \F/\pi}{\sqrt{1+(f/f_p)^2}} \,,
\end{equation}
where $L$ is the arm length, $\F$ is the finesse of the FP cavity, and $f_p = c/(4\F L)$ is the pole frequency. The additional factor $1/2$ accounts for the fact that $U_{\mathrm{lat}}$ is generated by laser with wavevector $\K/2$. The shift of phase accumulation $\Phi_{\X,\Y}$ of lasers in the $\X,\Y$ arms is given by:  
\begin{align}
    \Delta \Phi_{\X} &= \frac{1}{2} \T_{\Phi}(f) h_{0,+} \,,\label{Seqn:DeltaPhiX}\\
    \Delta \Phi_{\Y} &= -\frac{1}{2} \T_{\Phi}(f) h_{0,+} \,.\label{Seqn:DeltaPhiY}
\end{align}
The transfer function of a pure MI (without FP cavity) corresponding to our primary $\lambda$-laser is:
\begin{equation}\label{Seqn:T_Phi'}
    \T_{\Phi'}(f) = \dfrac{\Delta \Phi'}{h_{0,+}} = 2\K L' \, \sinc \left(2\pi f \dfrac{L'}{c}\right) \,.
\end{equation}
where we have assumed the $\lambda$-laser is cycling in a pure MI with arm length $L'\sim 1\mathrm{m}$ (see Fig.\ref{Sfig:MI}\textbf{c}).
The arm-length of LIGO interferometer is $L=4\mathrm{km}$, 
 for which the $\sinc$ terms in Eq.~\eqref{Seqn:T_Phi} and Eq.~\eqref{Seqn:T_Phi'} 
are well approximated by $1$ in the frequency region $f<10^3 \mathrm{Hz}$ of our interest in this study.

\subsection{ Lattice-deformation-mediated Interaction between GW and Bosons} \label{Ssubsec:Ulattice}
In this section we demonstrate the derivation of pseudo-magnetic field $B_x$ (defined in Eq.~\eqref{Seqn:Bx}), and analyze its behavior in the presence of a GW, under the condition where $V_0/U_0 \gg 1$.
We first divide the whole integral Eq.~\eqref{Seqn:JU} into contributions from different unit cells $\Omega_{\mathbf{R}}$ as:
\begin{align}\label{Seqn:JU'}
     J^{U}_{\alpha \beta} = \sum _{\mathbf{R}}(-1)^{R_{\alpha} +R_{\beta}}\int_{\Omega_{\mathbf{0}}} \d^2\mathbf{r} \ \psi^{*}_{\alpha}(\mathbf{r})  U_{\mathrm{lat}}(\mathbf{r}+\mathbf{R};\phi_U,\Phi_x,\Phi_y) \psi_{\beta}(\mathbf{r}) \,,
\end{align}
where $\psi_\alpha(\mathbf{r}+\mathbf{R}) = (-1)^{R_{\alpha}} \psi_\alpha(\mathbf{r})$ is used. Given that the lattice vectors of $U_{\mathrm{lat}}$ are $2\mathbf{a}_{1}$ and $2\mathbf{a}_{2}$, we decompose $\mathbf{R} = (2m+m_r)\mathbf{a}_1 + (2n+n_r)\mathbf{a}_2$ with $m_r,n_r \in \mathbb{Z}_2$ denoting the relative position in the enlarged unit cell. Using $\sum_{m,n} 1 = 1/4 \sum_\mathbf{R} 1 =  N_{\mathrm{lat}}/4$, we obtain:
\begin{align}
    J^U_{\alpha \beta} &= \frac{1}{4} \sum_{m_r,n_r \in \mathbb{Z}_2} (-1)^{R_{\alpha } +R_{\beta}}\int \d ^2 \mathbf{r} \ \psi _{\alpha }^{* }(\mathbf{r}) U_{\mathrm{lat}}(\mathbf{r} +m_r\mathbf{a}_{1} +n_r\mathbf{a}_{2}) \psi _{\beta}(\mathbf{r}) \,.
\end{align}
The summation of $m_r,n_r$ leads to:  
\begin{align}
    &\frac{1}{4} \sum_{m_r,n_r \in \mathbb{Z}_2}U_{\mathrm{lat}}(\mathbf{r} +m_r\mathbf{a}_{1} +n_r\mathbf{a}_{2}) = -U_0 \,, \\
    &\frac{1}{4} \sum_{m_r,n_r \in \mathbb{Z}_2} (-1)^{m_r +n_r}U_{\mathrm{lat}}(\mathbf{r} +m_r\mathbf{a}_{1} +n_r\mathbf{a}_{2}) = -U_0 \cos (kx/2+\Phi_x) \cos (ky/2+\Phi_y) \,.
\end{align}
 By inserting these into \eqref{Seqn:JU'}, and using the odd parity of $\psi_{x(y)}$ along $x(y)$ direction, we obtain:
\begin{align}
    J^U_{xx} = J^U_{yy} &= -U_0 \,, \\
    J^U_{xy} = J^U_{yx} &= -U_0 \sin \Phi_x \sin \Phi_y \eta(V_0,\phi) \,, \label{Seqn:JUxy}\\
    \eta(V_0,\phi) &= \int \d ^2 \mathbf{r} \ \psi _{x }^{*}(\mathbf{r};V_0,\phi) \sin(kx/2) \sin(ky/2) \psi _{\beta}(\mathbf{r};V_0,\phi) \,. \label{Seqn:eta_def}
\end{align}
The pseudo-magnetic field is then given by $B_x=2J^U_{xy}$, according to the definition in Eq.~\eqref{Seqn:Bx}.

In the harmonic limit $V_0/E_R \gg 1$,  we expand the Bloch functions in terms of the localized Wannier functions of $V_{\mathrm{lat}}$ at each site $\mathbf{R}$ as $\psi_{\alpha}(\mathbf{r}) = N_{\mathrm{lat}}^{-1/2}\sum_{\mathbf{R}} (-1)^{R_{\alpha}} \omega_{\alpha}(\mathbf{r}-\mathbf{R})$ (Fig.~\ref{Sfig:band structure}). The Wannier functions are given by $\omega_{\alpha}(\mathbf{r}) = \sqrt{\frac{2}{\pi l_H^2}}\frac{x_{\alpha}}{l_H}e^{-r^2/2l_H^2}$ with the characteristic length $l_H = (\frac{\hbar^2}{Mk^2V_0(1+\cos \phi)})^{1/4}$. 
The dimensionless parameter $\eta(V_0,\phi)$ takes the form: 
\begin{equation}
    \eta_{\mathrm{har}}(V_0,\phi) = \frac{\pi^2 l_H^2 }{2a^2} \exp(-\frac{\pi^2 l_H^2 }{2a^2}) \,. \label{Seqn:eta_har}
\end{equation}
The maximal value of $\eta(V_0,\phi)$ in the harmonic limit is $1/e$, which occurs when $\pi^2 l_H^2/2a^2 = 1$. Numerical results of $\eta(V_0,\phi)$ taking into account the unharmonicity of the lattice potential are shown in Fig.~\ref{Sfig:eta}. We observe that in general $\eta$ is not necessarily less than $1/e$.

We now consider the influence of a GW on the pseudo-magnetic-field $B_x$. For a continuous plus-polarized GW with frequency $f$, the change of arm length imparts an additional phase $\Delta \Phi _{\X,\Y} = \pm \Delta \Phi/2 = \pm \T_{\Phi} (f) h_{0,+}/2$ (see Eq.~\eqref{Seqn:T_Phi} for the definition of $\T_{\Phi}$) to the counter-propagating electromagnetic field with wavelength $\K=2\pi/\lambda$ upon reflection from the $\X,\Y$ arm ends. As a result, the optical lattice is modified by the GW from its flat spacetime form Eq.~\eqref{Seqn:Ulat} as follows:
\begin{align}\label{Seqn:Ulat in curved}
    U_{\mathrm{lat}}(\X,\Y;U_0,\phi_U) & \rightarrow -\frac{U_{0}}{4} \Big|\left( e^{i\K\X/2} + e^{-i(\Phi_\X+\Delta\Phi_\X)}  e^{-i\K\X/2}\right)  \notag \\
    & \quad \quad \quad \quad +e^{i(\phi_U+\Phi_{\Y}/2) } \left( e^{i\K\Y/2}+ e^{-i(\Phi_\Y+\Delta \Phi_\Y)} e^{-i\K\Y/2}\right) \Big|^{2} \notag \\
    &= U_{\mathrm{lat}}(\X+\Delta \X,\Y+\Delta \Y;U_0,\phi_U + \Delta \phi_U)  \,,
\end{align}
with $\Delta \X = -\Delta \Y =  \Delta \Phi /\K$ are the real space shift of the lattice caused by the GW, $ \Delta \phi_U =(\Delta\Phi_{\X}-\Delta\Phi_{\Y})/2 = \Delta \Phi/2$ accounts for the relative phase shift and $\Delta\Phi_x = (\Delta\Phi_\X+\Delta\Phi_\Y)/2=0,\ \Delta\Phi_y = (\Delta\Phi_\Y-\Delta\Phi_\X)/2 = -\Delta \Phi/2$. In the $(x,y)$ coordinate, the lattice deformation can be described as $U_{\mathrm{lat}}(\mathbf{r};\phi_U,\Phi_x,\Phi_y) \rightarrow U_{\mathrm{lat}}(\mathbf{r}+\mathbf{l}_U;\phi + \Delta \phi_U,\Phi_x,\Phi_y)$ in the presence of the GW with $\mathbf{l}_U = \frac{1}{\sqrt{2}}(\Delta \X+\Delta \Y,\Delta \Y-\Delta \X) = (0,-\Delta \Phi/k)$. 

The GW-induced lattice deformation can also be seen as a shift of phase parameters because $U_{\mathrm{lat}}(\mathbf{r}+\mathbf{l}_U;\phi + \Delta \phi_U,\Phi_x,\Phi_y) = U_{\mathrm{lat}}(\mathbf{r};\phi + \Delta \phi_U,\Phi_x,\Phi_y-\Delta\Phi/2)$. Consequently, the coefficients $J^U_{\alpha \beta}$ undergo a corresponding shift $J^U_{\alpha \beta} (\Phi_x,\Phi_y) \rightarrow J^U_{\alpha \beta} (\Phi_x,\Phi_y - \Delta\Phi/2)$ in the presence of a GW. Setting $\Phi_x=\pi/2, \Phi_y=0$ results in $J^U_{xy}=0$ in the flat spacetime, and it gains a finite value $J^U_{xy}=\eta U_0 \Delta\Phi/2$ when a GW is present.

For the same reason, the potential $V_{\mathrm{lat}}$ gets a real space shift $\mathbf{l}_V = (0,-\Delta \Phi'/(2k))$ and a phase shift $\Delta \phi = \T_{\Phi}(f)/\T_{\Phi'}(f) \times \Delta \phi_U$ (see Eq.~\eqref{Seqn:T_Phi} and Eq.~\eqref{Seqn:T_Phi'}). 

\subsection{ Direct Interaction between GW and Bosons}\label{Ssubsec:bosons}
This section analyzes the direct interaction between the GW and the trapped bosons. Note that in this section, $\Phi$ is used to denote a scalar field, not the interferometric phase accumulation.

Consider a scalar field $\Phi$ with $\Phi^4$ interaction minimally coupled to the gravity background $g^{\mu \nu}$~\cite{mukhanov2007introduction}:
\begin{equation}\label{Seqn:SGW}
    S[\Phi;h^{\mu \nu}] = \int \d^4x \sqrt{-g} \left(g^{\mu \nu} \partial_\mu \Phi^\dagger \partial_\nu \Phi -  \frac{m^2c^2}{\hbar^2} |\Phi|^2 - \kappa |\Phi|^4 \right)
\end{equation}
with $g\equiv \det(g_{\mu \nu})$ the determinant of the covariant metric tensor and $\kappa$ the interaction coefficient between bosons.
In the presence of a weak gravitational wave $h_{\mu \nu}$ described by Eq.~\eqref{Seqn:h_metric}, the spacetime metric is perturbed as $g_{\mu \nu} = \eta_{\mu \nu} + h_{\mu \nu}$. Its inverse is $g^{\mu \nu} = \eta^{\mu \nu} - h^{\mu \nu}$ with $h^{\mu \nu} = \eta^{\mu \rho} \eta^{\nu \lambda} h_{\rho \lambda}$. As a result, we have a perturbative expansion:
\begin{align}
    \sqrt{-g} &= \sqrt{1-h^2_{+}-h^2_{\times}} = 1 + O(h^2) \,, \\
    g^{\mu \nu} \partial_\mu \Phi^\dagger \partial_\nu \Phi &= \eta^{\mu \nu} \partial_\mu \Phi^\dagger \partial_\nu \Phi - \Big[ h_+|\partial_x \Phi |^2 +h_{\times} (\partial_x \Phi^\dagger \partial_y \Phi + \mathrm{h.c.})- h_+|\partial_y \Phi |^2 \Big ] \,.
\end{align}

We now take the non-relativistic limit. We first factor out the fast oscillating phase term, corresponding to the rest energy of $\Phi$, by defining the new field $\Psi$ as:
\begin{equation}
    \Phi(\mathbf{r},t) = \frac{\hbar}{\sqrt{2m}}\exp(-i\frac{mc^2}{\hbar}t)\Psi(\mathbf{r},t) \,. \label{Seqn:Phi1}
\end{equation} 
The new field $\Psi(\mathbf{r},t)$ now only has slowly time variations and satisfies $\partial_t \Psi \ll  -imc^2\Psi/\hbar$. By substituting Eq.~\eqref{Seqn:Phi1} into Eq.~\eqref{Seqn:SGW} and drop the $\partial_t \Psi^\dagger \partial_t \Psi$ term, we obtain the action of $\Psi$ in the GW background in the non-relativistic limit to the first order of $h$ as:
\begin{align}
    S[\Psi;h^{\mu \nu}] = \int \d^4x  \bigg\{ i\hbar \Psi^\dagger \partial_t \Psi &- \frac{\hbar^2}{2m}\Big[|\nabla \Psi|^2 + h_+ \left(|\partial_x \Psi |^2 - |\partial_y \Psi |^2 \right) + h_{\times} (\partial_x \Psi^\dagger \partial_y \Psi + \mathrm{h.c.}) \Big] \notag \\ &- \frac{\hbar^4}{4m^2}\kappa |\Psi|^4 \bigg \} \,.
\end{align}
The conjugate momentum field of $\Psi$ is described by:
\begin{equation}
    \pi = \frac{\partial \mathcal{L}_{\mathrm{GW}}}{\partial (\partial_t \Psi)} = i\hbar \Psi^\dagger \,.
\end{equation}
As a result, the Hamiltonian describing the coupling between bosons and gravitational wave is:
\begin{align}
    \hat{H} &= \int \d^3x  \left(\pi\partial_t\Psi-\mathcal{L}_{\mathrm{GW}} \right) = \hat{H}_B + \hat{H}_{\mathrm{GW-B}} \,,\\
    \hat{H}_B &= \int \d^3x \left(\frac{\hbar^2}{2m} |\nabla \hat{\Psi}|^2+ \frac{\hbar^4}{4m^2}\kappa |\hat{\Psi}|^4 \right) \,,\\
    \hat{H}_{\mathrm{GW-B}} &= \int \d^3x \Big[h_+ \left(|\partial_x \hat{\Psi} |^2 - |\partial_y \hat{\Psi} |^2\right) + h_{\times} (\partial_x \hat{\Psi}^\dagger \partial_y \hat{\Psi} + \mathrm{h.c.}) \Big]\,.
\end{align}

We now analyze the effect of GW under the two-mode approximation \eqref{Seqn:TMA}. The field operators $\hat{\Psi}$ in 3-dimensional space are now expanded as $ \hat{\Psi}(\mathbf{r},z) = \psi_{x}(\mathbf{r})\phi_0(z)\hat{p}_x + \psi_{y}(\mathbf{r})\phi_0(z)\hat{p}_y$ with $\phi_0(z)$ the normalized ground state of the trap potential along the $\mathbf{e}_z$ direction.
The direct interaction between the gravitational wave and the $p$-orbital bosons, $\hat{H}_{\mathrm{GW-B}}$ is obtained as:
\begin{align}
    \hat{H}_{\mathrm{GW-B}} &= \sum_{\alpha \beta} J^{\mathrm{GW-B}}_{\alpha \beta}  \hat{p}_{\alpha}^{\dagger} \hat{p}_{\beta} \,, \\
    J^{\mathrm{GW-B}}_{\alpha \beta} &= \int \d^2x \Big[h_+ \big(\partial_x \psi_\alpha^*  (\mathbf{r})\partial_x  \psi_{\beta}(\mathbf{r}) - \partial_y \psi_\alpha^* (\mathbf{r}) \partial_y \psi_\beta (\mathbf{r})\big) + h_{\times} \big(\partial_x \psi_\alpha^* (\mathbf{r})  \partial_y \psi_{\beta} (\mathbf{r}) + \mathrm{h.c.} \big) \Big] \,.
\end{align}
Due to the parity symmetries of the Bloch functions, the off-diagonal coupling term $J^{\mathrm{GW-B}}_{x y}$ vanishes. And for the diagonal terms $J^{\mathrm{GW-B}}_{xx}$ and $J^{\mathrm{GW-B}}_{yy}$,  we have: 
\begin{align}
    J^{\mathrm{GW-B}}_{xx} = -J^{\mathrm{GW-B}}_{yy} = h_+ \int \d^2x \big(|\partial_x \psi_x(\mathbf{r})|^2 - |\partial_y \psi_x(\mathbf{r}) |^2\big)\,,
\end{align}
considering the parity symmetry of the Bloch function $\psi_x(-x,y)=-\psi_x(x,y)$ and $\psi_y(x,-y)=-\psi_y(x,y)$. As a consequence, the direct coupling between GW and bosons leads to an energy shift between $p_{x,y}$ orbitals and generates a pseudo-magnetic field along the $z$-axis, as detailed in Sec.~\ref{Ssubsec:Rz}. Its magnitude is given by:
\begin{align}
    B^{\mathrm{GW-B}}_{z} = 2J^{\mathrm{GW-B}}_{xx} 
    \leq 2h_+ \int \d^2x \big(|\partial_x \psi_x(\mathbf{r})|^2 + |\partial_y \psi_x(\mathbf{r}) |^2\big)= 2h_+ \epsilon_k(\mathbf{Q}_x)\,,
\end{align}
where the inequality holds because $|\partial_\alpha \psi_\beta(\mathbf{r})|^2$ is non-negative. Here, \( \epsilon_k(\mathbf{Q}_x) \) represents the two-dimensional kinetic energy of the Bloch states and is typically of order \( O(E_R) \). The ratio between the direct GW-Boson coupling \( B^{\mathrm{GW-B}}_{z} \) and the \( 2\lambda \)-laser mediated \( B_x \) is given by:
\begin{align}
    \frac{B^{\mathrm{GW-B}}_{z}}{B_x} \leq \frac{2\epsilon_k(\mathbf{Q}_x)}{\eta U_0 \Delta \Phi} \sim \frac{1}{\K L\mathcal{F}} \sim \frac{1}{10^{10}\F}\,,
\end{align}
where we have used Eq.~\eqref{Seqn:Bx}. This ratio indicates that the direct coupling between GWs and bosons is approximately ten orders of magnitude smaller than the \( 2\lambda \)-laser mediated coupling between GWs and bosons. Therefore, $B^{\mathrm{GW-B}}_{z}$ is negligible when considering the interaction of GWs with our orbital optomechanical sensor.

\subsection{ Interaction between GW and the Primary Optical Lattice } \label{Ssec:lattice}
This section examines the effects of a gravitational wave (GW) on the primary optical lattice created by a $\lambda$-laser. As a GW passes through, it alters the laser's effective path length, inducing a phase modulation on the beam at the GW's frequency. This modulation induces a deformation in the optical lattice, denoted as $(\delta V_{\mathrm{lat}})_{\mathrm{GW}}$. Within the two-mode approximation, this deformation results in a modification of the single-body Hamiltonian, expressed as:
\begin{align}
    \hat{H}_{\mathrm{GW}}^{V} =  \sum_{\alpha \beta} J^{V,\mathrm{GW}}_{\alpha \beta} \hat{p}_{\alpha}^{\dagger} \hat{p}_{\beta} \,, \quad 
    J^{V,\mathrm{GW}}_{\alpha \beta}= \int \d^2\mathbf{r} \ \psi^{*}_{\alpha}(\mathbf{r}) \left(\delta V_{\mathrm{lat}}\right)_{\mathrm{GW}} \psi_{\beta}(\mathbf{r}) \,, \label{Seqn:JVArm0}
\end{align}
(Note that $J^{V,\mathrm{GW}}$ arises from the GW-induced lattice deformation $(\delta V)_{\mathrm{GW}}$ and represents the change in the coupling coefficients due to the GW. In contrast, $J^U$, as defined in the main text and Eq.~\eqref{Seqn:JU}, has non-zero components even in the flat spacetime.)

The matrix elements of this perturbation are given by:  
\begin{equation}
    J^{V,\mathrm{GW}}_{\alpha \beta} = \int \d^2\mathbf{r} \ 
    \psi^{*}_{\alpha}(\mathbf{r}) (\mathbf{l}_V \cdot \nabla + \Delta \phi \partial_\phi)V_{\mathrm{lat}}(\mathbf{r};\phi) \psi_{\beta}(\mathbf{r}) \,,
\end{equation}
where $\mathbf{l}_V = (0,-\Delta \Phi'/(2k))$ is the GW-induced lattice shift in the real space, as analyzed in Sec.~\ref{Ssubsec:Ulattice}. Due to the translation symmetry $\psi_\alpha(\mathbf{r}+\mathbf{R}) = (-1)^{R_{\alpha}} \psi_\alpha(\mathbf{r})$ and $V_{\mathrm{lat}}(\mathbf{r}+\mathbf{R}) = V_{\mathrm{lat}}(\mathbf{r})$, 
 we have $J^{V,\mathrm{GW}}_{x y} = (J^{V,\mathrm{GW}}_{y x})^{*} =0$.  The remaining two coefficients $J^{V,\mathrm{GW}}_{xx}, J^{V,\mathrm{GW}}_{yy}$ are non-vanishing. Due to the $D_4$ symmetry of $p_{x,y}$ orbitals, $J^{V,\mathrm{GW}}_{xx}=J^{V,\mathrm{GW}}_{yy}$, indicating that the $\lambda$-laser mediated coupling between GW and bosons only leads to an overall energy shift of the $p$-orbital states, which cannot be detected. 
 
 Under the harmonic approximation, these two coefficients are given by:  
\begin{align}
    J^{V,\mathrm{GW}}_{xx}=J^{V,\mathrm{GW}}_{yy} &= \int \d^2\mathbf{r} \ \omega^{*}_{x}(\mathbf{r})  (\mathbf{l}_V \cdot \nabla + \Delta \phi \partial_\phi)V_{\mathrm{lat}}(\mathbf{r};\phi) \omega_{x}(\mathbf{r}) \,, \notag \\
    &= 2V_0\sin\phi  (1-\frac{\pi^2 l_H^2 }{a^2}) \exp(-\frac{\pi^2 l_H^2 }{a^2})\Delta\phi \,.
\end{align}

\subsection{ Detection Protocol}\label{Ssubsec:protocol}
Our detection method is intrinsically designed for detecting persistent, nearly monochromatic signals. The protocol relies on integrating a signal over a long duration $T$ and using dynamical decoupling pulse sequences to act as a narrow-band filter~\cite{Cappellaro2017rev}. This makes the sensor an ideal instrument for searching for continuous gravitational waves (CWs)~\cite{CW2017LRR}, such as those expected from spinning neutron stars or other long-lived periodic sources. Unlike searches for transient, broadband events (e.g., black hole mergers), our method locks onto a specific target frequency, $f_{\mathrm{tar}}$, to coherently accumulate a signal over time, which significantly enhances the sensitivity at that particular frequency.

This section elaborates on the detection protocol illustrated in Fig.~\textcolor{magenta}{4} of the main text. 
Our detection strategy involves repeated measurement cycles. Each cycle consists of five steps, taking a total amount of time, $T=1.11 \mathrm{s}$.
The photons carrying the GW signal interact with the $p$-orbital bosons in the fourth step for a duration of one second, which is reasonably accessible to the current cold atom technology.  

For an alternating continuous GW described by $h_+(t) = h_{0,+} \cos(2\pi f t+\alpha)$, the induced pseudo-magnetic field $B_x(t)\propto \cos(2\pi f t+\alpha)$ oscillates, causing the net rotation angle $\theta$ to average to zero in the high-frequency limit $fT_{\mathrm{life}} \gg 1$. To restore our ideal detection sensitivity, we apply $\pi$-pulse sequences to the pseudo-spin, facilitating the accumulation of $\theta$ in different half cycles of the pseudo-magnetic field. 

\begin{figure}[htbp]
    \centering
    \subfloat[\textbf{a}][]{
        \includegraphics[width=0.47\textwidth]{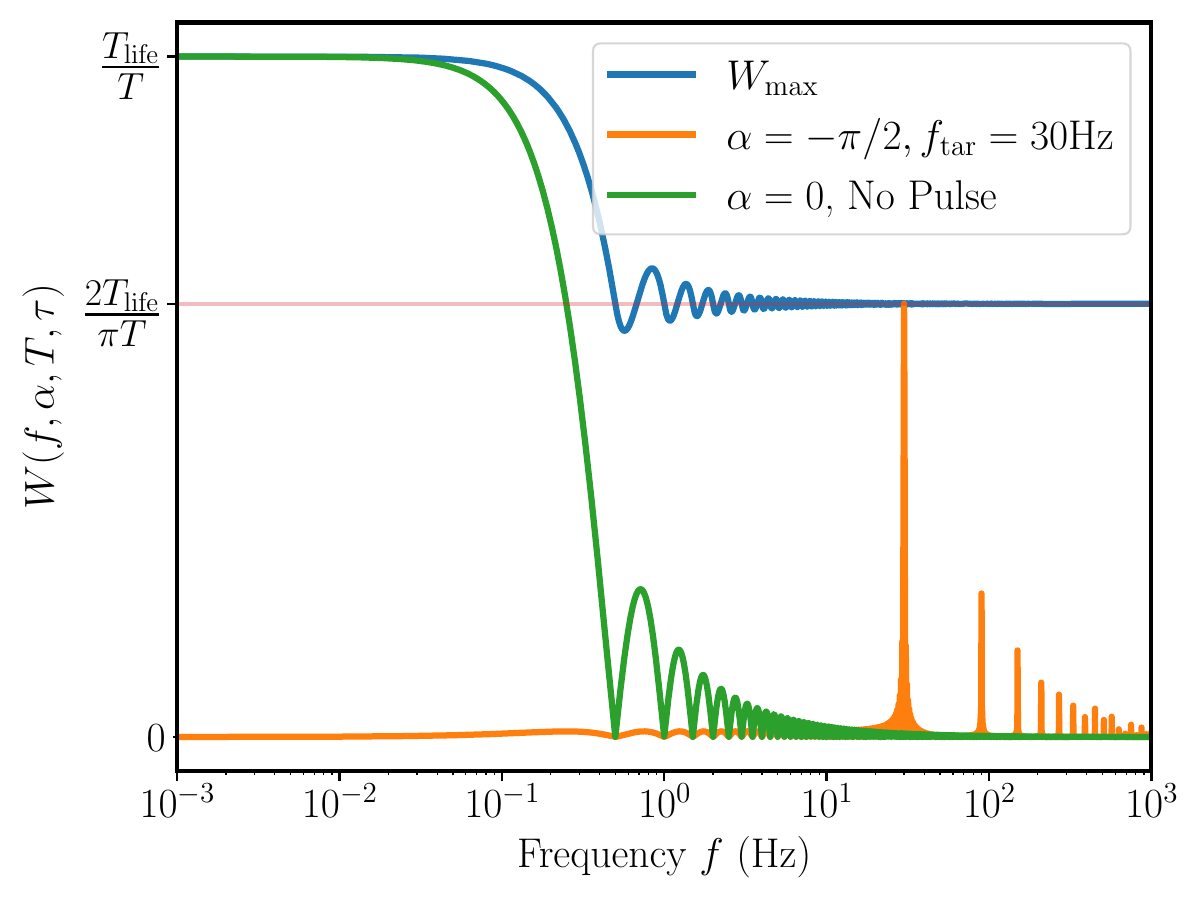}\label{Ssubfig:W1}
    }
    \subfloat[\textbf{b}][]{
        \includegraphics[width=0.47\textwidth]{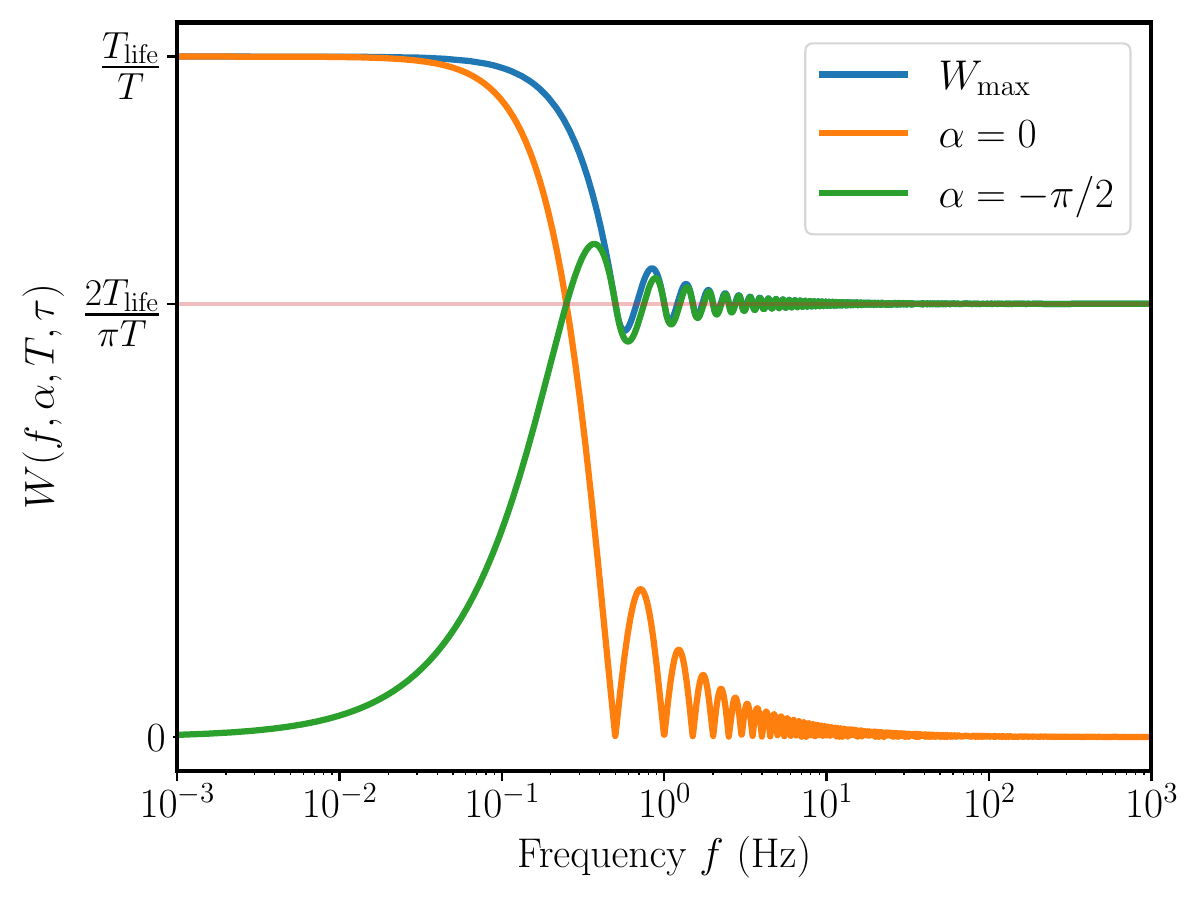}\label{Ssubfig:W2}
    }
    \caption{\fontsize{12pt}{14pt}\selectfont Weight function defined in Eq.~\eqref{Seqn:W}. \textbf{a}: Weight function without lock-in technique compared with the maximum weight function $W_{\max}$(blue line). $W$ with $\alpha = -\pi/2, f_\mathrm{tar} = 30\mathrm{Hz}$ (orange line) has a maximum sensitivity at $f=f_{\mathrm{tar}}$, while $W$ with $\alpha=0$ and no $\pi$ pulse (green line) achieves maximum weight function in low-frequency regime. \textbf{b}: Frequency locked-in weight function $W(f,\alpha,\tau=1/(2f))$. $W_{\max}$ corresponds to $\alpha=0$ in the low frequency regime $f<0.1\mathrm{Hz}$ and $\alpha=-\pi/2$ in the high frequency $f> 1\mathrm{Hz}$ regime.
   } 
    \label{Sfig:W}
\end{figure}

We adopt a simple periodic dynamical decoupling (PDD) multi-pulse sequence within our protocol. This sequence involves the application of evenly spaced $\hat{R}_z(\pi)$ pulses with duration $t_p$ at intervals of $\tau$, accompanied by a window function $y(t)$ that alternates in sign with each pulse application. As detailed in \ref{Ssubsec:Rz}, the duration of a single $\pi$ pulse is approximately $t_p \simeq 100 \, \mu\mathrm{s}$, which is significantly less than the minimum period $\tau_{\mathrm{min}} = 1/(2f_{\mathrm{max}})$ for our frequency range of interest between $10^{-3} \, \mathrm{Hz}$ and $10^3 \, \mathrm{Hz}$. The weight function in Eq.~\eqref{Seqn:Wdef}, is given as 
(Fig.~\ref{Sfig:W},~\cite{Cappellaro2017rev}):
\begin{equation}\label{Seqn:W}
    W(f,\alpha,T,\tau) = \dfrac{1}{2\pi f T}\left[\sin(\alpha)-(-1)^n\sin(2\pi f T_{\mathrm{life}}+\alpha)+2\sum_{j=1}^n(-1)^j\sin(2\pi f t_j+\alpha) \right] \,.
\end{equation}
Here, $t_j$ represents the time of the $j$-th $\pi$-pulse application with $0<t_j = j\tau<T_{\mathrm{life}}$ for $j=1,2,...,n$. From Eq.~\eqref{Seqn:W}, we observe: (1) 
 for a quasi-static signal with $fT_{\mathrm{life}} \ll 1$, no pulses are necessary, and the weight function simplifies to $W = -\dfrac{T_{\mathrm{life}}}{T} \sinc(\pi f T_{\mathrm{life}}) \cos(\pi f T_{\mathrm{life}} + \alpha)$. (2) 
 for an ac signal, the weight function peaks at $f = 1/(2\tau)$ with a bandwidth $\Delta f \simeq 1/T_{\mathrm{life}}$, and the peak value is $W \simeq -\dfrac{2T_{\mathrm{life}}}{\pi T} \sin(\alpha)$. As a consequence, to detect a signal at the target frequency $f_{\mathrm{tar}}$, we apply pulse sequences with spacing $\tau = 1/(2f_{\mathrm{tar}})$.

For both dc and ac signals, the accumulated phase $\theta$ is dependent on the frequency $f$ and the initial phase $\alpha$ of the signal, which are generally unknown a priori. According to Ref.~\cite{taylor2008high}, the sensitivity is moderately reduced by a factor of $\sqrt{2(1+\sqrt{2}/\pi^2)} \simeq 1.5$ due to this uncertainty. However, quantum (double) lock-in techniques, as detailed in~\cite{kotler2011Nature,Cappellaro22PRX,Lee2024NC}, enable the extraction of this information, thereby maximizing our detection sensitivity. The corresponding weight function is given by $W_{\max} = \underset{\alpha}{\max} \ W(f,\alpha,T,\tau=\dfrac{1}{2f}) = \dfrac{T_{\mathrm{life}}}{T}$ for $fT_{\mathrm{life}} \ll 1$ and $W_{\max} = \dfrac{2T_{\mathrm{life}}}{\pi T}$ for $fT_{\mathrm{life}} \gg 1$.

\medskip 
\section{ Quantum Noise of the Orbital Optomechanical Sensor}\label{Ssec:Noises}

To determine the ultimate sensitivity of our orbital optomechanical sensor, we must analyze its fundamental quantum noise sources. This section develops a comprehensive quantum optics model to derive the quantum-limited strain sensitivity. This model consistently incorporates the quantum fluctuations of both the atoms and the photons, revealing the interplay between atomic projection noise and photonic shot and radiation pressure noise.

In addition to the quantum noise, other classical noise such as technical laser noise and displacement noise  are also crucial to interferometric GW detectors. We deal with the classical noise in Sec.~\ref{Ssec:classical_noise}. The parameters of our orbital optomechanical sensor are summarized in Table~\ref{tab:parameters}.

\begin{figure}[ht]
    \centering
    \includegraphics[width=0.8\linewidth]{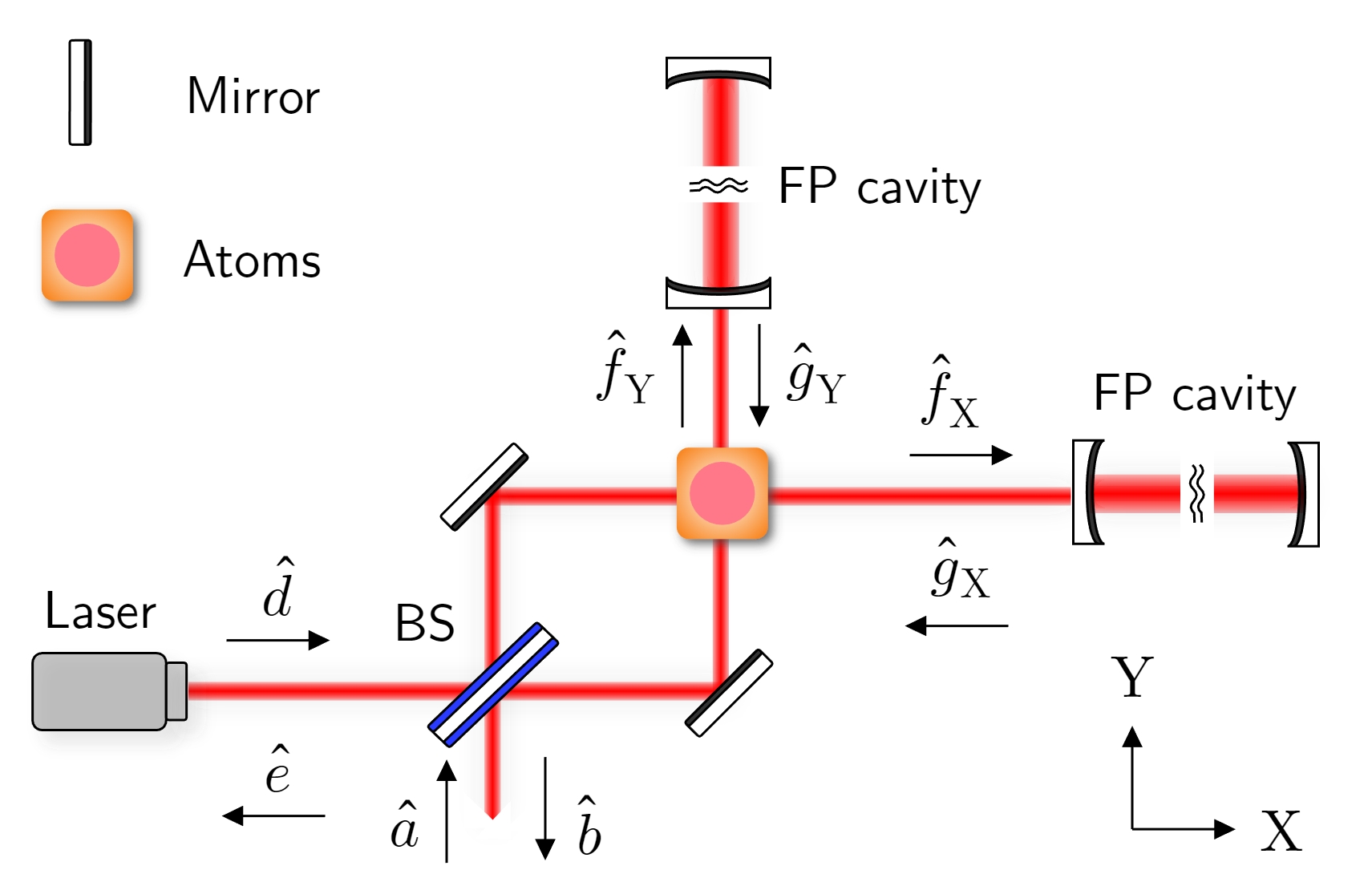}
    \caption{\fontsize{12pt}{14pt}\selectfont Electric field mode operators in the orbital optomechanical sensor setup, as defined in Eq.~\eqref{Seqn:fXdef} and Eq.~\eqref{Seqn:EU}. For clarity, only the  $2\lambda$-laser (1064 nm) is depicted. The distances from the beam-splitter (BS) to the Fabry-Perot cavity (FP cavity) mirrors and the cavity lengths are assumed to be integer multiples of $2\lambda$.}
    \label{Sfig:op_notation}
\end{figure}

\subsection{ Quantum Optical Description of the Orbital Optomechanical Sensor}\label{Ssubsec:quantum optics}
This section provides a quantum optics framework to describe the detection process of our orbital optomechanical sensor. 
A standard quantum optics notation~\cite{2001PRDligoQO} is adopted here. 

\subsubsection{ Quantized Electric Fields and Their Interfering Lattice}
We begin by quantizing the electric fields within the Fabry-Perot-Michelson interferometer. The positive frequency part of the quantized electric field in the $\X$ arm propagating to the right (as shown in Fig.~\ref{Sfig:op_notation}) is expressed as (For simplicity, in this first step we do not write down the operators with angular frequency near $2\omega_0$, corresponding to our 
 $\lambda$-laser, and we will discuss their contribution to the noise in Sec.~\ref{Ssubsubsec:uncertainty}):
\begin{align}
    \hat{E}^{(+)}_{R}(t) &= \sqrt{\frac{2\pi \hbar \omega_0}{\mathcal{A} c}} e^{-i \omega_0 t} \left[\hat{f}_{\X,0} + \int_0^{\infty} \left(\hat{f}_{\X,+} e^{-i \Omega t} + \hat{f}_{\X,-} e^{i \Omega t}\right) \frac{\d \Omega}{2\pi}\right]\,, \label{Seqn:fXdef}\\
    &= \sqrt{\frac{2\pi \hbar \omega_0}{\mathcal{A} c}} e^{-i \omega_0 t} \left(F_0 + \hat{F}_{\X}(t) \right) \,. \label{Seqn:ER}
\end{align}
Here, $\omega_0 = c\K/2$ is the angular frequency of the $2\lambda$-laser (also called the carrier frequency), and $\mathcal{A}=\pi w^2/2$ is the effective cross sectional area of the beam with $w$ the beam waist. The operator $\hat{f}_0$ is the annihilation operator for the $\omega_0$ mode, with $F_0= \sqrt{P_{\mathrm{arm},2\lambda}/\hbar \omega_0}$ being its expectation value. The power of light in each Michelson Interferometer arm is $P_{\mathrm{arm}}$, and the circulating power inside the cavity is $P_{\mathrm{cav}} = 2\mathcal{F}P_{\mathrm{arm}}/\pi $. The operators $\hat{f}_{\X,\pm}(\Omega)=\hat{f}_{\X,\omega_0 \pm \Omega}$ correspond to the side band  annihilation operators at angular frequency $\omega_0 \pm \Omega$. Given that our target frequency regime satisfies $\Omega/2\pi <10^3 \mathrm{Hz} \ll \omega_0/2\pi \sim 10^{15} \mathrm{Hz}$, we approximate all $\omega\pm \Omega$ with $\omega$ in the square root, and extend our integral over $\Omega$ to infinity. The classical amplitude $F_0$ of $\hat{a}_0$ is taken as real, and the operator $\hat{F}$ encapsulates the quantum fluctuations. These fluctuations are described by:
\begin{align}
    \hat{F}_{\X} &= \Delta \hat{f}_{\X} + \int_0^{\infty} \left(\hat{f}_{\X,+} e^{-i \Omega t} + \hat{f}_{\X,-} e^{i \Omega t}\right) \frac{\d \Omega}{2\pi} = \Her(\hat{F}) + i\AHer(\hat{F}) \,,
\end{align}
where $\Delta\hat{f}_{\X} = \hat{f}_{\X,0} - F_0$,  $\Her(\hat{F}_{\X})$ and $\AHer(\hat{F}_{\X})$ are the hermitian and anti-hermitian part of $\hat{F}$ respectively, given by:
\begin{align}
    \Her(\hat{F}_{\X}) &= \frac{1}{2} \left(\Delta \hat{f}_{\X}+\sqrt{2}\int_0^{\infty} \hat{f}_{\X,1} e^{-i \Omega t}  \, \frac{\d \Omega}{2\pi} \right) + \mathrm{h.c.} \,, \\
    \AHer(\hat{F}_{\X}) &= \frac{1}{2} \left(-i\Delta \hat{f}_{\X}+\sqrt{2}\int_0^{\infty} \hat{f}_{\X,2} e^{-i \Omega t}  \, \frac{\d \Omega}{2\pi} \right) + \mathrm{h.c.} \,, \label{Seqn:AHer}
\end{align}
with $\hat{f}_1=(\hat{f}_{+}+\hat{f}_{-}^{\dagger})/\sqrt{2}$ and $\hat{f}_2=(\hat{f}_{+}-\hat{f}_{-}^{\dagger})/\sqrt{2}i$ being the two-photon amplitude and phase quadrature operators~\cite{danilishin2012quantum}. (Subscripts have been omitted for simplicity). Their commutation relations are:
\begin{align}
\left[\hat{f}_1(\Omega),\hat{f}_1^\dagger(\Omega')\right] &= \left[\hat{f}_2(\Omega),\hat{f}_2^\dagger(\Omega')\right] = O(\frac{\Omega}{\omega_0})\simeq 0 \,, \\
\left[\hat{f}_1(\Omega),\hat{f}_2^\dagger(\Omega')\right] &= \left[\hat{f}_1^\dagger(\Omega),\hat{f}_2(\Omega')\right] = 2\pi i \delta(\Omega-\Omega') \,.
\end{align}

Following a similar approach to Eq.~\eqref{Seqn:ER}, the electric field in the 
$\Y$ arm propagating upwards is described by
\begin{equation}\label{Seqn:EU}
    \hat{E}^{(+)}_{U}(t) = \sqrt{\frac{2\pi \hbar \omega_0}{\mathcal{A} c}} e^{-i \omega_0 t} e^{i(\phi_U+\Phi_{\Y}/2)} \left(F_{0} + \hat{F}_{\Y}(t) \right) \,
\end{equation}
where the additional phase $\phi_U+\Phi_{\Y}/2$ represents the phase difference between the classical electric fields in $\Y$ and $\X$ arms. The total electric fields in the $\X,\Y$ arm are therefore:
\begin{align}
    \hat{E}^{(+)} (t,\mathbf{r}) &= \hat{E}_{\X}^{(+)} (t,\mathbf{r}) + \hat{E}_{\Y}^{(+)} (t,\mathbf{r}) \,, \\
    \hat{E}_{\X}^{(+)} (t,\mathbf{r}) &= e^{i \K \X/2} \hat{E}_{R}^{(+)} (t) + e^{-i \K\X/2} e^{-i\Phi_{\X}} \hat{E}_{L}^{(+)} (t) \,, \\
    \hat{E}_{\Y}^{(+)} (t,\mathbf{r}) &= e^{i \K \Y/2} \hat{E}_{U}^{(+)} (t) +  e^{-i \K\Y/2} e^{-i\Phi_{\Y}}\hat{E}_{D}^{(+)} (t) \,,
\end{align}
where $\Phi_{\X,\Y}$ are the tunable phases. The subscripts $R,L,U,D$ denote the right, left, up and down propagating fields, respectively. Their corresponding electric field operators are obtained by substituting $\hat{f}$ with $\hat{f}_{\X},\hat{g}_{\X}$ into Eq.~\eqref{Seqn:ER} and $\hat{f}_{\Y},\hat{g}_{\Y}$ into Eq.~\eqref{Seqn:EU}. Note that all four fields share the same classical amplitude $F_0$. 
A linear transformation is performed at the beam-splitter, described in terms of the two-photon operators as:
\begin{equation}
    \hat{f}_{\X,j} = \frac{\hat{d}_j - \hat{a}_j}{\sqrt{2}} , \ \  \ \ \hat{f}_{\Y,j} = \frac{\hat{d}_j + \hat{a}_j}{\sqrt{2}} \,,
\end{equation}
for $j=1,2$, where $\hat{a}$ and $\hat{d}$ represent annihilation operators of the dark port input and bright port input respectively, as shown in Fig.~\ref{Sfig:op_notation}. And the input-output relation of the FP cavities (Fig.~\ref{Sfig:op_notation}) are expressed as~\cite{2001PRDligoQO}:
\begin{align}
    \hat{g}_1(\Omega) &= \Delta\hat{g}_1= e^{2i\beta} \hat{f}_1 \,, \\
    \hat{g}_2(\Omega) &= \eta_{\Y\X} e^{i\beta} \frac{\tilde{h}}{h_{\mathrm{SQL}}}\sqrt{\k} + \Delta\hat{g_2} , \quad \Delta\hat{g}_2(\Omega) = e^{2i\beta}(\hat{f}_2 - \k \hat{f}_1) \, \label{Seqn:g2},
\end{align}
where $\eta_{\Y\X}$ takes the value $+1$ for $\Y$ arm and $-1$ for $\X$ arm. The standard quantum limit of a conventional interferometer is represented by $h_{\mathrm{SQL}}(\Omega) = \sqrt{8\hbar/M\Omega^2L^2}$. The Fourier transform of the GW strain $h(t)$ is denoted by $h(\Omega)$, satisfying:
\begin{equation}
    \tilde{h}(\Omega) = \int^{\infty}_{-\infty} h(t) e^{i\Omega t} \, \frac{\d \Omega}{2\pi} \,.
\end{equation}
Since $h(t)$ is real, it follows that $\tilde{h}^{*}(\Omega)=\tilde{h}(-\Omega)$, and thus $g_j^{\dagger}(\Omega)=g_j(-\Omega)$. The phase shift $\beta$ induced by the FP cavity and the optomechanical coupling strength $\k$ are defined as:
\begin{align}\label{Seqn:betaKappa}
    \beta(\Omega) = \arctan\left(\frac{\Omega}{\omega_c}\right) , \quad
    \k(\Omega) = \frac{16\omega_0 P_{\mathrm{arm},2\lambda}}{ML^2} \frac{1}{\Omega^2(\omega_c^2+\Omega^2)} \,,
\end{align}
with $\omega_c=\pi c/(2\F L)$ being the angular frequency of the pole frequency $f_p$ of FP cavity, as defined around Eq.~\eqref{Seqn:T_Phi}. Note that the cavity length is assumed to be adjusted to integer multiple of the carrier wavelength, ensuring no net phase shift for the carrier light traveling between two end mirrors of the cavity or upon reflection from the front mirror. 

Now, the optical lattice experienced by atoms is treated as an operator acting on the photon Hilbert space. This operator includes its classical value $U_{\mathrm{lat}}$ (see Eq.~\eqref{Seqn:Ulat}) and quantum fluctuations:
\begin{align}\label{Seqn:Ulat_withQuantumNoise}
    \hat{U}_{\mathrm{lat}}(t,\mathbf{r}) &= -\frac{1}{2}\alpha_{2\lambda} \overline{\hat{E}^{(-)}(t,\mathbf{r})\hat{E}^{(+)}(t,\mathbf{r})}\,, \\
    &\simeq -\frac{\hat{U}_{0}}{4} \abs{ \left( e^{i\K\X/2}e^{i\hat{\Phi}_R} +  e^{-i\K\X/2}e^{-i\Phi_\X}e^{i\hat{\Phi}_L}\right) +e^{i(\phi_U+\Phi_{\Y}/2)}\left( e^{i\K\Y/2}e^{i\hat{\Phi}_U} +  e^{-i\K\Y/2}e^{-i\Phi _\Y}e^{i\hat{\Phi}_D}\right) }^{2} \notag \,.
\end{align}
Here, $\alpha_{2\lambda}$ is the atomic polarizability at wavelength $2\lambda$, and $\overline{A(t)}$ represents the time average 
of $A(t)$ over a short period, typically of several multiples $2\pi/\omega_0$. In the second line we have omitted the second-order fluctuations $O(\hat{F}^2/|F_0|^2)$. The lattice depth is now expressed as $\hat{U}_0 = U_0 + \Delta\hat{U}_0$, where $\Delta \hat{U}_0$ accounts for quantum fluctuation. (As will be detailed in Sec.~\ref{Ssubsubsec:uncertainty}, $\Delta \hat{U}_0$ will not contribute to our sensitivity at the leading order of fluctuation, hence its expression is not elaborated here.) The differential accumulated phase after round trips in the X,Y arms is given by:
\begin{align}
    \Delta \hat{\Phi} &= \hat{\Phi}_{\X} - \hat{\Phi}_{\Y} \,, \label{Seqn:DeltaPhi_quantum}\\
    \hat{\Phi}_{\X} &= \Phi_{\X} +   \hat{\Phi}_{R} -   \hat{\Phi}_{L} = \Phi_{\X} + \frac{1}{F_0}\left(\AHer(\hat{F}_{\X}) - \AHer(\hat{G}_{\X})\right) \,, \label{Seqn:PhiX_quantum}\\
    \hat{\Phi}_{\Y} &= \Phi_{\Y} +   \hat{\Phi}_{U} -   \hat{\Phi}_{D} = \Phi_{\Y} + \frac{1}{F_0}\left(\AHer(\hat{F}_{\Y}) - \AHer(\hat{G}_{\Y})\right) \,. \label{Seqn:PhiY_quantum}
\end{align}
Using Eq.~\eqref{Seqn:AHer} and Eq.~\eqref{Seqn:g2}, the expression of $\hat{\Phi}_{\alpha}$ is simplified to: 
\begin{align}
     \hat{\Phi}_{\alpha} &= \Phi_{\alpha} + \frac{1}{2F_0} \left[\left(-i\Delta \hat{f}_{\alpha} + i\Delta\hat{g}_{\alpha} + \sqrt{2}\int^{\infty}_0 \left(\hat{f}_{\alpha,2}-\hat{g}_{\alpha,2}\right) e^{-i \Omega t} \, \frac{\d \Omega}{2\pi} \right)+\mathrm{h.c.}\right] , \notag \\
     &= \Phi_{\alpha} - \frac{\sqrt{2}}{2F_0} \int^{\infty}_{-\infty} \left[\left(e^{2i\beta}-1\right)\hat{f}_{\alpha,2}-\k e^{2i\beta} \hat{f}_{\alpha,1}+\eta_{\Y\X}e^{i\beta} \frac{\tilde{h}}{h_{\mathrm{SQL}}}\sqrt{\k} \right] e^{-i \Omega t} \, \frac{\d \Omega}{2\pi}  \,, \notag
\end{align}
with $\alpha=\X,\Y$. Consequently, the differential phase shift is given by:
\begin{align}
    \Delta \hat{\Phi} &= \Phi_{\X} - \Phi_{\Y} + \frac{1}{F_0} \int^{\infty}_{-\infty} \left[\left(e^{2i\beta}-1\right)\hat{a}_{2}-\k e^{2i\beta} \hat{a}_{1}+e^{i\beta} \frac{\tilde{h}}{h_{\mathrm{SQL}}}\sqrt{2\k} \right]e^{-i \Omega t}  \frac{\d \Omega}{2\pi} \,, \notag \\
    &= \Phi_{\X} - \Phi_{\Y} + \int^{\infty}_{-\infty} \T_{\Phi}(f)  e^{i\beta} (\tilde{h}+\hat{h}_n)  e^{-i 2\pi f t}  \, \d f \label{Seqn:DeltaHatPhi}\,, 
\end{align}
where the transfer function $\T_{\Phi}(f)=\T_{\Phi}(-f)$ takes the form of:
\begin{equation}
    \T_{\Phi}(f) = \frac{1}{F_0}\frac{\sqrt{2\k}}{h_{\mathrm{SQL}}} = 
    \dfrac{1}{2}\times 2\K L  \,
    \dfrac{2 \F/\pi}{\sqrt{1+(f/f_p)^2}} \,,
\end{equation}
which recovers Eq.~\eqref{Seqn:T_Phi}. The fluctuation $\hat{h}_n$ is expressed as:
\begin{equation}\label{Seqn:hn}
    \hat{h}_n(f) = h_{\mathrm{SQL}} \frac{2i\sin\beta \, \hat{a}_2 - e^{i\beta} \k \hat{a}_1}{\sqrt{2\k}} \,.
\end{equation}
It is important to note that the fluctuating term $\hat{h}_n(f)$ originating $(\hat{\Phi}_L-\hat{\Phi}_R)-(\hat{\Phi}_U-\hat{\Phi}_D)$ differs from the extra-cavity readout term $\hat{\Phi}_D-\hat{\Phi}_R$ of a conventional interferometer, which has the expression:
\begin{equation}\label{Seqn:hnLIGO}
    \hat{h}_{n,\mathrm{conv}}(f) = h_{\mathrm{SQL}}  \frac{\, e^{i\beta} \hat{a}_2 - e^{i\beta} \k \hat{a}_1}{\sqrt{2\k}} \,.
\end{equation}
Comparing Eqs.~\eqref{Seqn:hn} and~\eqref{Seqn:hnLIGO}, we can see that the photonic shot noise for our atomic sensor (corresponding to $\hat{a}_2$) is reduced in the low frequency regime $2\sin \beta < 1$, or equivalently $f < f_p/\sqrt{3}$ with $f_p$ the pole frequency of the FP cavity. This is because the shot noise of counter-propagating light fields partially cancels with each other. 

A more intuitive physical picture is as follows: unlike conventional interferometers that measure the phase of a single combined light field at a dark port~\cite{pitkin2011gravitational,danilishin2012quantum}, the atomic ensemble in this scheme is sensitive to the spatial position of the potential wells in the signal lattice, which is a standing wave formed by counter-propagating laser beams. For the incoming and reflected fields, their quantum phase fluctuations $\hat{f}_2$ and $\hat{g}_2$ are largely common when $f<f_p/\sqrt{3}$, therefore do not shift the spatial position of the standing wave pattern. In contrast, a gravitational-wave signal induces a differential phase shift, which directly causes a physical displacement of the standing wave pattern, and this is precisely what the atoms detect.  Therefore, our configuration naturally acts as a common-mode rejection system for phase noise, effectively suppressing shot noise while remaining highly sensitive to the differential GW signal.

\subsubsection{ Interaction between Atoms and Quantized Light}\label{Ssubsubsec:interation}
The Hamiltonian that describes the interaction between atoms and the quantized electric field, expanding upon the classical light case shown in Eqs.~\eqref{Seqn:H0U} and~\eqref{Seqn:JU}, is:
\begin{align}
    \hat{H}_{\mathrm{atom-photon}} &=  \sum_{\alpha \beta} \hat{J}^{U}_{\alpha \beta}  \hat{p}_{\alpha}^{\dagger} \hat{p}_{\beta} = \hat{J}^U_{xx}  \hat{N} + \hat{B}_x \hat{J}_x \,, \\
   \hat{J}^{U}_{\alpha \beta} &= \int \d^2\mathbf{r} \ \psi^{*}_{\alpha}(\mathbf{r})  \hat{U}_{\mathrm{lat}}(t,\mathbf{r}) \psi_{\beta}(\mathbf{r}) \,, 
\end{align}
where $\hat{J}^U_{\alpha \beta}$ is an operator that incorporates both classical expectation values and quantum fluctuations of the photon states. 
The term $\hat{J}^U_{xx} \hat{N}$ commutes with the operator $\hat{J}_y$ that we measure, and thus does not contribute to the signal or noise. 
The interacting Hamiltonian that contributes to our measurement is:
\begin{equation}\label{Seqn:HAP}
    \hat{H}(t) = y(t) \hat{B}_x \hat{J}_x \,,
\end{equation}
where $y(t)$ is the window function corresponding to our detecting protocol (see Sec.~\ref{Ssubsec:protocol}).

The pseudo-magnetic field operator,$\hat{B}_x$, is found by substituting the quantum expression for the differential phase Eq.~\eqref{Seqn:DeltaHatPhi} into its classical definition~Eq.~\eqref{Seqn:Bx}. For the chosen operating point, $\Phi_{\X}=\Phi_{\Y}=\pi/2$, this yields:
\begin{equation}
     \hat{B}_x(t)= \eta(V_0,\phi) \hat{U}_0 \int^{\infty}_{-\infty}  \T_{\Phi}(f)  e^{i\beta} (\tilde{h}+\hat{h}_n)  e^{-i 2\pi f t} \, \d f \,,
\end{equation}
The orbital rotation angle $\theta$ about the $\mathbf{e}_z$ axis, acquired for a measurement with duration $T$, now includes quantum fluctuations:
\begin{align}
    \hat{\theta} &= \int _0^T \d t \ y(t) \hat{B}_x(t)/\hbar \,, \\
    &= \frac{\eta \hat{U}_0T}{\hbar} \int^{\infty}_{-\infty} \T_{\Phi}(f)  Y(f,\beta) (\tilde{h}+\hat{h}_n)  \d f \,, \label{Seqn:hat_theta}
\end{align}
where $Y(f,\beta)$ is defined as the Fourier transform of $e^{i\beta}y(t)$:
\begin{equation}
    Y(f,\beta(f)) = \frac{1}{T}\int^{T}_{0} e^{i\beta}y(t) e^{-i 2\pi f t} \, \d f \,,
\end{equation}
whose real part is the weight function defined in Eq.~\eqref{Seqn:Wdef}.

\subsubsection{ Measurement Uncertainty}\label{Ssubsubsec:uncertainty}
In our protocol (as depicted in Fig.~\textcolor{magenta}{4} of the main text), we measure the orbital polarization $\langle J_y \rangle$ after a time evolution $T$ under the interacting Hamiltonian Eq.~\eqref{Seqn:HAP}, starting from the initial product state $|\psi\rangle = |\xi_R \rangle_{\mathrm{atom}} \otimes |\alpha\rangle_{\mathrm{photon}}$. Here, $|\xi_R \rangle_{\mathrm{atom}}$ is the Heisenberg-limited atomic squeezed state, and $|\alpha\rangle_{\mathrm{photon}} = |\alpha\rangle_{\omega_0} \otimes_\Omega |0_{\omega_0 \pm \Omega}\rangle$ is the coherent state of the $\hat{d}_0$ mode entering the beam-splitter (see Figure~\ref{Sfig:op_notation}). All dark port input modes are assumed to be in the vacuum state. After a time evolution of duration $T$, the product state becomes entangled between atoms and photons:
\begin{equation}
    |\psi_T \rangle = e^{-i\int_0^T \d t \hat{H}(t)/\hbar}|\psi\rangle = e^{-i\hat{\theta} \hat{J}_x} |\psi\rangle \,,
\end{equation}
with the orbital rotation angle $\hat{\theta}$ defined in Eq.~\eqref{Seqn:hat_theta}. The expectation value of our $\hat{J}_y$ and $\hat{J}_y^2$ measurement (to leading order) are:
\begin{align}
    \left\langle \hat{J}_y \right\rangle_T &= \left\langle \cos \hat{\theta} \hat{J}_y - \sin \hat{\theta} \hat{J}_z \right\rangle \simeq - \langle \hat{\theta} \rangle \langle \hat{J}_z \rangle  \,, \label{Seqn:Jy}\\
    \left \langle \hat{J}_y^2 \right \rangle_T &= \left \langle \cos^2 \hat{\theta} \hat{J}_y^2 -\sin\hat{\theta} \cos \hat{\theta} \{\hat{J}_y,\hat{J}_z\}+ \sin^2 \hat{\theta} \hat{J}_z^2 \right \rangle \simeq \langle \hat{J}_y^2 \rangle + \langle \hat{\theta}^2 \rangle \langle \hat{J}_z^2 \rangle \,, \label{Seqn:Jy2}
\end{align}
where $\left \langle \hat{A} \right \rangle_T$ and $\langle \hat{A} \rangle$ denote expectation values of $\hat{A}$ in the states $|\psi_T\rangle$ and $|\psi\rangle$ respectively. We have also used the fact that $\langle \hat{J}_y \rangle = \langle \hat{J}_y \hat{J}_z \rangle = 0$ for our orbital squeezed state. From Eq.~\eqref{Seqn:Jy2}, we identify two main sources of measurement noise: the atomic projection noise $\langle \hat{J}_y^2 \rangle = |\langle J_z \rangle|^2 (\Delta\theta)_{\mathrm{proj}}^2$, and the photonic quantum fluctuation $\langle \hat{\theta}^2 \rangle$. In the absence of a GW signal, the fluctuation $\Delta \hat{U}_0$ does not contribute to $\langle \hat{\theta}^2 \rangle$ at the leading order of fluctuation. The variance $\langle \hat{\theta}^2 \rangle$ is calculated according to Eq.~\eqref{Seqn:hat_theta},  
\begin{align}
    \langle \hat{\theta}^2 \rangle &=  (\frac{\eta U_0 T}{\hbar})^2 \int^{\infty}_{-\infty} \int^{\infty}_{-\infty} \T_{\Phi}(f) \T_{\Phi}(f') Y(f,\beta) Y(f',\beta') \left\langle \hat{h}_n(f) \hat{h}_n(f') \right\rangle \d f \d f' \,, \\
    &= \int^{\infty}_{0} |\T_{\theta}(f)|^2 S_{\mathrm{photon}}(f) \d f \,, \label{Seqn:hat_theta2}
\end{align}
where $\T_{\theta}(f)=\eta U_0 T W_{\max}(f) \T_{\Phi}(f)/\hbar$ and $W_{\max}(f) = |Y(f,\beta)|$. (For the definition of weight function $W$ and its maximum $W_{\max}$, please refer to Sec.~\ref{Ssubsec:protocol}). The single-sided noise spectral density $S_{\mathrm{photon}}(f)$ is defined by:
\begin{equation}\label{Seqn:Sdef}
    \left\langle \hat{h}_n(f) \hat{h}_n(f') \right\rangle = \left\langle \hat{h}_n(f) \hat{h}_n(-f')^{\dagger} \right\rangle = \frac{1}{2} S_{\mathrm{photon}}(f) \delta(f+f') \,.
\end{equation}
The factor $1/2$ accounts for the fact that $S_{\mathrm{photon}}(f)$ is single-sided (instead of double-sided) and satisfies $S_{\mathrm{photon}}(f) = S_{\mathrm{photon}}(-f)$. 
By inserting Eq.~\eqref{Seqn:hn} into Eq.~\eqref{Seqn:Sdef}, we obtain: 
\begin{align}\label{Seqn:Sphoton2}
    S_{\mathrm{photon}}(f) &= h^2_{\mathrm{SQL}} \frac{4\sin^2\beta+\k^2}{2\k} = S_{\mathrm{shot},2\lambda}(f)+ S_{\mathrm{rad},2\lambda}(f) \,, \\
    S_{\mathrm{shot},2\lambda}(f) &= h^2_{\mathrm{SQL}} \frac{2\sin^2\beta}{\k} = \frac{2\hbar(2\pi f)^2}{c\K \Ptwo} \,, \label{Seqn:Sshot2} \\
    S_{\mathrm{rad},2\lambda}(f) &= h^2_{\mathrm{SQL}} \frac{\k}{2} = \left(\frac{8}{ML(2\pi f)^2}\right)^2 \frac{\hbar \K \Ptwo}{2c} \left(\frac{2\F /\pi}{\sqrt{1+(f/f_p)^2}}\right)^2\, \label{Seqn:Srad2},
\end{align}
where $S_{\mathrm{shot},2\lambda}(f)$ (Fig.~\ref{Ssubfig:shot2}) originating from $(e^{2i\beta}-1)\hat{a}_1$ represents the quantum fluctuation of the phase of dark port input state, and $S_{\mathrm{rad},2\lambda}(f)$ (Fig.~\ref{Ssubfig:rad2}) stemming from $\k\hat{a}_2$ represents the fluctuation of the position of test masses induced by the quantum fluctuation of the radiation pressure. It is worth noting that the photonic noise $S_{\mathrm{photon}}(f)$can be further reduced by injecting photonic squeezed states~\cite{ma2017proposal,LIGOsqueezing1,LIGOsqueezing2,LIGOsqueezing3}.

We next consider the noise arising from the $\lambda$-laser, which is used to generate the primary lattice (see Fig.~\ref{Sfig:MI} for a detailed optical layout). Its interaction with gravitational waves also creates two side-bands at angular frequency $2\omega_0\pm \Omega$ (note that $2\omega_0 = c\K = 2\pi c/\lambda$), and these modes contribute to the electric field as:
\begin{equation}
    \hat{E}^{(+)}_{R,\lambda}(t) = \sqrt{\frac{2\pi \hbar \omega_0}{\mathcal{A} c}} e^{-2i \omega_0 t} \int_0^{\infty} \left(\hat{f}_{\X,2\omega_0+\Omega} e^{-i \Omega t} + \hat{f}_{\X,2\omega_0-\Omega} e^{i \Omega t}\right) \frac{\d \Omega}{2\pi}\,,
\end{equation}
and similarly for all the electric fields propagating along other directions as described in Sec.~\ref{Ssubsec:quantum optics}. The optical lattice generated by the $\lambda$-laser is given by:
\begin{equation}
    \hat{V}_{\mathrm{lat}}(t,\mathbf{r}) = -\frac{\hat{V}_{0}}{4} \abs{ \left( e^{i\K\X}e^{i\hat{\Phi}'_R} + e^{-i\K\X}e^{-i\Phi_\X'}e^{i\hat{\Phi}'_L}\right) +e^{i\phi }\left( e^{i\K\Y}e^{i\hat{\Phi}'_U} +  e^{-i\K\Y} e^{-i\Phi _\Y'}e^{i\hat{\Phi}'_D}\right) }^{2} \,, \notag
\end{equation}
where $\hat{V}_0$, $\Delta \hat{\Phi}'$, $\hat{\phi}$ now all contain quantum corrections. However, these corrections do not contribute noise to Eq.~\eqref{Seqn:Jy2}. Although they do influence $\eta(V_0,\phi)$, the fluctuation of $\eta$ does not produce noise in the absence of a GW signal.

\begin{figure}[htp]
    \centering
    \subfloat[\textbf{a}][]{
        \includegraphics[width=0.47\textwidth]{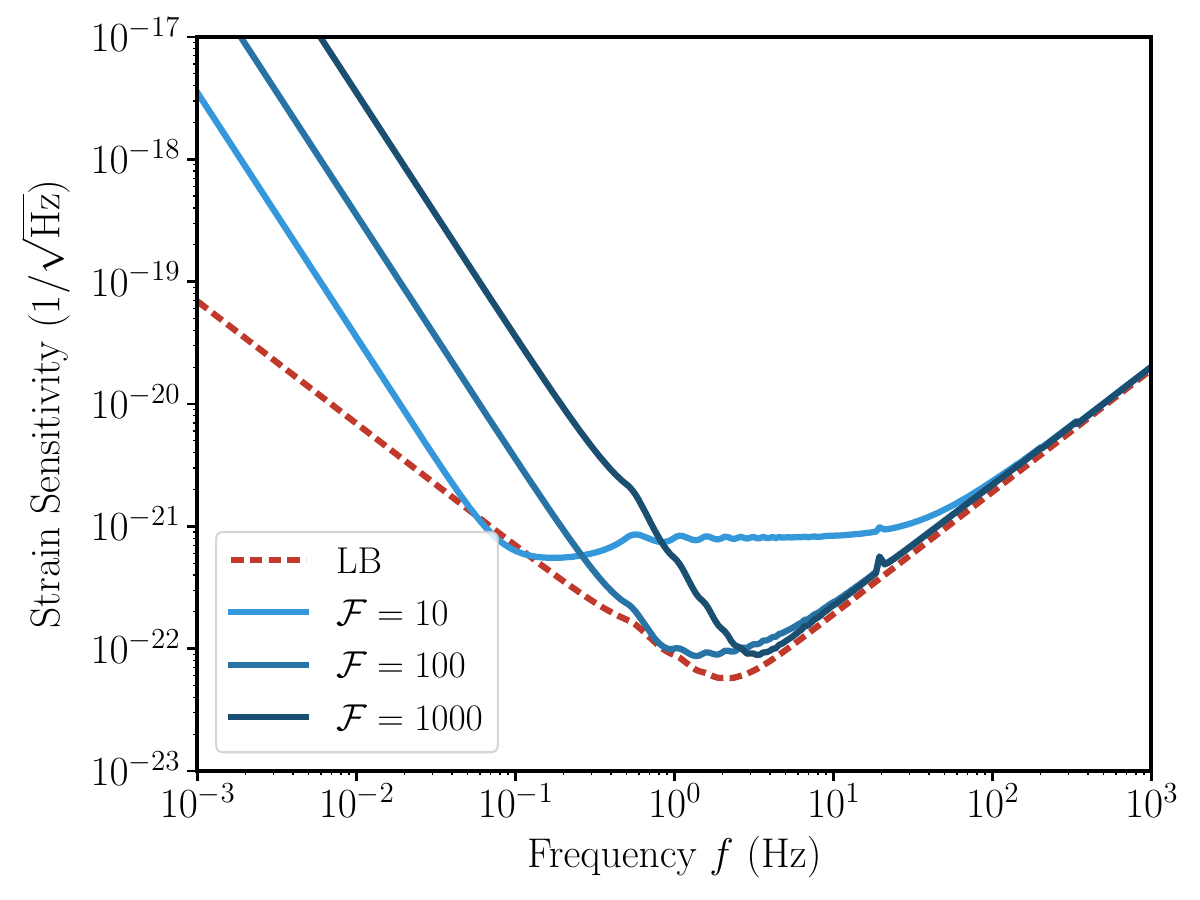}
    }
    \subfloat[\textbf{b}][]{
        \includegraphics[width=0.47\textwidth]{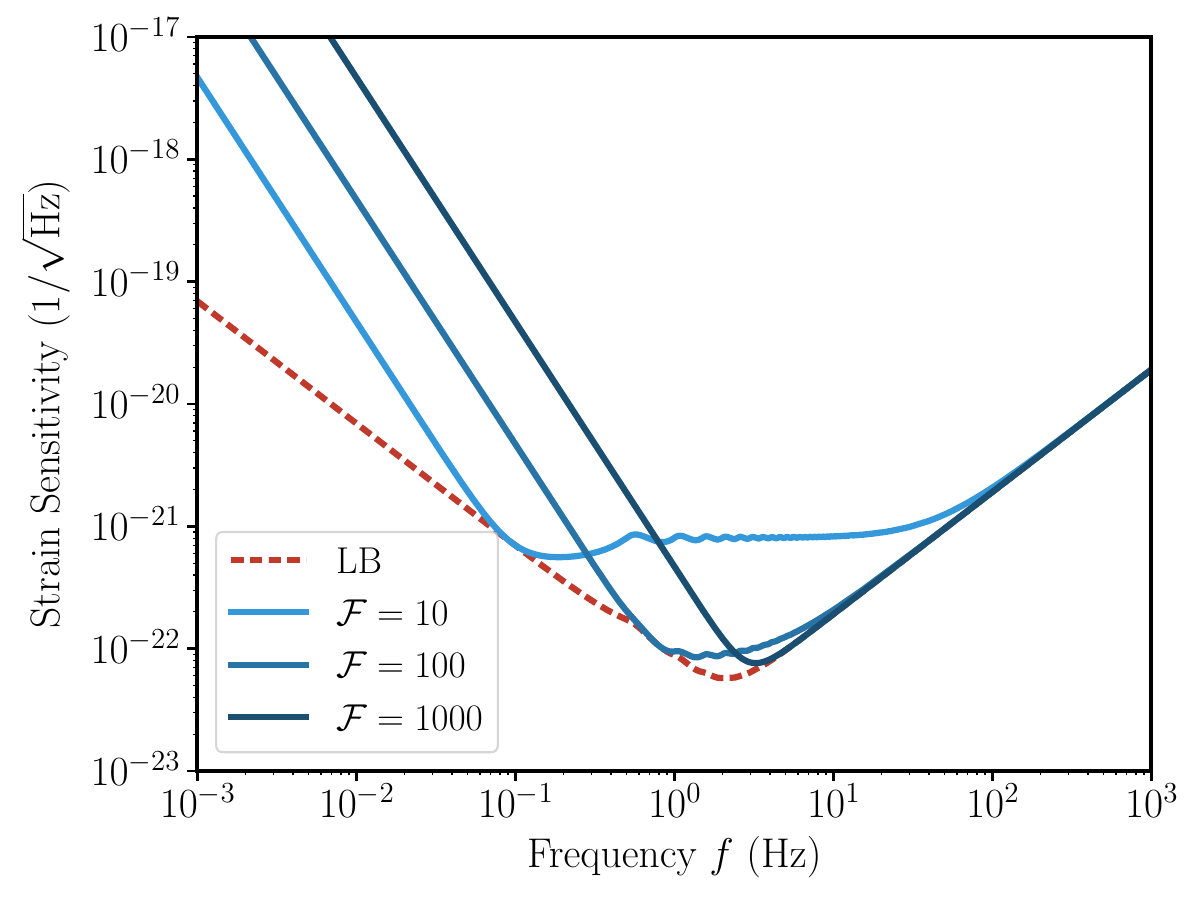}
    }
    \caption{\fontsize{12pt}{14pt}\selectfont Comparison of sensitivity curves: \textbf{a}, derived from the exact formula Eq.~\eqref{Seqn:Sexact}; \textbf{b}, derived from the approximate formula Eq.~\eqref{Seqn:Sapp}. The difference between the sensitivities in these two plots is less than $0.2$ dB, which  justifies the approximation in   Eq.~\eqref{Seqn:Sapp}. $\mathcal{F}$ stands for the finesse of cavities. LB represents the optimal sensitivity  obtained by varying cavity finesse.} 
    \label{Sfig:approximation}
\end{figure}

Now suppose a plus-polarized continuous GW, described by $h_+(t)=h_{0,+} \cos\left(2\pi f t+\alpha\right)$ with $\tilde{h}(f') = h_{0,+} \left[e^{i\alpha} \delta(f'-f) + e^{-i\alpha} \delta(f'+f)\right]/2$, interacts with our orbital optomechanical sensor, then Eq.~\eqref{Seqn:Jy}  is reduced to:  
\begin{align}
 \left\langle \hat{J}_y \right\rangle_T 
    = -|\langle J_z \rangle| \frac{\eta U_0T}{\hbar} W(f,\alpha+\beta)h_{0,+} = -|\langle J_z \rangle| \T_{\theta}(f) h_{0,+} \,,
\end{align}
where we assume lock-in techniques are used to maximize $W(f,\alpha+\beta)$ (see Section~\ref{Ssubsec:protocol}).  The variance of our $\hat{J}_y$ measurement when $h_{0,+}=0$ is given by:
\begin{align}\label{Seqn:Jy2Noise}
    (\Delta J_y)^2 = \left \langle \hat{J}_y^2 \right \rangle_T &= |\langle J_z \rangle|^2 \left( (\Delta\theta)_{\mathrm{proj}}^2 +  (\Delta \theta)_{\mathrm{photon}}^2\right) \,.
\end{align}
Here, the photonic quantum fluctuation is represented by $(\Delta \theta)_{\mathrm{photon}}^2$, which is defined to be $\langle \hat {\theta}^2 \rangle$, whose expression is given by Eq.~\eqref{Seqn:hat_theta2}. $|\langle J_z^2\rangle|$ in Eq.~\eqref{Seqn:Jy2} are approximated with $|\langle J_z \rangle|^2$ because that $(\Delta J_z)^2/|\langle J_z \rangle|^2 \sim 1/N$ vanishes when $N$ is large.  The signal-to-noise-ratio (SNR) of our measurement is given by:
\begin{equation}\label{Seqn:SNR}
    \mathrm{SNR} = \frac{\left\langle \hat{J}_y \right\rangle_T}{(\Delta J_y)} = \dfrac{|\T_{\theta}(f)| h_{0,+}}{\left[(\Delta\theta)_{\mathrm{proj}}^2 + \int^{\infty}_{0} |\T_{\theta}(f)|^2 S_{\mathrm{photon}}(f) \d f \right]^{1/2}} \,.
\end{equation}
For a periodic GW with frequency $f$, the SNR is given in terms of the strain sensitivity $\sqrt{S_n(f)}$ as~\cite{2007BookMaggiore},
\begin{equation}\label{Seqn:SnDef}
    \mathrm{SNR} = \sqrt{\frac{T}{S_n(f)}} h_{0,+} \,.
\end{equation}
Combining Eq.~\eqref{Seqn:SNR} and Eq.~\eqref{Seqn:SnDef}, the square of strain sensitivity of our orbital optomechanical sensor is derived as:
\begin{align}
    S_n(f) &= T \left(\frac{(\Delta\theta)_{\mathrm{proj}}}{\T_{\theta}(f)}\right)^2 + T \,\frac{\int^{\infty}_{0} |\T_{\theta}(f')|^2 S_{\mathrm{photon}}(f')\d f'}{|\T_{\theta}(f)|^2} \,, \label{Seqn:Sexact}\\
    &\simeq S_{\mathrm{atom}}(f) + \frac{T}{T_{\mathrm{life}}} S_{\mathrm{photon}}(f) \,, \notag \\
    &\simeq S_{\mathrm{atom}}(f) + S_{\mathrm{photon}}(f) \,, \label{Seqn:Sapp} 
\end{align}
where the noise spectral density contributed by atoms is $S_{\mathrm{atom}}(f) = T\left((\Delta\theta)_{\mathrm{proj}}/\T_{\theta}(f)\right)^2$. In the second line, we have used the fact that $W_{\max}$ has a bandwidth $\Delta f =1/T_{\mathrm{life}}$ (see Sec.~\ref{Ssubsec:protocol}) and so does $\T_{\theta}(f)$. 
For our current settings, with $T=1.11\mathrm{s}$ and $T_{\mathrm{life}}=1\mathrm{s}$, the ratio  $T/T_{\mathrm{life}}$ 
leads to a reduction of  the photon noise sensitivity $S^{1/2}_{\mathrm{photon}}$ by $0.2\mathrm{dB}$, 
which is thus negligible. 
This is further confirmed by our numerical simulation results, as shown in Figure~\ref{Sfig:approximation}. 
By substituting Eq.~\eqref{Seqn:T_theta} into equation \ref{Seqn:Sapp}, we obtain a expression of $S_{\mathrm{atom}}$ as:
\begin{equation}
    S^{1/2}_{\mathrm{atom}}(f) =\sqrt{T}\frac{(\Delta\theta)_{\mathrm{proj}}}{\T_{\theta}(f)}
    = \frac{\hbar}{\eta U_0\sqrt{T}} \frac{(\Delta \theta)_{\mathrm{proj}}}{ W_{\max}(f)\K L} \frac{\sqrt{1+(f/f_p)^2}}{2\F /\pi} \,.\label{Seqn:Satom}
\end{equation}
The frequency-dependent behavior of  $S_{\mathrm{atom}}$ is depicted in Figure \ref{Ssubfig:proj}.

\subsection{ Calculation of the Quantum Limited Strain Sensitivity Curve}
Considering the fundamental quantum noises---atomic projection noise and photonic quantum fluctuations---the full, quantum-limited strain sensitivity from Eq.~\eqref{Seqn:Sapp} is given by:
\begin{align}\label{Seqn:Sn}
S_n(f) &= S_{\mathrm{atom}}(f) + S_{\mathrm{photon}}(f) \,, \\
&= \frac{S_0(f)}{2} \left(\frac{1}{\mathcal{A}(f,\F)}+\mathcal{A}(f,\F)\right) + S_{\mathrm{shot},2\lambda}(f) \,.
\end{align}
where the dependence on cavity finesse $\F$ is isolated in the dimensionless parameter $\mathcal{A}(f,\F)$. The parameter $\mathcal{A}$ quantifies the ratio of radiation pressure noise to atomic projection noise; when $\mathcal{A} \gg 1$, radiation pressure dominates, and when $\mathcal{A} \ll 1$, atomic projection noise dominates. The expressions of $S_0(f)$ and $\mathcal{A}$ are given by:
\begin{align}
    S_0(f) &= \frac{16\hbar}{M L^2 (2\pi f)^2} \frac{(\Delta \theta)_{\mathrm{proj}}}{\eta U_0 W(f,T)}\sqrt{\frac{\hbar  \Ptwo}{2 \K c T}} \label{Seqn:S0} \,, \\
    \mathcal{A}(f,\F) &= \frac{4}{M(2\pi f)^2} \frac{\eta U_0 W(f,T)}{\DTP} \sqrt{\frac{2T}{\hbar c}\K^3\Ptwo} \left(\frac{2\F /\pi}{\sqrt{1+(f/f_p)^2}}\right)^2 \,.
\end{align}
\begin{figure}[htp]
    \centering
    \includegraphics[width=0.6\linewidth]{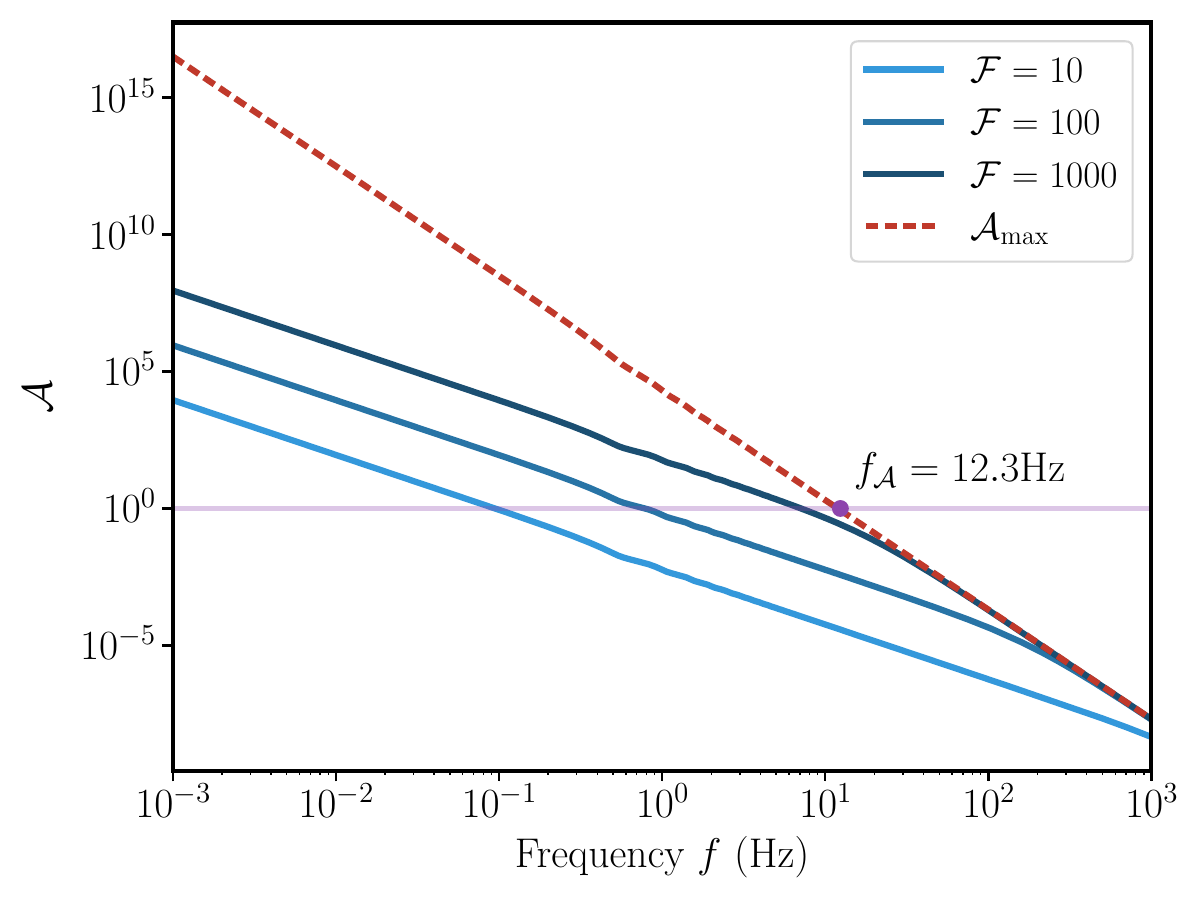}
    \caption{\fontsize{12pt}{14pt}\selectfont $\mathcal{A}(f,\F)$ as a decreasing function of $f$ and increasing function of $\mathcal{F}$. The critical frequency at which $\mathcal{A}_{\max} = 1$, which is defined in Eq.~\eqref{Seqn:Kmax}, is given by $f_{\mathcal{A}}= 12.3 \mathrm{Hz}$. When $f<f_{\mathcal{A}}$, the lower bound of the sum of radiation pressure of the $2\lambda$-laser 
    and the quantum projection noise of atom equals to $S_0(f)$, as described by Eq.~\eqref{Seqn:SLB}. }
    \label{Sfig:A}
\end{figure}
It is evident from Fig.~\ref{Sfig:A} that, in the low frequency region $f\ll f_{\mathcal{A}}$, $\mathcal{A} \gg 1$, and the radiation pressure dominates the noise; while in the high frequency region $f\gg f_{\mathcal{A}}$, $\mathcal{A} \ll 1$, the quantum projection noise dominates.

Since $S_0(f)$ and $S_{\mathrm{shot},2\lambda}(f)$ are independent of $\mathcal{F}$, we optimize the strain sensitivity at each frequency by varying the finesse. 
The envelop of these optimized curves will be denoted as $S_{\mathrm{LB}}(f) = \underset{\F}{\min} S_n(f)$, which has the expression:
\begin{equation}
    S_{\mathrm{LB}}(f) = \begin{cases} 
S_0(f) + S_{\mathrm{shot},2\lambda}(f) & \mathrm{if } f\leq f_{\mathcal{A}},  \\[8pt]
S_0(f)\dfrac{\mathcal{A}_{\max}^{-1}+\mathcal{A}_{\max}}{2} + S_{\mathrm{shot},2\lambda}(f) & \mathrm{if } f > f_{\mathcal{A}}. 
\end{cases} \label{Seqn:SLB}
\end{equation}
Here, $f_{\mathcal{A}}=12.3\mathrm{Hz}$ is the critical frequency at which $\mathcal{A}_{\max} = 1$, where $\mathcal{A}_{\max}(f) = \underset{\F}{\max} \ \mathcal{A}(f,\F)$ is plotted in Fig.~\ref{Sfig:A} and its expression is given by:
\begin{equation}\label{Seqn:Kmax}
     \mathcal{A}_{\max}(f) = \frac{4c^2}{ML^2(2\pi f)^4}\frac{\eta U_0 W(f,T)}{\DTP} \sqrt{\frac{2T}{\hbar c}\K^3\Ptwo} \,.
\end{equation}

\begin{figure}
    \centering
    \includegraphics[width=0.6\linewidth]{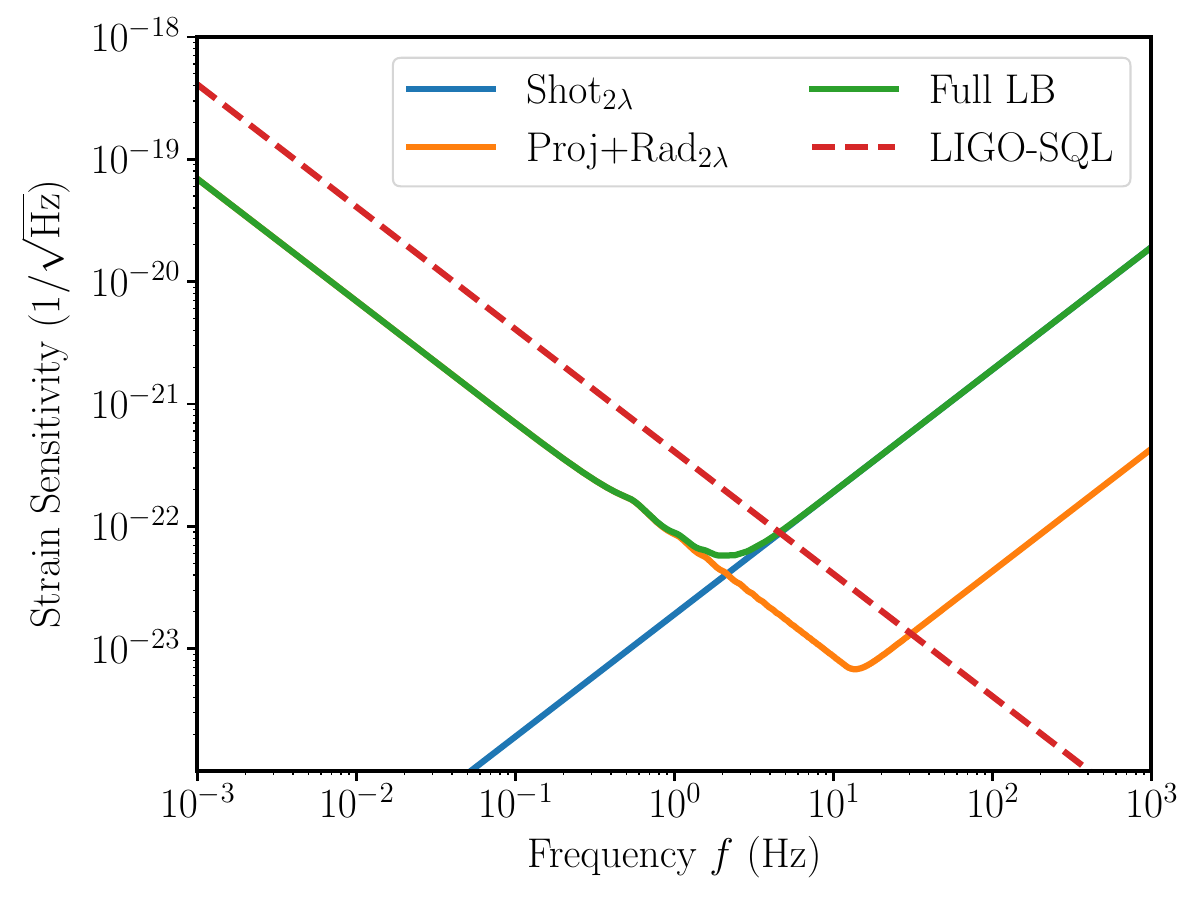}
    \caption{\fontsize{12pt}{14pt}\selectfont Shot noise of the $2\lambda$-laser compared with the full lower bound. For frequencies in the regime $f<f_c=1.87\mathrm{Hz}$, the ratio $S_{\mathrm{shot},2\lambda}(f)/S_{\mathrm{LB}}(f)$ is less than $1\mathrm{dB}$, and the full lower bound is well approximated by $S_{\mathrm{atom}} + S_{\mathrm{rad},2\lambda}$.}
    \label{Sfig:radiation pressure}
\end{figure}
As shown in Fig.~\ref{Sfig:A} and Fig.~\ref{Sfig:radiation pressure}, for $f <f_c=1.87\mathrm{Hz}$, the lower bound is well approximately by $S_{\mathrm{LB}}(f) = S_0(f)$, which takes a form of:
\begin{equation}
    S_{\mathrm{LB}}(f) = S_{\mathrm{SQL}}(f)\frac{(\Delta \theta)_{\mathrm{proj}}}{\eta U_0 W(f,T)}\sqrt{\frac{2\hbar  P_{\mathrm{arm},2\lambda}}{ \K c T}} \, . 
\end{equation}
Here, $S_{\mathrm{SQL}}(f) = 8\hbar/[ML^2(2\pi f)^2]$, as defined in the main text, represents the SQL of a conventional interferometric GW detector with test mass $M$ and arm length $L$. 

To investigate the scaling relationship of $S_{\mathrm{LB}}(f)$ with atomic parameters, we observe that $U_0 = \alpha_{2\lambda} I_U/2$. Here, $\alpha_{2\lambda}$ denotes the dynamic polarizability of atoms at the $2\lambda$ wavelength, and $I_U$ is the intensity of the  $2\lambda$-laser used to generate $U_{\mathrm{lat}}$ in each Michelson arm. Assuming a laser waist radius of $w = 50 \, \mu\mathrm{m}$, the power of the 
$2\lambda$-laser in each arm is calculated as $\Ptwo = \pi w^2 I_U/2$. Consequently, the lower bound takes a form of:
\begin{equation}
    S_{\mathrm{LB}}(f) = S_{\mathrm{SQL}}(f)\frac{(\Delta \theta)_{\mathrm{proj}}}{\eta  \sqrt{U_0 T}} \frac{1}{W(f,T)} \sqrt{\frac{2 \pi \hbar w^2}{\K c \alpha_{2\lambda}}} \,,
\end{equation}
which recovers Eq.~\textcolor{magenta}{(3)} in the main text with $C_{\mathrm{atom}}$ given by:
\begin{equation}
    C_{\mathrm{atom}}(f,T) = \frac{1}{W(f,T)} \sqrt{\frac{2 \pi \hbar w^2}{\K c \alpha_{2\lambda}}} \,.
\end{equation}

\begin{figure}
    \centering
    \includegraphics[width=\linewidth]{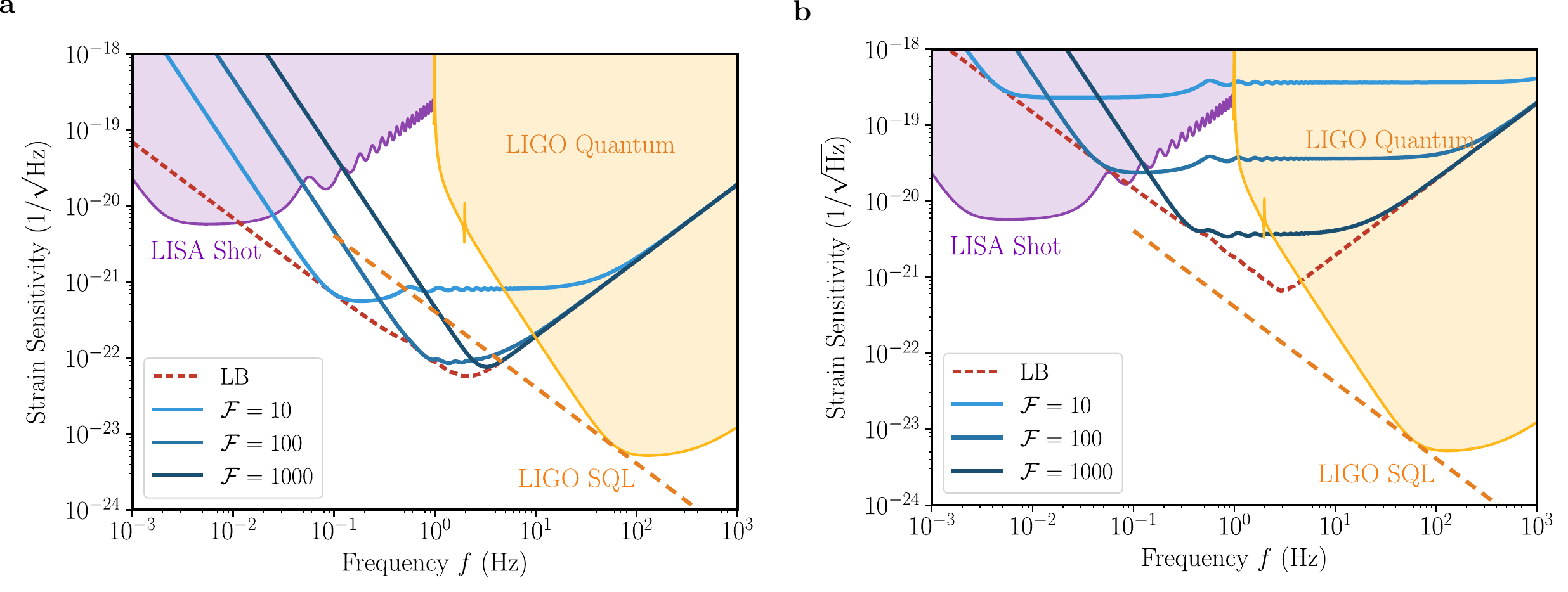}
    \caption{Comparisons of squeezed orbital states and $p$-orbital condensed state. \textbf{a}: Orbital optomechanical sensor achieves a $8$ dB improvement over LIGO's SQL if the initial state for detection is $|\xi_R \rangle_{\rm atom} \otimes |\rm{vac}\rangle_{\rm photon}$. Here, $|\xi_R\rangle_{\rm atom}$ represents an orbital squeezed state of atoms with a squeezing parameter $\xi_R$, and $|\rm{vac} \rangle_{\rm photon}$ denotes the vacuum dark port input state. \textbf{b}:Orbital optomechanical sensor behaves worse than LIGO's SQL if the initial state for detection is a orbital condensed state $|p_x \rangle^{\otimes N}_{\rm atom} \otimes |\rm{vac}\rangle_{\rm photon}$.}
    \label{Sfig:no_squeezing}
\end{figure}

\begin{figure}
    \centering
    \includegraphics[width=0.5\linewidth]{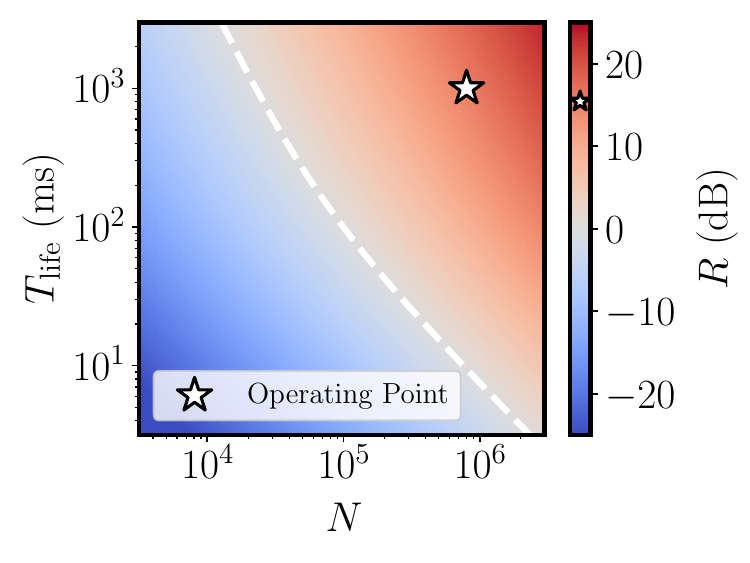}
    \caption{\textbf{Performance Advantage of the OOM Sensor.} The plot shows the ratio ($R\equiv\min_f \sqrt{S_{\mathrm{LB}}(f)/S_{\mathrm{SQL}}(f)}$ ) in dB of the OOM sensor's sensitivity to the conventional SQL, as a function of the total atom number $N$, and the $p$-orbital state lifetime $T_{\mathrm{life}}$. The red region ($R > 0$) indicates the parameter space where the OOM sensor surpasses the SQL. The white dashed line marks the break-even contour ($R=0$ dB). The white star indicates the sensor's operating point used for the main analysis ($N=8\times 10^5$, $T_{\mathrm{life}}=1000$ ms), which corresponds to a performance enhancement of $R \approx 16$ dB, as indicated by the marker on the colorbar.}
    \label{fig:R_improvement}
\end{figure}

For current experimental parameters, orbital squeezed states are crucial to ensure that $C_{\mathrm{atom}}(f,T) (\Delta \theta)_{\mathrm{proj}}/\eta \sqrt{U_0 T} < 1$. From Fig.~\ref{Sfig:no_squeezing}, we can see that without orbital squeezing, the atomic sensor behaves worse than the SQL of a conventional sensor. Conversely, Fig.~\ref{fig:R_improvement} shows the broad parameter space of atom number $N$ and $p$-orbital lifetime $T_{\mathrm{life}}$ where our sensor does surpass the SQL, highlighting the robustness of the advantage provided by orbital squeezing.

\subsection{ Influence of Atomic Species on Sensitivity}\label{Ssubsec:species}
Our previous calculations assume an atomic sensor composed of $^{7}\mathrm{Li}$. This section investigates how different atomic species affect sensor performance,
mainly considering the following three aspects:

\begin{figure}[htp]
    \centering
    \subfloat[\textbf{a}][]{
        \includegraphics[width=0.40\textwidth]{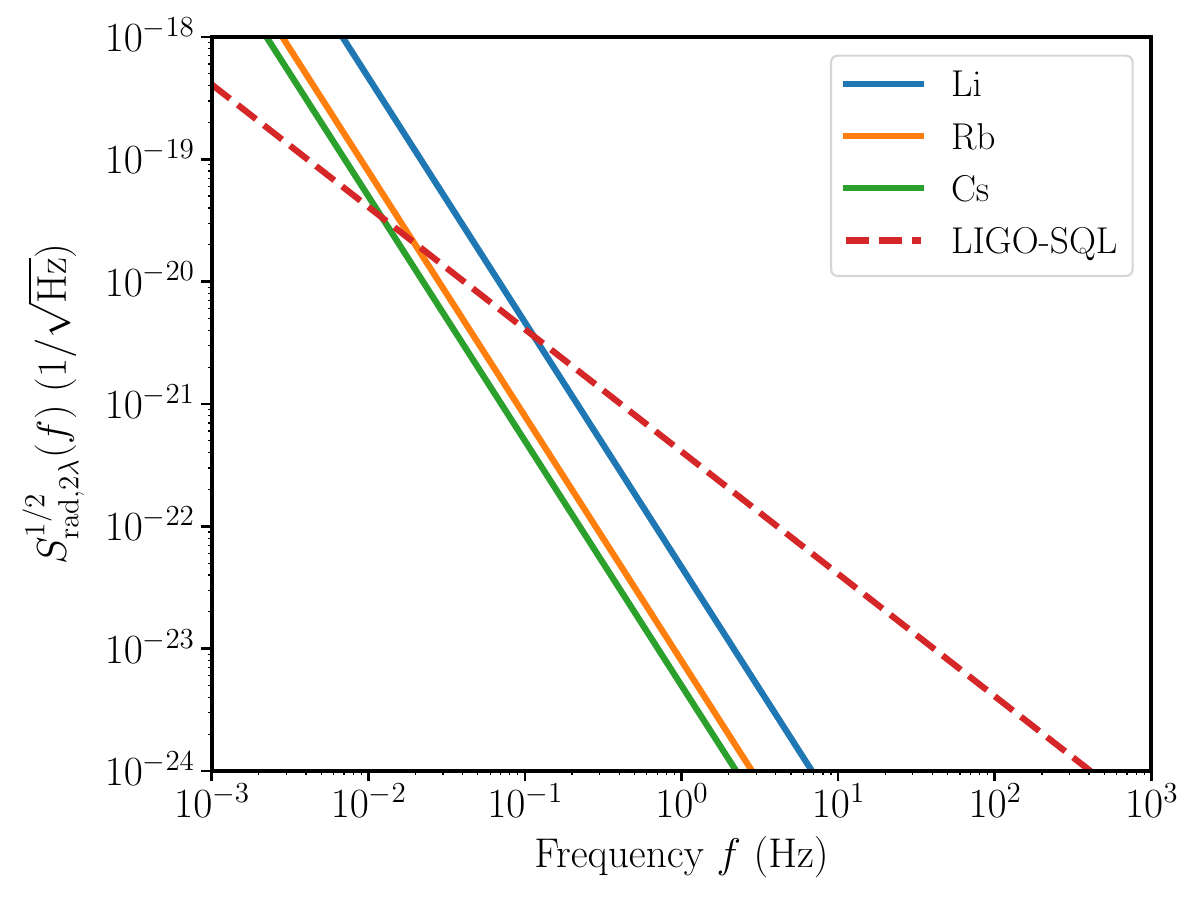}\label{Ssubfig:rad2}
    } 
    \subfloat[\textbf{b}][]{
        \includegraphics[width=0.40\textwidth]{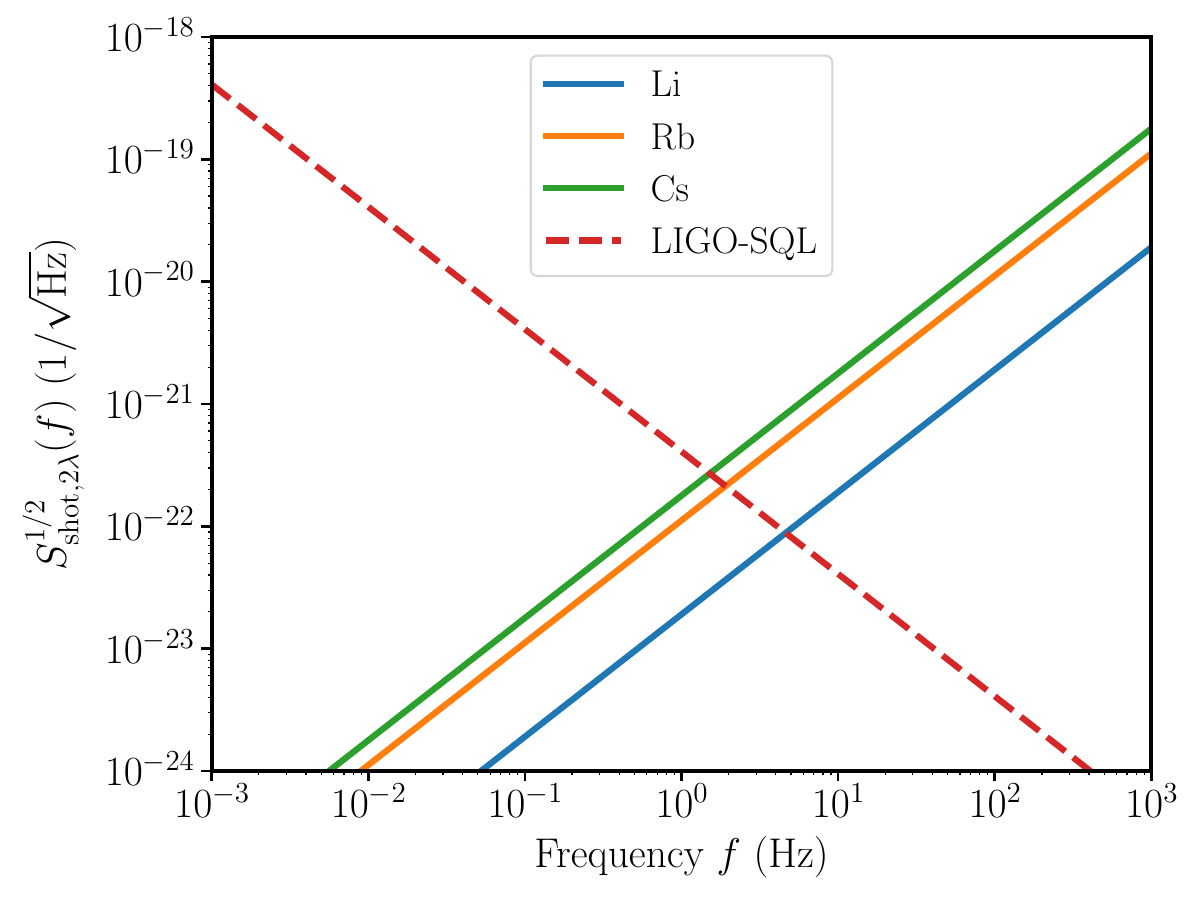}\label{Ssubfig:shot2}
    }\\
    \subfloat[\textbf{c}][]{
        \includegraphics[width=0.40\textwidth]{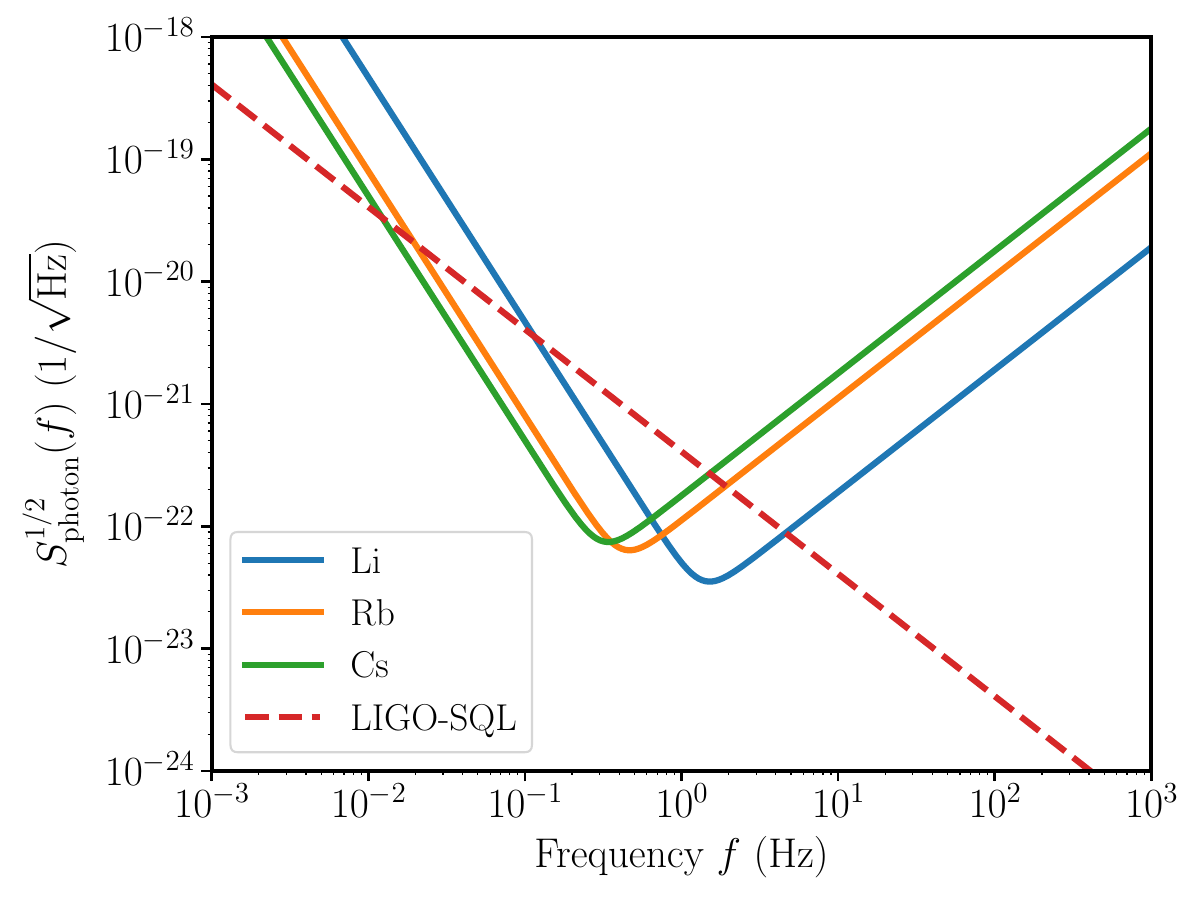}\label{Ssubfig:photon}
    }
    \subfloat[\textbf{d}][]{
        \includegraphics[width=0.40\textwidth]{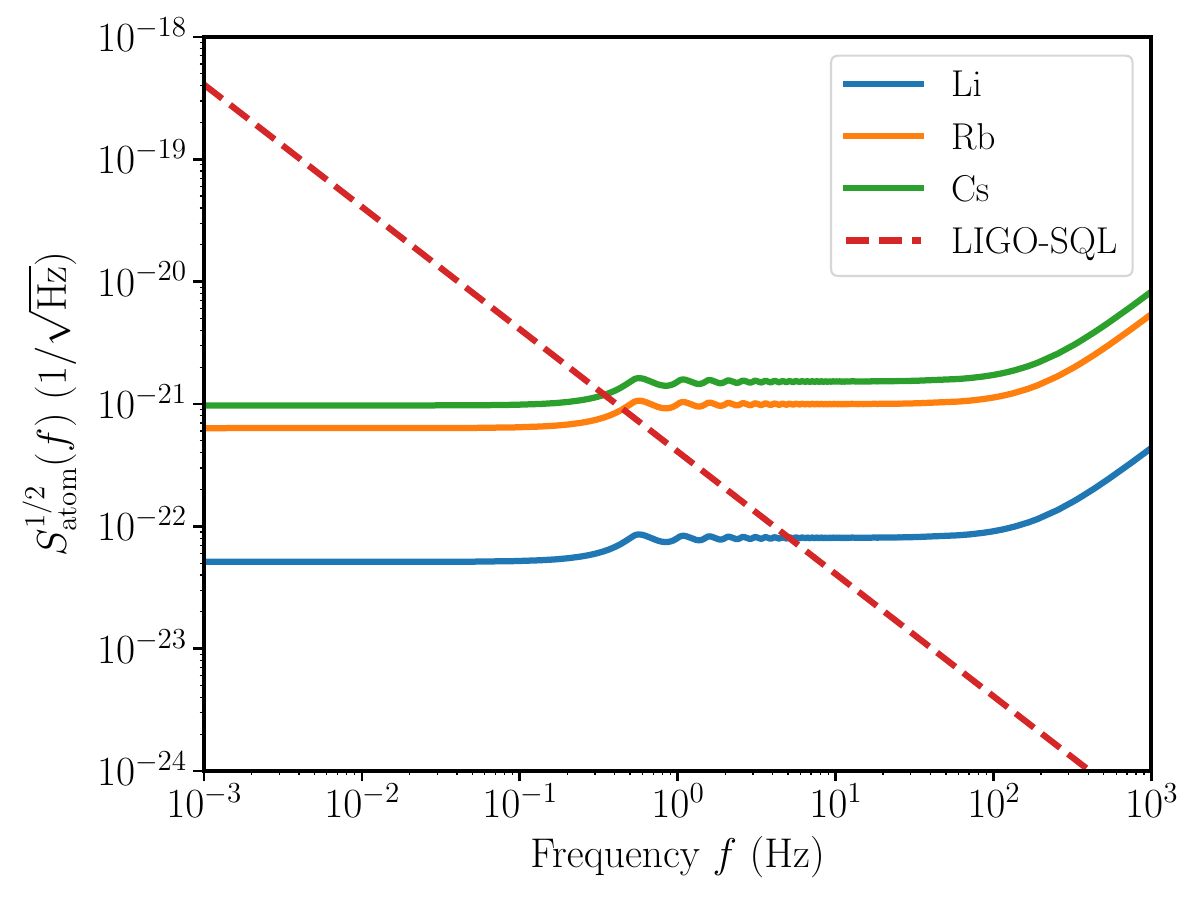}\label{Ssubfig:proj}
    }\\
    \subfloat[\textbf{e}][]{
        \includegraphics[width=0.40\textwidth]{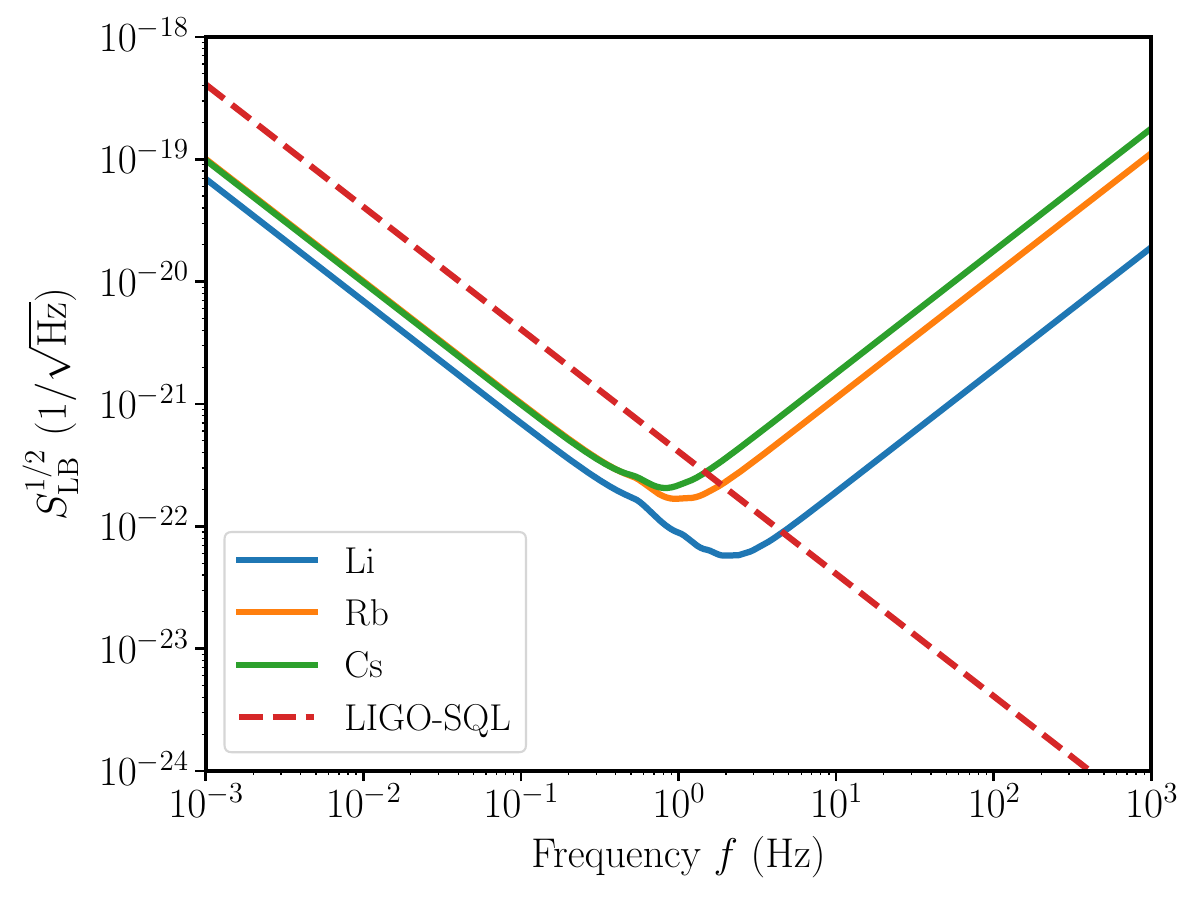}\label{Ssubfig:LB}
    }
    \caption{\fontsize{12pt}{14pt}\selectfont  Detection sensitivity with different atomic species such as   $^{7}\mathrm{Li}$, $^{87}\mathrm{Rb}$ and $^{133}\mathrm{Cs}$. 
\textbf{a}: Radiation pressure noise of the 
$2\lambda$-laser in Eq.~\eqref{Seqn:Srad2} with $\mathcal{F}=100$,
    \textbf{b}: Shot noise of the $2\lambda$-laser in Eq.~\eqref{Seqn:Sshot2} with $\mathcal{F}=100$, 
    \textbf{c}: Full photon noise in Eq.~\eqref{Seqn:Sphoton2} with $\mathcal{F}=100$,  
    \textbf{d}: Quantum projection noise of atoms in Eq.~\eqref{Seqn:Satom} with $\mathcal{F}=100$ ,
    \textbf{e}: Full lower bound of the strain sensitivity in Eq.~\eqref{Seqn:SLB}. $^7\mathrm{Li}$ performs the best near $1\mathrm{Hz}$ because that $S_{\mathrm{LB}}$ scales as the recoil energy as  $E_R^{-1/2}$ in this regime and $^7\mathrm{Li}$ has the largest recoil energy among the alkali metal atoms.}
    \label{Sfig:atom species}
    \vspace{-10pt}
\end{figure}

1. The signal lattice depth $U_0$
would vary with different atoms. The projection noise $S_{\mathrm{atom}}^{1/2}(f)$ is inversely proportional to the signal lattice depth $U_0$, as indicated by Eq.~\eqref{Seqn:Satom}. However, 
$U_0$ cannot be arbitrarily large for the experimental constraints in optical lattices.  Specifically, the trap lattice depth $V_0$ must be of the order of magnitude $O(E_R)$ to maintain the stability of the $p$-orbital condensate. Moreover, the two-mode-approximation condition requires $V_0 \gg U_0$. For a consistent comparison, we set $V_0=6E_R$ and $U_0 = 0.1E_R$ for all atomic species. Given that $E_R=\hbar^2 \K^2/2m$ is inversely proportional to the atomic mass, a lighter atom such as $^7 \mathrm{Li}$ results in reduced quantum projection noise, as illustrated in Figure~\ref{Ssubfig:proj}.

2. The laser power $P_{\mathrm{arm}}$ used to generate optical lattices varies with atoms. The radiation pressure noise amplitude, $S_{\mathrm{rad}}^{1/2}(f)$, is proportional to the square root of the laser power $\sqrt{P_{\mathrm{arm}}}$. Since the required power to achieve a given lattice depth $U_0$ depends on the atomic polarizability ($P_{\mathrm{arm}} =\pi w^2 U_0 /\alpha_{\mathrm{atom}}$), this noise source is also species-dependent. For convenience, we list the polarizability of Li, Rb, Cs~\cite{UDportal} as : $\alpha_{\mathrm{Li}}(532\mathrm{nm}) = 284.36 \mathrm{ a.u.}$, $\alpha_{\mathrm{Li}}(1064\mathrm{nm}) = 260.66 \mathrm{ a.u.}$, $\alpha_{\mathrm{Rb}}(532\mathrm{nm}) = 258.78 \mathrm{ a.u.}$, $\alpha_{\mathrm{Rb}}(1064\mathrm{nm}) = 722.37 \mathrm{ a.u.}$, $\alpha_{\mathrm{Cs}}(532\mathrm{nm}) = 220.07 \mathrm{ a.u.}$, $\alpha_{\mathrm{Cs}}(1064\mathrm{nm}) = 1372.9 \mathrm{ a.u.}$, with the atom unit given by $1\mathrm{ a.u.} = 1.6479\times 10^{-41} \ \mathrm{Hz}/(\mathrm{V}/\mathrm{m})^2$.

3. The interaction control largely depends on atoms: The interaction between $^{7}\mathrm{Li}$ atoms is more manageable and experimentally accessible compared to other alkali atoms. Specifically, the preparation of cold atomic $p$-orbital squeezing begins with a weakly interacting BEC in the excited band of a bipartite optical lattice~\cite{wirth2011evidence,2021_Zhou_PRL,2021_Xu_Nature}. Squeezing in the $p$-orbital is achieved by an interaction quench, where the interaction strength is abruptly increased. 
This is accessible to the current cold atom technology using Feshbach resonances. 
For instance, in $^7$Li atoms, a broad Feshbach resonance is present at approximately 737~G~\cite{2009_Hulet_PRL,2020_Hulet_Lithium}. The initial $p$-orbital BEC is formed with the weakly interacting state $|F=1,m_F =-1\rangle$. The interaction quench is realized by driving a Raman transition that converts the atomic internal state to $|F=1, m_F = 1\rangle$, which exhibits strongly repulsive interactions.

As shown in Figure~\ref{Ssubfig:LB}, $^{7}\mathrm{Li}$ exhibits the highest sensitivity, considering the feasible parameters in atomic optical lattice experiments.  
We expect $^{7}\mathrm{Li}$ to be the optimal choice for the atomic species in our orbital optomechanical sensor.

\subsection{ Sky- and Polarization-averaging}\label{Ssubsubsec:average}
This section addresses the pattern function and angular sensitivity of our orbital optomechanical sensor. In the above calculations of sensitivity curves, a plus-polarized gravitational wave propagating along the $\mathbf{e}_z$ direction is assumed. Now consider a generic gravitational wave $h_{ij}(t) = h_{+}(t) \mathbf{e}^{+}_{ij} + h_{\times}(t) \mathbf{e}^{\times}_{ij}$ propagating along $\mathbf{e}_n=(\theta,\phi)$ direction, where $\mathbf{e}^{+,\times}_{ij}$ are the polarization tensors of $+,\times$ polarization, and $(\theta,\phi)$ the sky-angles. The GW-induced phase shifts in the $\X$, $\Y$ arms are generalized from Eq.~\eqref{Seqn:DeltaPhiX} and Eq.~\eqref{Seqn:DeltaPhiY} to~\cite{2007BookMaggiore}:
\begin{align}
    \Delta \Phi_{\X} &= \frac{1}{2} \mathcal{T}_{\Phi}h_{\X \X} = \frac{1}{2} \mathcal{T}_{\Phi} \Big[ h_{+} \left(\cos^2\theta \cos^2\phi-\sin^2\phi \right) + h_{\times}\cos\theta \sin (2\phi)\Big]\,,\\
    \Delta \Phi_{\Y} &= \frac{1}{2} \mathcal{T}_{\Phi}h_{\Y \Y} = \frac{1}{2} \mathcal{T}_{\Phi} \Big[ h_{+} \left(\cos^2\theta \sin^2\phi-\cos^2\phi \right) - h_{\times}\cos\theta \sin (2\phi) \Big] \,,
\end{align} 
where the expressions for $h_{\X \X}$ and $h_{\Y \Y}$ incorporate the directional dependence of the GW. Consequently, the GW-induced pseudo-magnetic field, previously given by Eq.~\eqref{Seqn:Bx}, must be revised. 
Notably, in a general scenario where $\Phi_{x,y}$ are not set to $0$ or $\pi$, the pseudo-magnetic field $B_x$ can be non-zero even in the absence of a GW. The additional pseudo-magnetic field $\Delta B_x$ induced by the GW is:
\begin{align}\label{Seqn:DeltaBx}
    \Delta B_x = \frac{\partial B_x}{\partial \Phi_x}\Delta \Phi_x + \frac{\partial B_x}{\partial \Phi_y}\Delta \Phi_y = \frac{1}{2} \eta(V_0,\phi) U_0 \T_{\Phi}(f) \big(\sin \Phi_{\X} h_{\X\X} - \sin \Phi_{\Y} h_{\Y\Y} \big) \,,
\end{align}
where $\Phi_{\X,\Y}$ are the phase accumulations in $\X,\Y$ arms and relates to the phases $\Phi_{x,y}$ in Eq.~\eqref{Seqn:Ulat} by $\Phi_{\X}=\Phi_x-\Phi_y$ and $\Phi_{\Y}=\Phi_x+\Phi_y$. The input GW signal is derived by comparing Eq.~\eqref{Seqn:Bx} and Eq.~\eqref{Seqn:DeltaBx}:
\begin{equation}
    h(t) = \frac{1}{2}\big(\sin \Phi_{\X} h_{\X\X} - \sin \Phi_{\Y} h_{\Y\Y} \big) = h_{+}(t)F_{+}(\theta,\phi;\psi=0) + h_{\times}(t)F_{\times}(\theta,\phi;\psi=0) \,.
\end{equation}
Here, $F_{+,\times}$ are the pattern functions of our sensor, and $\psi$ is the polarization angle, indicating the orientation of the $\X,\Y$ axes relative to the axis to whom GW polarization is defined. The pattern functions are then given by:
\begin{align}
F_{+}(\theta, \phi ; 0) &= \frac{1}{2} \Big[\left(\cos^2\theta \cos^2\phi-\sin^2\phi \right) \sin \Phi_{\X} - \left(\cos^2\theta \sin^2\phi-\cos^2\phi \right) \sin \Phi_{\Y} \Big] \,, \\
F_{\times}(\theta, \phi ; 0) &=  \cos \theta \sin \phi \cos \phi  \big(\sin \Phi_{\X} + \sin \Phi_{\Y}\big) \,, \\
F_{+}(\theta, \phi ; \psi) &=F_{+}(\theta, \phi ; 0) \cos 2 \psi-F_{\times}(\theta, \phi ; 0) \sin 2 \psi \,, \\
F_{\times}(\theta, \phi ; \psi) &=F_{+}(\theta, \phi ; 0) \sin 2 \psi+F_{\times}(\theta, \phi ; 0) \cos 2 \psi \,.
\end{align}
These pattern functions describe how our orbital optomechanical sensor response to GWs with different propagating direction and polarizations. The sky- and polarization- averaged signal response function is further defined as the average of pattern functions, and have the expressions~\cite{2007BookMaggiore,2019LISA}:
\begin{align}
    \mathcal{R}(f) &= \langle |F_{+}|^2 \rangle = \langle |F_{\times}|^2 \rangle = \frac{1}{8\pi^2}  \int^{\pi}_0 \sin \theta \d \theta \int^{2\pi}_0 \d \phi \int^{2\pi}_0 \d \psi |F_{+}(\theta,\phi;\psi)|^2 \,, \notag \\
    &= \frac{1}{15} \big( \sin \Phi_{\X}^2+ \sin \Phi_{\X}\sin \Phi_{\Y}+ \sin \Phi_{\Y}^2 \big) \leq \frac{1}{5} \,.
\end{align}
The maximum response is achieved when $\Phi_{\X}=\Phi_{\Y}=\pi/2$, or equivalently $\Phi_x=\pi/2,\Phi_y=0$ (up to $\pi$ phases), which also maximizes the pseudo-magnetic field induced by a plus-polarized GW (see Eq.~\eqref{Seqn:Bx}).  It is also remarkable that under these parameters, there is no pseudo-magnetic field in the flat spacetime. 
The sky- and polarization- averaged sensitivity is obtained as~\cite{2019LISA}:
\begin{equation}
    S_n^{\mathrm{avg}}(f) = \frac{S_n(f)}{\mathcal{R}(f)} = 5S_n(f) \ \ (\mathrm{when } \Phi_x=\pi/2,\Phi_y=0) \,.
\end{equation}
All the sensitivity curves plotted in figures of the main text or Supplementary Materials represent the sky- and polarization- averaged sensitivity. 

\medskip 
\section{ Classical and Technical Noise Limits}\label{Ssec:classical_noise}
Achieving the quantum-limited sensitivity described in Sec.~\ref{Ssec:Noises} requires a thorough understanding and suppression of classical noise sources. The ultimate performance of any terrestrial or space-based interferometer is first constrained by fundamental displacement noise from its environment. To approach this boundary, technical noise arising from the instrument itself must be meticulously controlled. This section analyzes these critical noise contributions, beginning with the fundamental displacement limits (\ref{Ssubsec:displacement_noise}) and then detailing the technical requirements on the signal laser (\ref{Ssubsec:signal_noise}), primary laser (\ref{Ssubsec:primary_noise}), and atomic readout system (\ref{Ssubsec:readout_noise}). By quantifying these noise budgets, we establish the key engineering requirements for the orbital optomechanical sensor to operate at its full potential.

\subsection{ Fundamental Displacement Noise Limits}\label{Ssubsec:displacement_noise}
Beyond the quantum noise that defines the interferometer's intrinsic sensitivity, the sensor's performance is constrained by fundamental sources of classical displacement noise. The nature of these noises and the requirements for their mitigation differ significantly between terrestrial and space-based operation.

\begin{figure}[htp]
    \centering
    \subfloat[\textbf{a} Displacement Noise Budget][]{\includegraphics[width=0.47\textwidth]{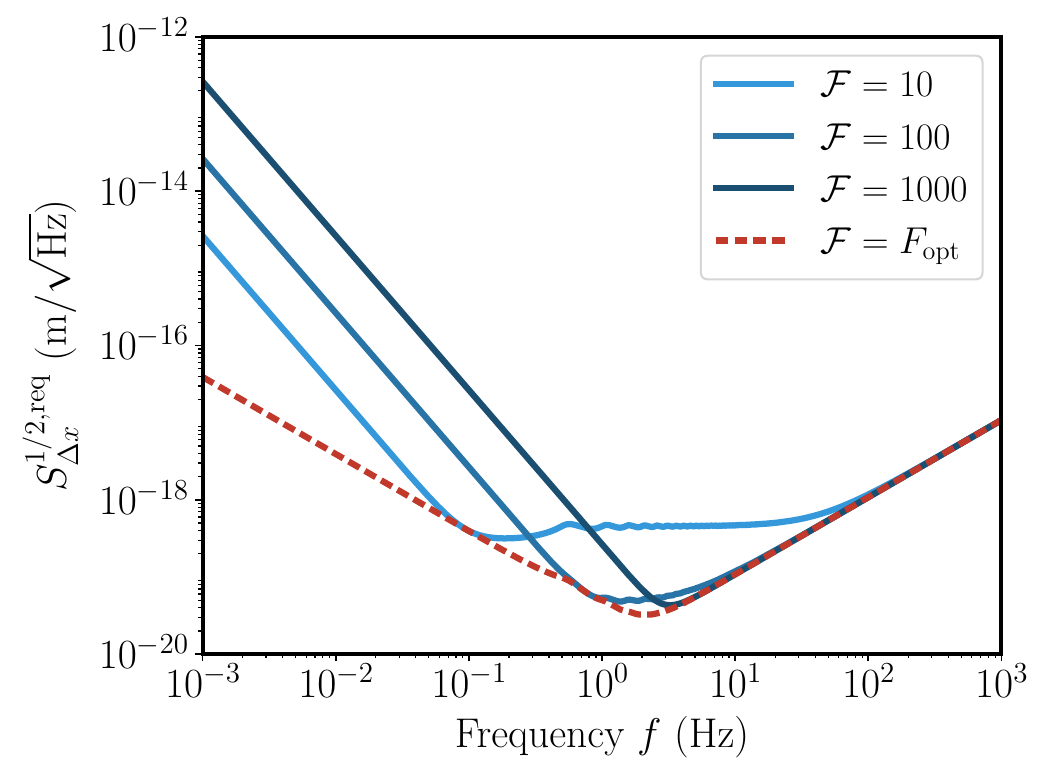}
        \label{Ssubfig:displacement_noise}
    }
    \subfloat[\textbf{b} Acceleration Noise Requirement][]{\includegraphics[width=0.47\textwidth]{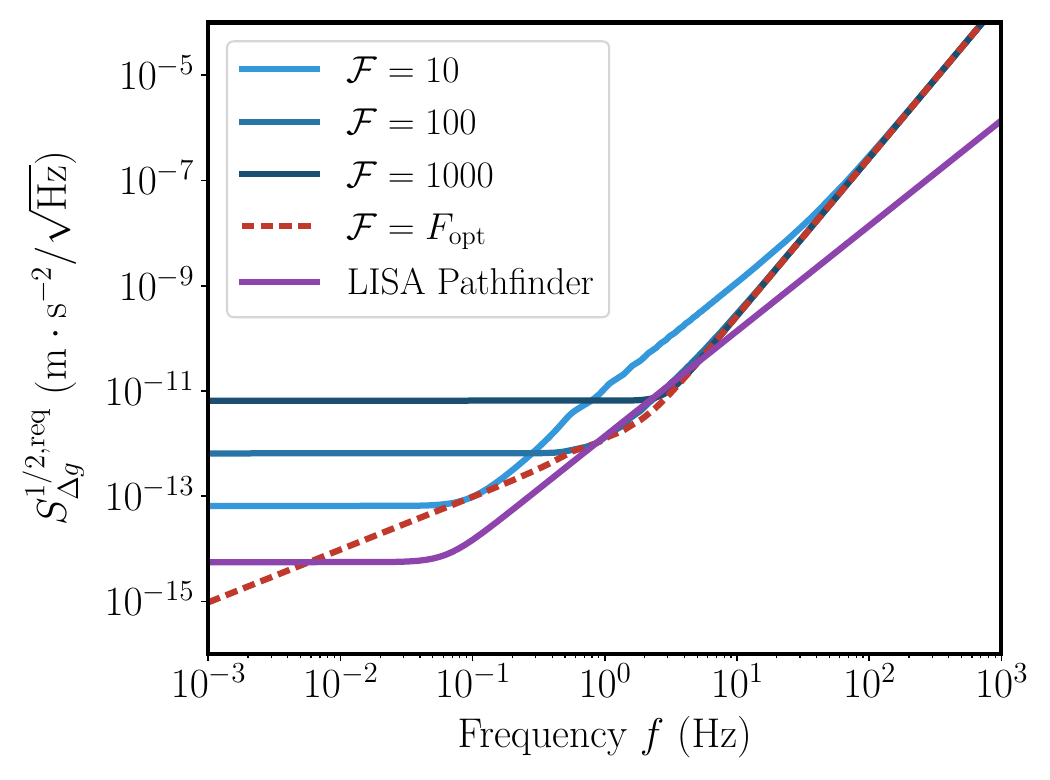}
        \label{Ssubfig:acceleration_noise}
    }
    \caption{Requirements on displacement and acceleration noise. (\textbf{a}) Required differential displacement noise budget, $S_{\Delta x,\text{req}}^{1/2}$. This limit applies to the sum of all classical displacement noises for a ground-based sensor, including seismic, Newtonian, and thermal noise. (\textbf{b}) Required differential acceleration noise budget, $S_{\Delta a,\text{req}}^{1/2}$, for a space-based sensor. The achieved performance of the LISA Pathfinder mission~\cite{2016LISA_pathfinder} is shown for comparison. In both panels, solid lines are for a fixed finesse ($\mathcal{F}$) and the dashed red line is for an optimized, frequency-dependent finesse ($F_{\mathrm{opt}}$).
}
    \label{Sfig:other_noises}
\end{figure}

\subsubsection{ Ground-Based Operation}
A ground-based implementation of the orbital optomechanical sensor would be limited by dominant environmental displacement noises. These include seismic and Newtonian noise at low frequencies (typically below 10~Hz), and thermal noise from sources like mirror coatings and suspensions at higher frequencies~\cite{ligo2015cqg,cahillane2021controlling,2025LIGO_O4}. To achieve quantum-noise-limited sensitivity, the power spectral density (PSD) of the total differential displacement noise from all classical sources, $S_{n,\Delta x}$, must be suppressed well below the sensor's quantum noise limit, $S_n(f)$. We establish this requirement by demanding $S_{n, \Delta x}(f) \le 0.1 \cdot L^2 S_n(f)$.

The resulting required noise amplitude spectral density, $S_{\Delta x,\text{req}}^{1/2}$, is plotted in Fig.~\ref{Ssubfig:displacement_noise}. The plot reveals that in the most sensitive band (around $2$~Hz), environmental noise must be mitigated to an exceptional level of $10^{-19}$--$10^{-20} \ \mathrm{m}/\sqrt{\mathrm{Hz}}$. While this is approximately two orders of magnitude more stringent than the isolation in current Advanced LIGO detectors~\cite{2020LIGO_O3,2025LIGO_O4}, it is a key design target for next-generation observatories~\cite{ET_design,ET_ScienceCase,ETandCE,CE_horizon}. Facilities like the Cosmic Explorer and the Einstein Telescope aim to achieve such levels by utilizing deep underground locations and cryogenic technologies to drastically reduce seismic, Newtonian, and thermal noise.

\subsubsection{ Space-Based Operation}
Deploying the sensor in space effectively eliminates terrestrial noise sources, offering a clear path to higher sensitivity. However, this introduces a different fundamental noise limit: spacecraft acceleration noise~\cite{danilishin2012quantum,2019LISA}. This residual noise, arising from factors like solar radiation pressure fluctuations, micrometeoroid impacts, and thruster activity, can impart non-gravitational forces on the test masses and mimic a GW signal.

The displacement noise budget from the quantum limit (Fig.~\ref{Ssubfig:displacement_noise}) is translated into a requirement on the maximum allowable differential acceleration noise, since their spectra are related by $S_a(f) = (2\pi f)^4 S_x(f)$. The requirement $S_{\Delta a,\text{req}}^{1/2}$, is presented in Fig.~\ref{Ssubfig:acceleration_noise}. To make a fair comparison with existing space-based instruments, this acceleration noise budget is calculated assuming an arm length equivalent to that of the LISA mission $L_{\mathrm{LISA}} = 2.5\times10^{6} \ \mathrm{km}$. This requirement is benchmarked against the quantum noise limit, $S_n(f)$, which corresponds to the $4$ km baseline of our proposed sensor.

Crucially, the plot compares this requirement to the performance already achieved by the LISA Pathfinder mission~\cite{2016LISA_pathfinder}, which serves as a benchmark for state-of-the-art inertial control in space. For a significant portion of the target frequency band (below approximately 1~Hz), the required inertial stability for our sensor technology is less stringent than the performance already achieved by LISA Pathfinder. This provides strong evidence that the challenge of mitigating spacecraft acceleration noise is well within the capabilities of current, proven technologies. We therefore conclude that a space-based deployment is a highly promising route for this sensor, with acceleration noise representing a significant but solvable engineering challenge.

\subsection{ Signal Laser Noise}\label{Ssubsec:signal_noise}
This subsection examines the coupling of classical (technical) signal laser noise to the atomic readout signal, specifically how classical laser fluctuations affect the orbital rotation angle $\theta$ defined in Eq.~\eqref{Seqn:theta_def}. Our analysis, which assumes ideal plane-wave electric fields and neglects effects like mode-mismatch and misalignment~\cite{izumi2016frequency1,izumi2016frequency2,izumi2016frequency3}, focuses on two primary noise types: frequency noise and intensity noise~\cite{2006PRDLaserNoise,cahillane2021controlling}.

In a perfectly balanced Michelson interferometer, these classical fluctuations are common-mode in both arms and cancel within the differential phase $\Delta \Phi = \Delta \Phi_{\X} - \Delta \Phi_{\Y}$. However, any practical implementation will feature small asymmetries between the arms that break this cancellation. These imbalances include differences in~\cite{izumi2016frequency3,cahillane2021controlling}:
\begin{itemize}
    \item Arm cavity reflectivity $r_a$
    \item Cavity pole frequency $\omega_c$ (or equivalently the finesse $\F$)
    \item Reduced mass of the test mass pair
          (ITM and ETM) $\mu = m_i m_e /(m_i + m_e)$
    \item Arm cavity power gain $g_{\mathrm{cav}}$
\end{itemize}
These asymmetries couple laser noise to $\Delta \Phi$, as detailed in Secs.~\ref{Ssubsec:signal_freq} and~\ref{Ssubsec:signal_intensity}. Notably, while conventional interferometers require an intentional differential arm length (DARM) offset ($\Delta L_\epsilon$) for DC readout~\cite{izumi2016frequency2},  our orbital optomechanical sensor can, in principle, operate in a perfectly balanced configuration.

\subsubsection{ Signal Lattice with Arm Imbalances}
We analyze the signal lattice and its associated pseudo-magnetic field under conditions of arm imbalance. For a resonant carrier field, the amplitude reflectivity $r_a^{(\X,\Y)}$ of the Fabry-Perot arm cavities is given by~\cite{izumi2016frequency1}:
\begin{equation}
    r_a^{(\X,\Y)} = \frac{r_e(t_i^2+r_i^2)-r_i}{1-r_ir_e} \equiv r_a \pm \delta r_a \,,
\end{equation}
where $r_i,t_i$ are the reflectivity and transmissivity of ITM and $r_e$ is the reflectivity of ETM. The terms $r_a^{(\X)}$ and $r_a^{(\Y)}$ denote the reflectivities for the X and Y arms, respectively. We use this notation for all cavity properties (e.g., $\omega_c \pm \delta\omega_c$). The specific imbalance parameters used in our simulations are listed in Table~\ref{tab:noise_parameters}.

The signal lattice potential in the presence of imbalance is modified from Eq.~\eqref{Seqn:Ulat} and Eq.~\eqref{Seqn:Ulat_withQuantumNoise} to become:
\begin{align}
    U_{\mathrm{lat}}(\mathbf{r};r_a,\delta r_a) =  -&\frac{U_{0}+\delta U_0}{4} \Big|\left[(\mathbf{e}_{z}\cos \alpha_U +\mathbf{e}_{\Y}\sin \alpha_U )  e^{i\K\X/2}e^{i\delta \Phi_R}+ r_a^{(\X)}\mathbf{e}_{z} e^{-i\K\X/2}e^{-i\Phi_\X}e^{i\delta \Phi_L}\right] \notag \\
    & \ \ \quad \quad \quad \, \ +\mathbf{e}_{z} e^{i(\phi_U+\Phi_{\Y}/2)}\left( e^{i\K\Y/2}e^{i\delta \Phi_U} + r_a^{(\Y)}  e^{-i\K\Y/2}e^{-i\Phi _\Y}e^{i\delta \Phi_D}\right) \Big|^{2}  \,.
\end{align}
The polarization rotation angle $\alpha_U = \arccos{r_a}$ maintains the degeneracy between $p_x$ and $p_y$ orbitals. The phase $\Phi_{\X,\Y}=\pi/2$ is chosen to maximize the pseudo-magnetic field generated by gravitational waves, as described in Eq.~\eqref{Seqn:Bx}. The differential phase in this configuration becomes $\Delta \Phi = \delta \Phi_\X - \delta \Phi_\Y = (\delta \Phi_R - \delta \Phi_L) - (\delta \Phi_U - \delta \Phi_D)$, where $\delta \Phi_{R,L,U,D}$ represent the phase modulations of the right-, left-, up-, and down-propagating electric field components, respectively.

The phase noise contributions have distinct physical origins. The terms $\delta \Phi_{R,U}$ originate from the input laser noise; in this context, we have treated the beam splitters as perfect optical elements following references~\cite{izumi2016frequency1,izumi2016frequency2,izumi2016frequency3}. The phase modulations of the reflected fields $\delta \Phi_{L,D}$ differ due to arm imbalances. Additionally, the term $\delta U_0$ accounts for amplitude fluctuations of the optical lattice potential caused by laser intensity noise. This formulation captures all relevant noise sources and imbalance effects in the signal lattice potential.

The contribution of $U_{\mathrm{lat}}$ to the two-mode pseudo-spin Hamiltonian $\hat{H}^U_0$ can be calculated using the same procedure outlined in Sec.~\ref{Ssubsec:Ulattice}. The resulting coupling terms are:
\begin{align}
    J^U_{xx} = J^U_{yy} &= -(U_0+\delta U_0) \left(1+r_a^2+\delta r_a^2\right)/2 \,, \label{Seqn:JUxx_new}\\
    J^U_{xy} = J^U_{yx} &= -(U_0+\delta U_0)\left[r_a \left(r_a+\delta r_a\right) \sin(\delta \Phi_{\X}) - (r_a-\delta r_a) \sin(\delta \Phi_{\Y}) \right] \eta(V_0,\phi)/2 \notag \,, \\
    & \simeq  \eta(V_0,\phi) U_0 (\delta \Phi_{L} - \delta \Phi_{D})/2 \label{Seqn:JUxy_new}\,,
\end{align}
where the approximation assumes $r_a \approx 1$ and neglects second-order terms $\mathcal{O}(\delta r_a \cdot \delta \Phi)$ and $\mathcal{O}(\delta U_0 \cdot \delta \Phi)$.
From Eq.~\eqref{Seqn:H0U}, we obtain the pseudo-magnetic field induced by the signal laser's phase modulation:
\begin{equation}
    \delta B_x^{\rm{sig}} = 2 J^U_{xy} = \eta(V_0,\phi) U_0 (\delta \Phi_{L} - \delta \Phi_{D}), \quad \delta B_z^{\rm{sig}} = 0.
\end{equation}
Arm imbalance thus has two crucial consequences. First, the essential $p_{x,y}$ orbital degeneracy is preserved. Second, and more critically, it creates a noise field $\delta B_x$ that directly contaminates the measurement channel for gravitational waves. The following sections quantify the stability required to suppress this effect.
\subsubsection{ Signal Laser Frequency Noise}\label{Ssubsec:signal_freq}
\begin{figure}[htp]
    \centering
    \subfloat[\textbf{a}][]{
        \includegraphics[width=0.47\textwidth]{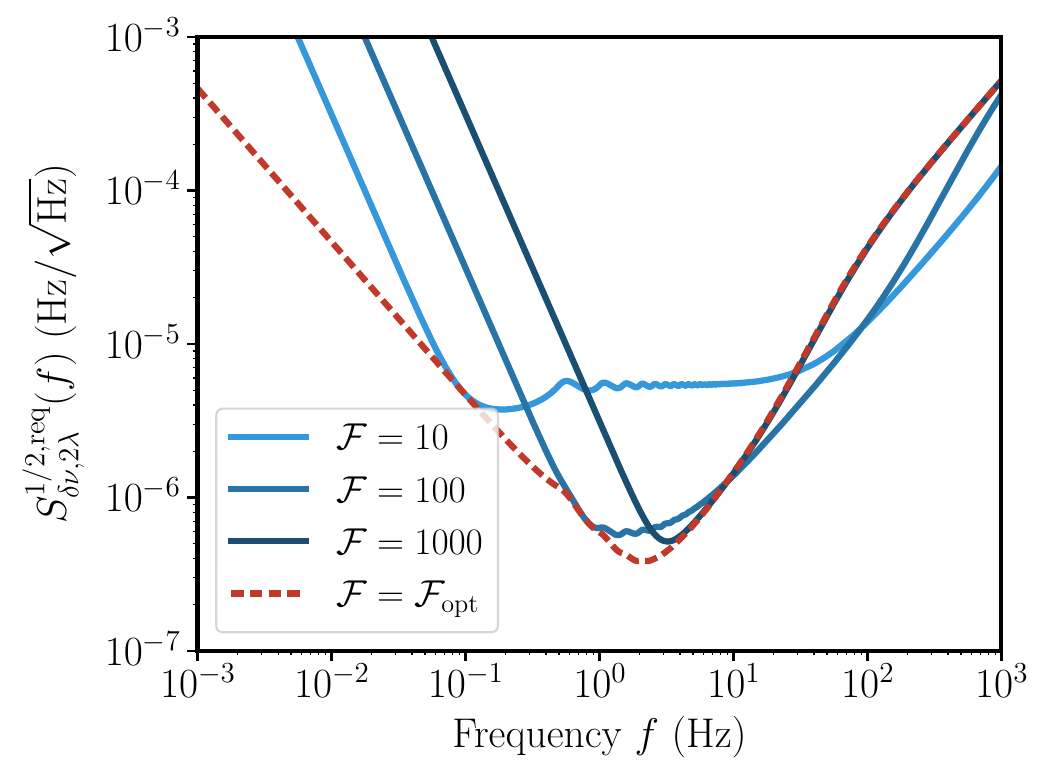}\label{Ssubfig:signal_freq}
    }
    \subfloat[\textbf{b}][]{
        \includegraphics[width=0.47\textwidth]{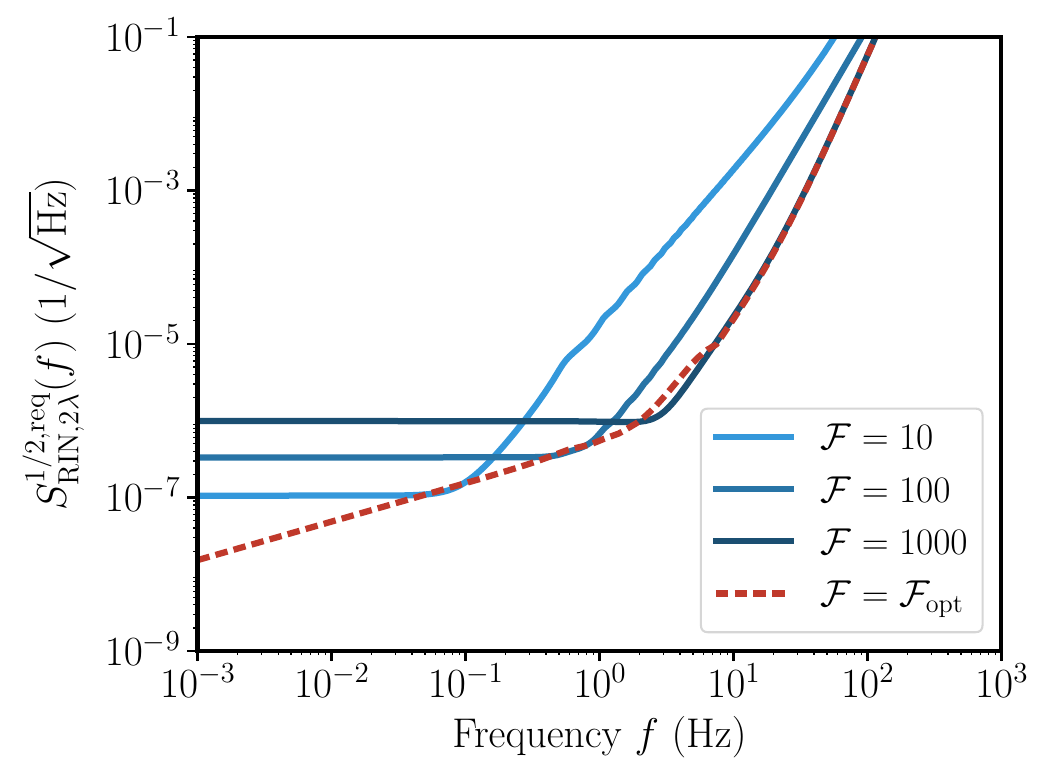}\label{Ssubfig:signal_int}
    }
    \caption{Technical noise requirements for the signal laser. (\textbf{a}) Required amplitude spectral density of the laser frequency noise. (\textbf{b}) Required amplitude spectral density of the laser's relative intensity noise (RIN). Solid curves correspond to fixed finesse ($\mathcal{F}$) values, while the dashed red curve represents the requirement for an optimized, frequency-dependent finesse that maximizes overall detector sensitivity.} 
    \label{Sfig:signal_laser_noise}
\end{figure}

The frequency noise of the signal laser can be analyzed by considering an input carrier field $E_0 e^{-i\omega_0 t}$. The frequency noise $\omega_{\delta \nu} = \omega_0 + 2\pi \delta \nu \cos{\omega t}$ manifests as phase modulation of the carrier field:
\begin{align}
    E_{\delta \nu} =  E_0 e^{-i\int \omega_{\delta \nu} dt} =  E_0 e^{-i \omega_0 t} \left(1+\frac{\pi \delta \nu}{\omega}e^{i\omega t} - \frac{\pi \delta \nu}{\omega}e^{-i\omega t}\right) \,,
\end{align}
where $E_0$ is the carrier amplitude, $\omega_0=c\K/2$ is the carrier frequency, and $\delta \nu$ and $\omega$ represent the amplitude and angular frequency of the frequency noise. Under the assumption $\delta \nu/\omega \ll 1$, we neglect higher-order sidebands. When this field reflects from a Fabry-Perot cavity resonant with the carrier, the phase modulation $\delta \Phi_r$ of the reflected field transforms to:
\begin{align}
    E_r &= E_0 e^{-i \omega_0 t} \left(r_a\cdot 1  +r_{\rm{arm}}(-\omega)\cdot\frac{\pi \delta \nu}{\omega}e^{i\omega t} - r_{\rm{arm}}(\omega) \cdot \frac{\pi \delta \nu}{\omega}e^{-i\omega t}\right)  \notag \,, \\[4pt]
    &=r_a E_0 e^{-i\omega_0 t} e^{-i\delta \Phi_r}\,,  \\[4pt]
    \delta \Phi_r& \equiv \frac{\pi \delta \nu}{i \omega}\frac{r_{\rm{arm}}(\omega)}{r_a} e^{-i\omega t} + \rm{h.c.} \,, \label{Seqn:deltaPhi_twosidebands}
\end{align}
The cavity reflectivity for sidebands $r_{\rm{arm}}$ is given by~\cite{izumi2016frequency2}:
\begin{equation}
    r_{\rm{arm}}(\omega;r_a,\omega_c) = \frac{r_a-s_c}{1+s_c} \,,
\end{equation}
where $s_c = -i\omega/\omega_c$ and $\omega_c = \pi c/(2\F L)$ represents the cavity pole. From Eq.~\eqref{Seqn:deltaPhi_twosidebands}, the Fourier component of $\delta \Phi$ becomes:
\begin{equation}
    \delta \tilde{\Phi}(\omega;r_a,\omega_c) = \frac{\pi \delta \tilde{\nu}(\omega)}{i \omega}\frac{1-s_c/r_a}{1+s_c} \,,
\end{equation}
with $\delta \tilde{\nu}(\omega)$ denoting the Fourier transform of $\delta \nu$. The differential phase modulation between arms is:
\begin{align}
    \delta \tilde{\Phi}_{L}(\omega) - \delta \tilde{\Phi}_{D}(\omega) &= \delta \tilde{\Phi}(\omega;r_a+\delta r_a,\omega_c+\delta \omega_c) - \delta \tilde{\Phi}(\omega;r_a-\delta r_a,\omega_c-\delta \omega_c) \,, \notag \\[4pt] 
    &= -\frac{2\pi\delta\tilde{\nu}(\omega)}{\omega_c} \left\{ \frac{\delta r_a}{1+s_c} + \frac{\delta \omega_c}{\omega_c} \frac{1+1/r_a}{(1+s_c)^2}\right\} \,, \notag \\[4pt]
    &= - \frac{4\F L}{c} \delta\tilde{\nu}(\omega)\cos{\beta}e^{i\beta} \left\{\delta r_a + \frac{\delta \omega_c}{\omega_c}(1+1/r_a)\cos \beta e^{i\beta} \right\}  \,. 
\end{align}
Here, $\beta(\omega) = \arctan{(\omega/\omega_c)}$ follows the definition in Eq.~\eqref{Seqn:betaKappa}. Using the transfer function from GW strain to differential phase shift (Eq.~\eqref{Seqn:T_Phi}), we define the effective strain caused by phase modulation $\tilde{h}_{\rm{freq},2\lambda}(\omega)$ as:
\begin{align}
    \tilde{h}_{\rm{freq},2\lambda}(\omega) &\equiv  (\delta \tilde{\Phi}_{L}(\omega) - \delta \tilde{\Phi}_{D}(\omega))/\T_{\Phi}(\omega) \,, \\
    &=\frac{\pi\delta\tilde{\nu}(\omega)}{\omega_0}e^{i\beta} \left\{\delta r_a + \frac{\delta \omega_c}{\omega_c}(1+1/r_a)\cos \beta e^{i\beta} \right\} \,.
\end{align}
The corresponding strain noise spectrum density is:
\begin{equation}
    S_{\rm{freq},2\lambda}(f) = \left(\frac{\pi}{\omega_0}\right)^2 \left\{ \left(\delta r_a\right)^2 + \left(\frac{\delta \omega_c}{\omega_c}\right)^2(1+1/r_a)^2 \frac{\omega^2}{\omega^2+\omega_c^2} \right\}S_{\delta \nu,2\lambda}(f) \,,
\end{equation}
where $S_{\delta \nu,2\lambda}$ represents the frequency noise spectral density of the $2\lambda$ signal laser. To ensure that the induced strain noise $S_{\rm{freq},2\lambda}(f)$ remains below 10\% of the quantum-noise-limited sensitivity $S_n(f)$ (Eq.~\eqref{Seqn:Sn}), we derive the required signal laser frequency noise spectrum $S_{\mathrm{freq},2\lambda}^{\mathrm{req}}$ shown in Fig.~\ref{Ssubfig:signal_freq}. The most stringent requirement, $5\times10^{-7} \ \rm{Hz}/\sqrt{\rm{Hz}}$ in the 1-10 Hz band, is demanding but has been achieved by state-of-the-art laser systems developed for the Advanced LIGO detectors~\cite{2025LIGO_O4}.

\subsubsection{ Signal Laser Intensity Noise}\label{Ssubsec:signal_intensity}
The intensity noise is described as modulation of the carrier field in the amplitude quadrature, as:
\begin{equation}
    E_{\delta P} = E_0 \left(1 + \frac{\delta P}{2 P} \cos(\omega t) \right)e^{-i\omega_0 t} = E_0 e^{-i\omega_0 t} \left(1 + \frac{\delta P}{4 P} e^{-i\omega t} + \frac{\delta P}{4 P} e^{i\omega t} \right) \label{Seqn:E_deltaP}
\end{equation}
where $\delta P/P$ represents the relative intensity noise (RIN) at frequency $\omega$. When this field reflects from a Fabry-Perot cavity, the intensity modulation will couple to phase modulation through the radiation pressure effect described by Eq.~\eqref{Seqn:g2}. 

For perfectly balanced $\X$ and $\Y$ arms, the radiation pressure effects cancel exactly, leaving no net effect on either the differential arm length $L_{-} = L_{\X}-L_{\Y}$ or the differential phase shift $\delta \Phi_{\X} - \delta \Phi_{\Y} = 2(\omega_0/c)L_{-}$. However, arm imbalances introduce coupling between intensity noise and DARM~\cite{izumi2016frequency3}:
\begin{equation}
    L_{-} =  -\frac{4P_{\mathrm{cav},2\lambda}}{\mu\omega^2c} \left[\frac{\delta g_{\mathrm{cav}}}{g_{\mathrm{cav}}} + \frac{\delta \omega_c}{\omega_c} \frac{s_c}{1+s_c} - 2\frac{\delta\mu}{\mu}\right]\frac{\delta P}{2P} \,.
\end{equation}
Here, $g_{\mathrm{cav}}^2 = P_{\mathrm{cav},2\lambda}/P_{\mathrm{arm},2\lambda} = 2\F/\pi$ denotes the arm cavity power gain, and $\mu$ is the reduced mass of the ITM-ETM system. This analysis assumes free test masses, where the mechanical spring constant $k_m$ from mirror suspensions is neglected, consistent with the treatment in~\cite{izumi2016frequency3}. This model can be extended to include suspension effects by replacing the $-\mu\omega^2$ term (in the denominator of the mechanical susceptibility) with $(k_m-\mu\omega^2)$.

The resulting effective strain and strain noise spectral density are given by:
\begin{align}
    \tilde{h}_{\rm{int},2\lambda}(\omega) &\equiv  L_{-}/L 
    =-\frac{2P_{\mathrm{arm},2\lambda}}{\mu\omega^2cL} g_{\mathrm{cav}}^2
    \left[\frac{\delta g_{\mathrm{cav}}}{g_{\mathrm{cav}}} + \frac{\delta \omega_c}{\omega_c} \frac{s_c}{1+s_c} - 2\frac{\delta\mu}{\mu}\right]\frac{\delta P}{P} \,. \\
    S_{\mathrm{int},2\lambda}(f) &= \left(\frac{2P_{\mathrm{arm},2\lambda}}{\mu \omega^2cL} g_{\mathrm{cav}}^2 \right)^2
    \left[\left(\frac{\delta g_{\mathrm{cav}}}{g_{\mathrm{cav}}}\right)^2 + \left(\frac{\delta \omega_c}{\omega_c}\right)^2 \frac{\omega^2}{\omega^2+\omega_c^2} + 4\left(\frac{\delta\mu}{\mu}\right)^2\right] S_{\mathrm{RIN,2\lambda}}(f) 
\end{align}
The requirements for intensity noise suppression, derived from the condition that $S_{\mathrm{int},2\lambda}(f) < 0.1 S_n(f)$, are shown in Fig.~\ref{Ssubfig:signal_int}. The analysis indicates that these requirements can be satisfied, as state-of-the-art lasers for LIGO O4 have achieved a stabilized relative intensity noise (RIN) with an amplitude spectral density of approximately $4\times 10^{-9} \ \mathrm{Hz}^{-1/2}$~\cite{2025LIGO_O4}.

\subsection{ Primary Laser Noise}\label{Ssubsec:primary_noise}
Noise from the primary laser, which generates the atom-trapping lattice, couples to the system through different and generally weaker mechanisms than signal laser noise. As we show below, this leads to significantly more relaxed stability requirements.
\subsubsection{ Primary Lattice with Arm Imbalances}
We analyze the primary lattice generated by the optical setup in Fig.~\ref{Sfig:MI}\textcolor{magenta}{(a)}, incorporating mirror reflectivity imbalances \( r_m^{(\X,\Y)} = r_m \pm \delta r_m \) and primary laser noise. The lattice potential is modified from Eq.~\eqref{Seqn:Vlat} to become:
\begin{align}
    V_{\rm{lat}}(\mathbf{r};r_m,\delta r_m) 
    =  -&\frac{V_{0}+\delta V_0}{4} \Big|\left[(\mathbf{e}_{z}\cos \alpha +\mathbf{e}_{\Y}\sin \alpha )  e^{i\K\X}+ r_m^{(\X)}\mathbf{e}_{z} e^{-i\K\X}e^{-i\delta \Phi'_{\X}}\right] \notag \\
    & \ \ \quad \quad \quad \, \, +\mathbf{e}_{z} e^{i\phi}\left( e^{i\K\Y} + r_m^{(\Y)}  e^{-i\K\Y/2}e^{-i\delta \Phi'_{\Y}}\right) \Big|^{2} \label{Seqn:Vlat_imbalanced} \,.
\end{align}
Here, $\delta \Phi'_{\X} = \delta \Phi'_{R}-\delta \Phi'_{L},\ \delta \Phi'_{R} = \delta \Phi'_{U} - \delta \Phi'_{D}$ and $\delta \Phi'_{R,L,U,D}$ represent the phase modulations of the right-, left-, up-, and down-propagating electric field components, respectively. The term $\delta V_0$ represents the amplitude fluctuation of the primary lattice depth caused by the primary laser's intensity fluctuation.

The primary lattice retains translational symmetry $ V_{\mathrm{lat}}(\mathbf{r}+\mathbf{a}_{1,2}) = V_{\mathrm{lat}}(\mathbf{r})$. The anisotropy, defined as $\delta V_{\rm{lat}}(x,y) \equiv \left[V_{\rm{lat}}(x,y)-V_{\rm{lat}}(y,x) \right]\propto \delta r_m$ can now be treated as perturbation to the original $p$-orbital states with $D_4$ symmetry and translational symmetry $\psi_\alpha(\mathbf{r}+\mathbf{R}) = (-1)^{R_{\alpha}} \psi_\alpha(\mathbf{r})$.
The corresponding pseudo-spin Hamiltonian coefficients (Eq.~\eqref{Seqn:JVArm0}) are:
\begin{equation}
    J^{V}_{\alpha \beta}(r_m,\delta r_m)= \int \d^2\mathbf{r} \ \psi^{*}_{\alpha}(\mathbf{r}) V_{\mathrm{lat}}(\mathbf{r};r_m,\delta r_m)\psi_{\beta}(\mathbf{r}) \,.
\end{equation}
The term $J^V_{xy} = 0$ due to the mismatch of translational symmetry of $\psi_x,\psi_y$ and $V_{\rm{lat}}$. On the other hand, the energy anisotropy between $p_{x,y}$ orbitals $J^V_{xx}-J^V_{yy}$ is proportional to $\delta r_m$ and can be expressed as:
\begin{align}
    J^V_{xx}-J^V_{yy} &= \int \d^2\mathbf{r} \ \left(\left|\psi_{x}(\mathbf{r})\right|^2-\left|\psi_{y}(\mathbf{r})\right|^2\right) V_{\mathrm{lat}}(\mathbf{r};r_m,\delta r_m) \,, \\
    &= \int\d^2\mathbf{r} \ \left|\psi_{x}(\mathbf{r})\right|^2 \delta V_{\mathrm{lat}}(\mathbf{r};r_m,\delta r_m) \,, \\
    &= \int\d^2\mathbf{r} \ \left|\psi_{x}(\mathbf{r})\right|^2 \delta \bar{V}_{\mathrm{lat}}(\mathbf{r};r_m,\delta r_m)\,,
\end{align}
where $\delta \bar{V}_{\mathrm{lat}}$ is the even-symmetrization of the potential anisotropy $\delta V_{\mathrm{lat}}$, given by:
\begin{align}
    \delta \bar{V}_{\mathrm{lat}}(x,y) &= \left[\delta V_{\mathrm{lat}}(x,y)+\delta V_{\mathrm{lat}}(-x,y)+\delta V_{\mathrm{lat}}(x,-y)+\delta V_{\mathrm{lat}}(-x,-y)\right]/4 \,,  \\ 
    &= \frac{V_0}{2} \Bigl[\cos (k x)-\cos (k y)\Bigr] \Bigl \{r \cos (\phi ) - (\delta r+r) \cos \left(\phi+\delta \Phi' _\X\right) \notag \\
    &\quad \quad \quad  \, - (r-\delta r) \left[r \cos \left(\phi -\delta \Phi' _\Y\right) - (\delta r+r) \cos \left(\phi+\delta \Phi' _\X-\delta \Phi'_\Y \right)\right]\Bigr\} \notag \\
    &\simeq \frac{V_0}{2} \sin\phi \Big[\cos (k x)-\cos (k y)\Big]\cdot \delta r_m (\delta\Phi'_\X + \delta \Phi'_\Y) + O(\delta r_m^2, \delta r_m^2\cdot \delta \Phi') \,, \label{Seqn:delta_bar_V}
\end{align}
where we have approximated $r_m\simeq 1$ in Eq.~\eqref{Seqn:delta_bar_V}. The pseudo-magnetic fields generated by primary laser noise are characterized by:
\begin{align}
    B_x^{\rm{pri}} &= 0 \,, \\
    B_z^{\rm{pri}} &= 2(J^V_{xx}-J^V_{yy}) = \xi V_0 \delta r_m (\delta\Phi'_\X + \delta\Phi'_\Y) \,, \label{Seqn:Bz_pri}
\end{align}
where the dimensionless coupling coefficient $\xi(V_0,\phi)$ is defined as:
\begin{equation}
    \xi(V_0,\phi) \equiv \sin\phi \int \d^2\mathbf{r} \, |\psi_{x}(\mathbf{r})|^2 \left[\cos (k x) - \cos (k y)\right] \,.
\end{equation}
For the specific configuration with $V_0 = 6E_R$ and $\phi = 0.44857\pi$, numerical evaluation yields $\xi(V_0,\phi) = 0.1807$.

In summary, the $B_z^{\rm{pri}}$ field breaks the $p_{x,y}$ orbital degeneracy. This field depends on the sum of phase noise contributions from both arms ($\delta\Phi'_\X + \delta\Phi'_\Y$), rather than their differential. Successful measurement requires that noise-induced field fluctuations satisfy $\Delta B_z^{\mathrm{pri}} \ll B_x^{\mathrm{GW}}$, where $\Delta B_z^{\mathrm{pri}}$ is the standard deviation of $B_z^{\mathrm{pri}}$, and $B_x^{\mathrm{GW}}$ represents the gravitational wave-induced signal field. This condition translates into specific limits on the allowable laser phase noise $\delta\Phi'_{\X,\Y}$.
These limits will be analyzed in subsequent
subsections, considering both frequency and
intensity noise of the primary laser.

\subsubsection{ Primary Laser Frequency Noise}\label{Ssubsec:primary_freq} 
The impact of frequency noise from the primary laser on the system differs notably from that of the signal laser. Unlike the signal laser, whose relevant beams interact with high-finesse Fabry-Perot arm cavities (as detailed in Sec.~\ref{Ssubsec:signal_freq}), the primary laser's beams, including its left- and down-moving fields, are reflected by retroreflectors. 

For the primary lattice, the phase modulations experienced by its counter-propagating components (e.g., $\delta \Phi'_L$ relative to $\delta \Phi'_R$, and $\delta \Phi'_D$ relative to $\delta \Phi'_U$) are expected to be nearly identical. As a result, the differential phase terms, $\delta \Phi'_{\X} = \delta \Phi'_{R}-\delta \Phi'_{L}$ and $\delta \Phi'_{\Y} = \delta \Phi'_{U} - \delta \Phi'_{D}$, exhibit strong common-mode rejection of such laser-induced fluctuations and are suppressed. Therefore, the contribution of primary laser frequency noise to $B_z^{\rm{pri}}$ is not a dominant concern, particularly when the mirror reflectivity imbalance $\delta r_m$ is also minimized.

This effective common-mode rejection means that the stability requirements for the primary laser's frequency are significantly relaxed compared to those for the signal lasers. The demands are more aligned with those typical for lasers in standard cold atom experiments, where intensity stability is often more critical than ultra-high frequency stability far from atomic resonances~\cite{Grimm2000OpticalTraps}.

\subsubsection{ Primary Laser Intensity Noise}\label{Ssubsec:primary_int} 
Intensity noise from the primary laser, however, can couple to the system via radiation pressure on the retroreflectors. The fluctuating laser power, quantified by the relative intensity noise (RIN), exerts a fluctuating force $\delta F_{\mathrm{rp}}(t)$ on each retroreflector mirror. This force induces a mechanical displacement $\delta x(t)$, which in turn creates a phase modulation on the reflected light, $\delta\Phi_r = 2 k_L \delta x$, where $k_L = 2\omega_0/c$ is the primary laser's wavenumber.

Modeling the retroreflector as a damped, high-Q mechanical oscillator driven by the fluctuating radiation pressure force $\delta F_{\mathrm{rp}}$, with an effective mass $m_{\mathrm{eff}}$, mechanical resonance angular frequency $\omega_m$, and damping coefficient $\gamma = \omega_m/Q$ with $Q\gg1$, its equation of motion is:
\begin{equation}
    m_{\mathrm{eff}}\left({\delta \ddot x}(t) + \gamma {\delta \dot x}(t) + \omega_m^2 \delta x(t)\right) = \delta F_{\mathrm{rp}}(t) \,.
\end{equation}
In the Fourier domain, the displacement is $\delta \tilde x(\omega) = \chi(\omega) \delta \tilde{F}_{\mathrm{rp}}(\omega) $, where the mechanical susceptibility $\chi^{-1}(\omega) = m_{\mathrm{eff}}(\omega_m^2 - i\gamma\omega -\omega^2)$. The fluctuating radiation pressure force is given by $\delta \tilde{F}_{\mathrm{rp}}(\omega) = 2 {\delta \tilde P}(\omega)/c = 2P_{\mathrm{arm,\lambda}} \mathrm{RIN}(\omega)/c$
, this using the relative intensity noise (RIN) of the primary laser $\mathrm{RIN}(\omega) = \delta \tilde{P}(\omega)/P_{\mathrm{arm,\lambda}}$.
The resulting displacement leads to a phase modulation for the reflected light:
\begin{equation}
    \delta\tilde{\Phi}_r(\omega) = 2k_L {\delta \tilde x}(\omega)
    = \frac{4 P_{\mathrm{arm,\lambda}} \omega_0}{c^2} \chi(\omega) \mathrm{RIN}(\omega) \,. \label{Seqn:deltaPhi_pri_int}
\end{equation}
This phase noise contributes to the pseudo-magnetic field. Following Eq.~\eqref{Seqn:Bz_pri}, the spectrum of the noise field is:
\begin{align}
    \tilde{B}_z^{\mathrm{pri}}(\omega) &= -\xi V_0 \delta r_m (\delta\tilde{\Phi}_L(\omega) + \delta\tilde{\Phi}_D(\omega)) \notag \approx -2 \xi V_0 \delta r_m \delta\tilde{\Phi}_r(\omega) \notag \\
    &= -8 \xi V_0 \delta r_m \frac{P_{\mathrm{arm},\lambda} \omega_0}{c^2} \chi(\omega) \mathrm{RIN}(\omega) \,. \label{Seqn:Bz_pri_RIN_spectrum}
\end{align}
The root-mean-square (RMS) fluctuation of this field, $\Delta B_z^{\mathrm{pri}}$, can be found by integrating its power spectral density, due to the Wiener–Khinchin theorem~\cite{2007BookMaggiore}. If $S_{\mathrm{RIN},\lambda}(\omega)$ is the single-sided power spectral density of the RIN, then the
variance of $B_z^{\mathrm{pri}}$ is:
\begin{align}
    (\Delta B_z^{\mathrm{pri}})^2 &= \int_0^{\infty} S_{B_z^{\mathrm{pri}}}(\omega) \frac{d\omega}{2\pi} \notag \\
    &= \left(8 \xi V_0 \delta r_m \frac{P_{\mathrm{arm},\lambda} \omega_0}{c^2} \right)^2 \int_0^{\infty} |\chi(\omega)|^2 S_{\mathrm{RIN},\lambda}(\omega) \frac{d\omega}{2\pi} \,. \label{Seqn:DeltaBz_pri_RIN_integral}
\end{align}
In the high-Q limit, the kernel $|\chi(\omega)|^2$ behaves like a delta function~\cite{rubiola2008phase}, and the integral can be evaluated as:
\begin{equation}
    \int_0^{\infty} |\chi(\omega)|^2 S_{\mathrm{RIN},\lambda}(\omega) \frac{d\omega}{2\pi} = \frac{Q}{4m_{\mathrm{eff}}^2\omega_m^3} S_{\mathrm{RIN},\lambda}(\omega_m) \,,
\end{equation}
which implies the noise is dominated by the RIN at the mechanical resonance frequency $\omega_m$.
The RMS noise field is thus:
\begin{equation}
    \Delta B_z^{\mathrm{pri}} = 4 \xi V_0 \delta r_m \frac{P_{\mathrm{arm},\lambda} \omega_0}{c^2} \sqrt{\frac{Q}{m^2_{\mathrm{eff}}\omega_m^3} S_{\mathrm{RIN},\lambda}(\omega_m)} \,.
\end{equation}

\begin{figure}
    \centering
    \includegraphics[width=0.7\linewidth]{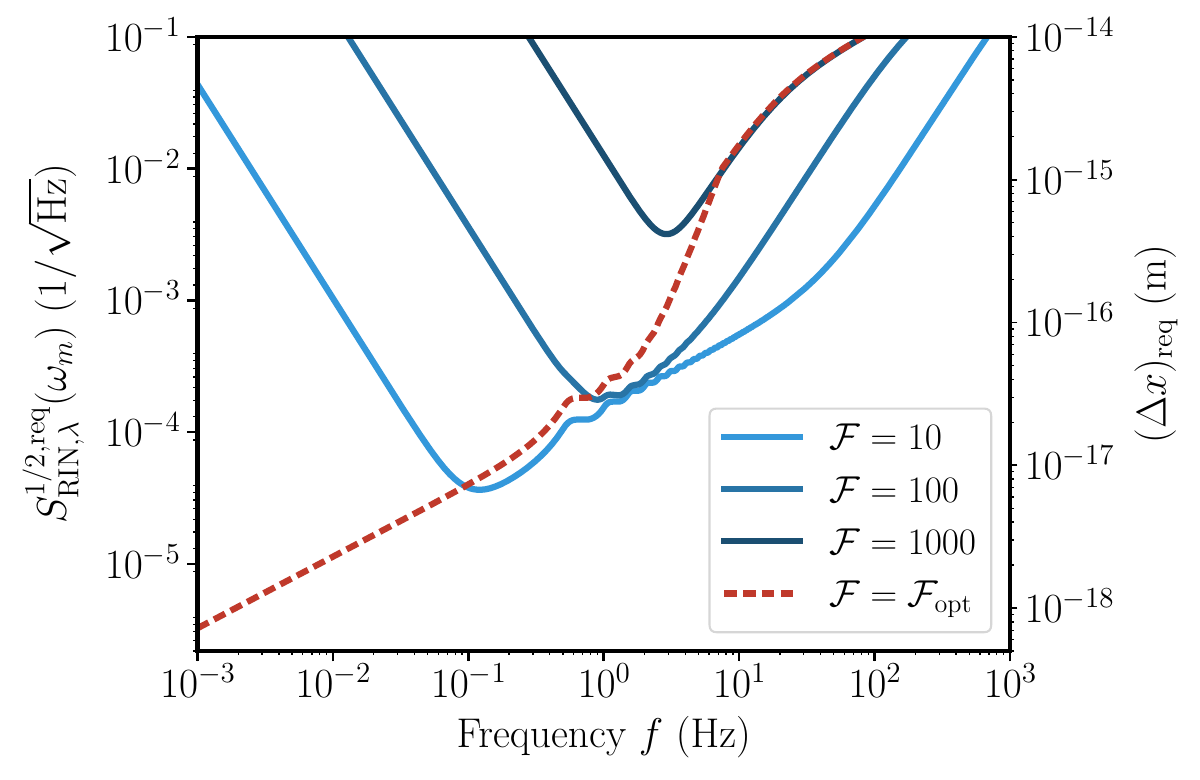}
    \caption{Requirements on the primary laser's relative intensity noise (RIN). The plot shows the required RIN amplitude spectral density, $S_{\text{RIN},\lambda}^{1/2, \text{req}}$ (left axis), and the corresponding RMS mirror displacement fluctuation, $(\Delta x)_{\text{req}}$ (right axis), as a function of target continuous GW frequency $f$. }
    \label{Sfig:primary_int}
\end{figure}

To maintain sensitivity to a continuous
GW signal at a given frequency $f$, $\Delta B_z^{\mathrm{pri}} \ll B_x(f)$ is required. The target signal strength can be defined using the detector's quantum-noise-limited strain sensitivity, $h_n(f) = \sqrt{fS_n(f)}$. This characteristic noise strain produces a signal field $B_x(f) = \eta U_0 \T_{\Phi}(f) h_n(f)$ (Eq.~\eqref{Seqn:Bx}). By setting a detection threshold where the noise contribution is a small fraction of the signal from a target strain, e.g., $\Delta B_z^{\mathrm{pri}} = 0.01 B_x(f)$ , we can establish a requirement on the primary laser's RIN at the mirror resonance frequency.

The required RIN spectral density, $S_{\mathrm{RIN},\lambda}^{\mathrm{req}}(\omega_m)$, and the corresponding limit on the RMS mirror displacement are shown in Fig.~\ref{Sfig:primary_int}.
The requirement—a RIN below $1\times10^{-6} \ \mathrm{Hz}^{-1/2}$ at the mechanical resonance—is readily achievable with standard commercial 532\,nm lasers~\cite{Paschotta_2007_relative_intensity_noise}. This confirms that the primary laser's intensity noise does not pose a significant obstacle to reaching the sensor's design sensitivity.

\subsection{ Atomic Readout Noise}\label{Ssubsec:readout_noise}
The detection protocol concludes with a destructive, time-of-flight (TOF) measurement to determine the final state of the atomic ensemble~\cite{wirth2011evidence, kock2016orbital}. As depicted in the overall time sequence (Fig.~\textcolor{magenta}{4} in the main text), this final step reads out the GW-imparted orbital rotation angle $\theta$, which is encoded in the pseudo-spin expectation value $\langle \hat{J}_y\rangle \approx |J_z|\theta$. This section analyzes the noise sources inherent to this readout procedure.

The measurement of $\langle \hat{J}_y\rangle$ is performed using a standard Ramsey-like technique. First, a global $\pi/2$ pulse is applied to the atomic ensemble, rotating the pseudo-spin vector. This rotation maps the signal from $\langle \hat{J}_y\rangle$ into the population imbalance $\langle \hat{J}_z\rangle = (n_x-n_y)/2$ between the two orbital states.

\begin{figure}[htbp]
    \centering
    \includegraphics[width=1.0\linewidth]{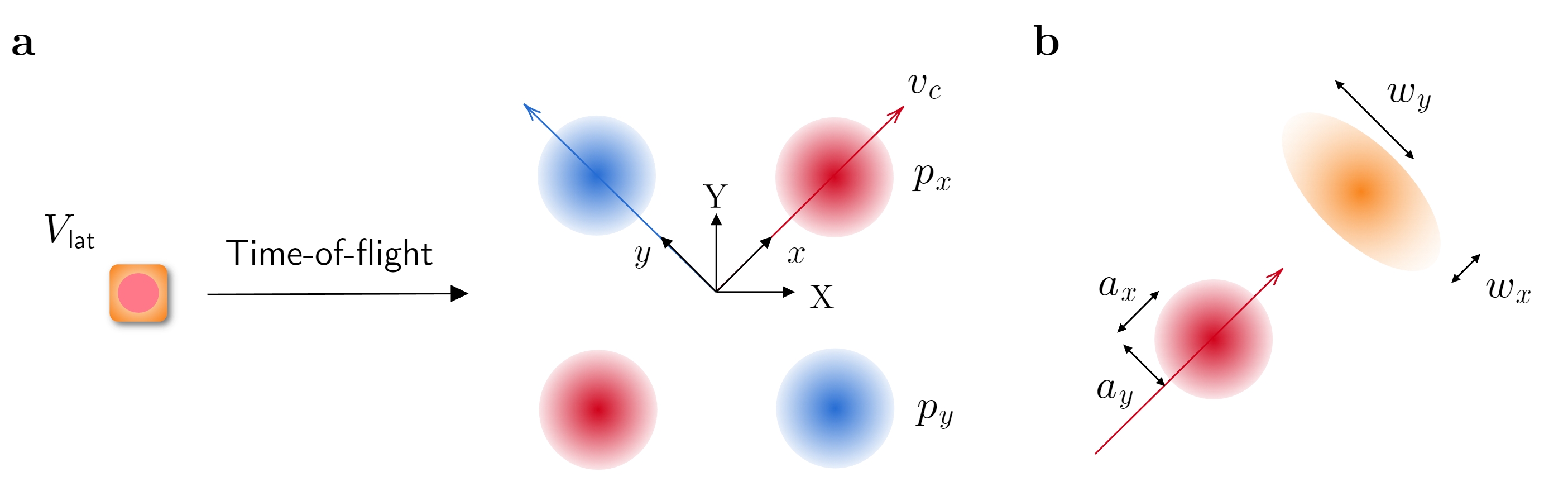}
    \caption{Schematic of the time-of-flight (TOF) measurement. (\textbf{a}) After the trapping potential is turned off, atoms in the $p_{x,y}$ orbitals separate spatially due to their different center-of-mass velocities, mapping the orbital population difference onto a spatial separation for imaging. (\textbf{b}) Geometry for optical imaging, where the separated atom clouds (with radii $a_{x,y}$) pass through a detection laser beam (with waists $w_{x,y}$).}
    \label{Sfig:tof_schematic}
\end{figure}

Immediately following this pulse, all trapping potentials are switched off, and the atomic cloud expands ballistically for a duration of $t_{\mathrm{TOF}}=30$ ms. During this expansion, the $|p_x\rangle$ and $|p_y\rangle$ states separate spatially due to their distinct quasi-momenta, expanding with center-of-mass velocities of $\pm v_c$ along the $x$ and $y$ axes, respectively. As depicted in Fig.~\ref{Sfig:tof_schematic}\textcolor{magenta}{a}, this process maps the orbital populations onto spatially distinct clouds. The center-of-mass velocity is $v_c = \pi\hbar/(Ma)$, where $M$ is the atomic mass and $a$ is the lattice constant. For a sufficiently long $t_{\mathrm{TOF}}$, the cloud size is given by $a_{x,y}=\Delta v\cdot t_{\mathrm{TOF}}$, where $\Delta v = \sqrt{k_BT/M}$ is the thermal velocity width.

After expansion, the number of atoms in each separated clouds is counted via optical imaging (Fig.~\ref{Sfig:tof_schematic}\textcolor{magenta}{b}). The precision of this final atom-counting step introduces an imaging noise variance, $\mathrm{Var}(\theta)_{\mathrm{imaging}}$, to the measurement. The total variance of the measured angle $\theta$ is thus:
\begin{equation}
    \mathrm{Var}(\theta) = \mathrm{Var}(\theta)_{\mathrm{quantum}} + \mathrm{Var}(\theta)_{\mathrm{imaging}} \,,
\end{equation}
where $\mathrm{Var}(\theta)_{\mathrm{quantum}} = (\Delta \theta)^2_{\rm{proj}} + (\Delta \theta)^2_{\rm{photon}}$ is the fundamental quantum noise of the sensor, comprising atomic projection noise and quantum fluctuation of signal laser (Eq.~\eqref{Seqn:Jy2Noise}). The dominant contribution to $\mathrm{Var}(\theta)_{\mathrm{imaging}}$ is typically the photoelectron shot noise of the photodetector~\cite{2014PRL_entanglement,2024PRXZhangSqueezing}. In the following, we evaluate this shot-noise-limited detection noise for two methods: fluorescence imaging (FI) and frequency modulation imaging (FMI).

\subsubsection{Comparison of FI and FMI Methods}
Fluorescence imaging (FI) measures the number of photons scattered by the atoms. Its signal-to-noise ratio (SNR) is highly dependent on the photon collection efficiency, $\eta_{\mathrm{solid}}$, which is typically less than $0.1$. FI is often the standard method for thermal atomic sources with low optical depth (OD).

Frequency modulation imaging (FMI), in contrast, is an absorption-based method. It uses resonant frequency modulation spectroscopy to convert the atomic absorption signal into a radio-frequency heterodyne beat note. This allows the signal to be moved to a frequency domain with low technical noise, avoiding the $1/f$ noise that can affect simple absorption measurements. FMI can offer a superior SNR for optically dense atomic sources, such as the BEC used in our proposed sensor, especially when the solid angle efficiency $\eta_{\mathrm{solid}}$ for FI is limited.

\subsubsection{Imaging Noise and Detectable Squeezing}
A low-noise imaging technique is crucial for resolving our atomic orbital squeezed state. We can quantify the performance of a detection method by its ability to resolve a given level of atomic squeezing. This is characterized by the minimum detectable squeezing parameter, $\xi_{\mathrm{min}}^2\equiv N\cdot\mathrm{Var}(\theta)_{\mathrm{imaging}}$, where a smaller $\xi_{\mathrm{min}}$ indicates a better detection system. The photoelectron-shot-noise-limited $\xi_{\mathrm{min}}$ for both FI and FMI methods can be calculated as~\cite{2024PRA_optimal}:
\begin{align}
    \xi_{\mathrm{min, FI}} &= \sqrt{\frac{4eF}{\eta_{\mathrm{solid}} \sigma I_{\mathrm{sat}} R_0 f_x f_y \tau_s}}\,, \\
    \xi_{\mathrm{min, FMI}} &= \sqrt{\frac{2e}{\sigma I_{\mathrm{sat}} R_0 f_x f_y \tau_s\mathrm{OD}}}\,,
\end{align}
where the parameters are defined as follows: $e$ is the elementary charge; $R_0$ and $F$ are the photodetector responsivity and excess noise factor; $\sigma$ is the on-resonance absorption cross section; $I_{\mathrm{sat}}$ is the saturation intensity; $f_{x,y} \equiv w_{x,y}/a_{x,y}$ is a geometric factor comparing the laser waist to the atom cloud size; and $\tau_s$ is the effective atom-light interaction time. While a larger laser waist is generally better for FI, FMI performance is optimized with a horizontal waist of $w_{x}\approx1.5a_{x}$~\cite{2024PRA_optimal}.

For the numerical evaluation, we model specific commercial photodetectors based on the analysis in~\cite{2024PRA_optimal}. For FMI, we use parameters for the Thorlabs DET02A photodetector with $R_{0,\mathrm{FMI}} = 0.40~\mathrm{A/W}$. For FI, we model the Thorlabs APD440A photodetector with $R_{0,\mathrm{FI}} = 0.46~\mathrm{A/W}$
and an excess noise factor of $F = 2.4$.

 \begin{figure}[ht]
    \centering
    \includegraphics[width=0.7\linewidth]{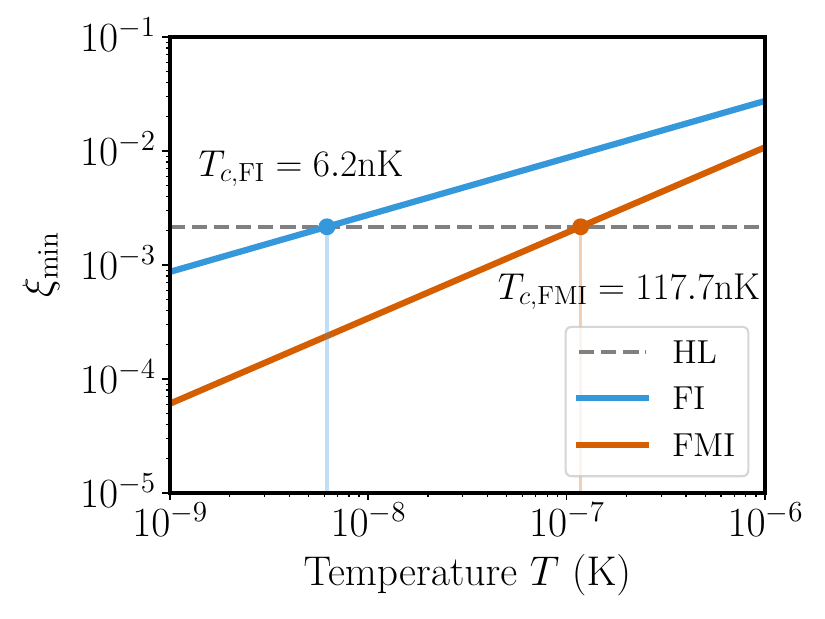}
    \caption{Minimum detectable squeezing parameter $\xi_{\mathrm{min}}$ of fluorescence imaging (FI) and frequency modulation imaging (FMI) methods as a function of atomic cloud temperature. The critical temperature, $T_c$, is defined as the temperature at which the imaging noise equals the Heisenberg-limited (HL) quantum projection noise of our sensor, $(\xi^2_R)_{\mathrm{HL}}=3.738/N$ (gray dashed line, from Fig.~\ref{Sfig:Xi2}).}
    \label{Sfig:xi_min}
\end{figure}

To evaluate the suitability of these methods, we calculated $\xi_{\mathrm{min}}$ for both as a function of the atomic cloud temperature (Fig.~\ref{Sfig:xi_min}). As shown in the plot, FMI consistently outperforms FI for the cold, dense ensembles relevant to our sensor. The critical temperature, $T_c$, below which our target Heisenberg-level squeezing can be resolved is $T_{c,\mathrm{FMI}}=117.7 \ \mathrm{nK}$ for FMI. This is well within the reach of modern ultracold atom experiments. In contrast, FI would require a much lower temperature of $T_{c,\mathrm{FI}}=6.2 \ \mathrm{nK}$, which is significantly more challenging. We therefore conclude that FMI provides a robust and effective method for reading out the state of our orbital optomechanical sensor, ensuring that the overall sensitivity is not limited by detection noise.

\medskip
\section{ Detecting Dark Photon Dark Matter}\label{Ssec:dark_matter}
Beyond their primary mission of observing gravitational waves, interferometric sensors are powerful probes for new fundamental physics, including searches for Dark Photon Dark Matter (DPDM)~\cite{zhao2018PRL,ligo2020PRD,ligo2022PRD,2021PRDImproved}. Our orbital optomechanical sensor, with its exceptional sensitivity, can therefore provide a new window into this compelling dark matter candidate. This section details the mechanism by which our sensor detects a DPDM signal and derives the projected upper limits on its coupling to ordinary matter.

\subsection{ Dark Photon Electrodynamics}
\label{Ssubsec:DPDM}
This section examines the fundamental properties of DPDM, mainly following Refs.~\cite{Guo:2019NC,2021PRD:lightDM,carney2021ultralight}. The dark photon is 
a  spin-$1$ gauge boson beyond the standard model, and a potential ultralight candidate for dark matter. Unlike heavier candidates like Weakly Interacting Massive Particles (WIMPs), the ultralight dark photon is characterized by its classical, wave-like nature due to its high phase space density $n_{\mathrm{phase}}$, given by:
\begin{equation}
    n_{\mathrm{phase}}=\frac{\rho_{\DM}}{m_A c^2} \lambda_A^3 \simeq 10^6 \left(\frac{1\mathrm{eV/c^2}}{m_A}\right)^4 \,,
\end{equation}
where $\lambda_A = 2\pi \hbar/(m_A v_{\DM})$ and $m_A$ are the de Broglie wavelength and mass of the dark photon, respectively. The virial velocity $v_{\DM}$ and local energy density $\rho_{\DM}$ of dark matter are observed to be $v_{\DM} = 230 \mathrm{km/s}$ and $\rho_{\DM} = 0.4 \mathrm{GeV/cm^3}$~\cite{Guo:2019NC}, respectively. For $m_A \lesssim 0.1 \mathrm{eV/c^2}$, $n_{\mathrm{phase}} \gg 1$, allowing the dark photon field to be treated as a classical condensate wave, whose spatial components are expressed as:
\begin{equation}
    \mathbf{A}(t,\mathbf{x}) = \mathbf{A}_0 \cos(\omega t-\mathbf{k}\cdot\mathbf{x}+\phi_A) \,,
\end{equation}
where $\mu_0$ is the magnetic permeability in the vacuum, $\mathbf{A}_0, \omega, \mathbf{k}$ are the polarization vector, angular frequency and wave vector of the dark photon field, respectively, and $\phi_A$ is its random phase. The magnitude of $\mathbf{A}_0$ is determined by the local density $|\mathbf{A}_0|= \dfrac{1}{\omega} \sqrt{\dfrac{2\rho_{DM}}{\epsilon_0}}$. The dispersion relation of dark photon is $\hbar \omega = \sqrt{(\hbar kc)^2+(m_A c^2)^2} \simeq m_A c^2 + \frac{1}{2}m_A v_{\DM}^2$. This velocity-dependent kinetic energy term leads to a frequency spread of the signal $\Delta f = \dfrac{1}{2} (\dfrac{v_{\DM}}{c})^2 f$. The central frequency, $f$, is related to the dark photon mass by $f = \dfrac{m_A c^2}{2\pi \hbar}$. This further results in the coherence time of the dark matter $T_{\mathrm{coh}}$ given by:
\begin{equation}\label{Seqn:Tcoh}
    T_{\mathrm{coh}}(f) = \frac{1}{\Delta f} = \frac{2}{f}\left(\frac{c}{v_{\DM}}\right)^2 \,.
\end{equation}
In the Lorentz gauge $\partial_{\mu} A^{\mu} = 0$, the temporal component of $A^{\mu}$ is of magnitude $\dfrac{v_{\DM}}{c} |\mathbf{A}_0|$, which is negligible in calculating the dark electric field $\mathbf{E}$. 
The dark electric and magnetic fields generated by the dark photon are described by: 
\begin{align}
    \mathbf{E}(t,\mathbf{x}) &\simeq -\partial_t \mathbf{A} = \omega \mathbf{A}_0 \sin(\omega t-\mathbf{k}\cdot\mathbf{x}+\phi_A) \,, \label{Seqn:darkE}\\
    \mathbf{B}(t,\mathbf{x}) &=\nabla \times \mathbf{A} =\mathbf{k}\times \mathbf{A}_0 \sin(\omega t-\mathbf{k}\cdot\mathbf{x}+\phi_A) \,.
\end{align}
The ratio of the dark magnetic field to the dark electric field is $\dfrac{|\mathbf{B}|}{|\mathbf{E}|} = \dfrac{v_{\DM}}{c^2}\simeq 10^{-3}/c$, suggesting that the dark magnetic field is negligible in the interaction between dark photons and ordinary matter. The magnitude of $\mathbf{E}_0$ is given by $|\mathbf{E}_0| = \omega |\mathbf{A}_0| = \sqrt{\dfrac{2\rho_{\DM}}{\epsilon_0}} = 3.80\times 10^3 \mathrm{kg\cdot m/(C\cdot s^2)}$.

\subsection{ Coupling between DPDM and the Orbital Optomechanical Sensor}\label{Ssubsec:DMCoupling}
The dark photon is expected to couple to conserved currents in the standard model. The Lagrangian describing the coupling of $A^{\mu}$ to current $J^\mu_{\mathrm{D}}$ ($\mathrm{D}=\mathrm{B}-\mathrm{L}$ for baryon-lepton and $\mathrm{D}=\mathrm{B}$ for baryon) is given by:
\begin{equation}\label{Seqn:DM Lagrangian}
    \mathcal{L} = -\frac{1}{4\mu_0}F^{\mu \nu} F_{\mu \nu} + \frac{1}{2\mu_0}(\frac{m_A c}{\hbar})^2 A^{\mu} A_{\mu} - eg_\mathrm{D} J^\mu_\mathrm{D} A_\mu \,,
\end{equation}
where $\mu_0$ is the vacuum magnetic permeability, $F_{\mu \nu} = \partial_\mu A_\nu - \partial_\nu A_\mu$ is the dark photon field strength tensor, and $g_{\mathrm{D}}$ is the $U(1)_{\mathrm{D}}$ coupling constant normalized to the electromagnetic coupling constant $e$. As a consequence, dark photon exerts force $\mathbf{F}$ on ordinary matter, as described by:
\begin{equation}
    \mathbf{F}(t,\mathbf{x}) = e g_\mathrm{D} q_{\mathrm{D}} \mathbf{E}(t,\mathbf{x})
\end{equation}
In this equation, $q_{\mathrm{D}}$ denotes the $U(1)_{\mathrm{D}}$ charge of the test masses, with $q_{\mathrm{B}}$ being the sum of the number of protons and neutrons, and $q_{\mathrm{B-L}}$ being the number of neutrons. 

For our orbital optomechanical sensor, we must consider two potential coupling channels: (1) direct coupling to the atoms in the BEC, and (2) coupling to the macroscopic test masses of the interferometer.

\subsubsection{ Coupling between DPDM and Atoms}
As detailed in~\ref{Ssubsec:species}, the atomic sensor is most sensitive when composed of $^{7}\mathrm{Li}$. This section focuses on the interaction between DPDM and $^{7}\mathrm{Li}$ atoms. A single $^{7}\mathrm{Li}$ atom contains 3 protons and 4 neutrons (along with 3 electrons). Consequently, the dark photon exerts a force on a single $^{7}\mathrm{Li}$ atom, the magnitude of which is given by:
\begin{equation}
    F_{\DM} = 4eg_{\mathrm{B-L}}|\mathbf{E}_0| = 2.43 \times 10^{-39} \bigg(\frac{g_{\mathrm{B-L}}}{10^{-24}}\bigg) \ \mathrm{kg\cdot m/s^2} \,.
\end{equation}
This force creates a differential potential energy across a single lattice site of size $a = \lambda/\sqrt{2}$. The magnitude of this potential is:
\begin{equation}
    V_{\DM} = F_{\DM}a = 9.17 \times 10^{-46} \left(\frac{g_{\mathrm{B-L}}}{10^{-24}}\right) \ \mathrm{kg\cdot m^2/s^2} = 1.38 \times 10^{-17} \bigg(\frac{g_{\mathrm{B-L}}}{10^{-24}}\bigg) E_R \,,
\end{equation}
where $E_R = h^2/2m\lambda^2$ is the photon recoil energy of $^{7}\mathrm{Li}$, with $m$ being its mass. Current experimental constraints from torsion balance and equivalence principle tests already limit $g_{\mathrm{B-L}} \lesssim 10^{-24}$ in the relevant mass range~\cite{EotWash1,EotWash2,MICROSCOPE1,MICROSCOPE2,MICROSCOPE3,MICROSCOPE4}. This implies $V_{\DM} \lesssim 10^{-16} E_R$. As this potential is many orders of magnitude smaller than the relevant energy scales of the atomic system, its effect is entirely negligible.
\subsubsection{ Coupling between DPDM and Test Masses}
The dominant interaction channel is the force exerted by the DPDM field on the four macroscopic test masses of the interferometer. This force induces a time-varying acceleration and a corresponding displacement for each test mass:
\begin{align}
    \mathbf{a}(t) &= eg_\mathrm{D} \frac{q_{\mathrm{D}}}{M} \omega \mathbf{A}_0 \sin(\omega t-\mathbf{k}\cdot\mathbf{x}+\phi_A) \,, \\
    \delta \mathbf{x}(t) &= -eg_\mathrm{D} \frac{q_{\mathrm{D}}}{M} \frac{\mathbf{A}_0}{\omega} \sin(\omega t-\mathbf{k}\cdot\mathbf{x}+\phi_A) \,.
\end{align}
For silica test masses, similar to those in LIGO, the charge-to-mass ratios are $q_{\mathrm{B}}/M = 5.61 \times 10^{26} \ \mathrm{kg}^{-1}$ and $q_{\mathrm{B-L}}/M = 2.80 \times 10^{26} \ \mathrm{kg}^{-1}$~\cite{ligo2022PRD}. The $\mathrm{B-L}$ coupling induces oscillations of the order of $|\delta \mathbf{x}_0|$, calculated as:
\begin{equation}
    |\delta \mathbf{x}_0| = eg_\mathrm{B-L} \frac{q_{\mathrm{B-L}}}{M} \frac{\mathbf{A}_0}{\omega} = 4.32 \times 10^{-15} \left(\frac{1 \mathrm{Hz}}{f}\right)^2 \bigg(\frac{g_{\mathrm{B-L}}}{10^{-24}}\bigg) \ \mathrm{m} \,.
\end{equation}
Such oscillations result in a non-zero phase accumulation $\Phi_{\X,\Y}$ of lasers in the $\X,\Y$ arms. Our sensor is capable of detecting displacements $\delta \mathbf{x}$ of the order of $h_{0,+} L \simeq 10^{-23} \times 4 \mathrm{km} \simeq 10^{-20} \mathrm{m}$. Therefore, it is expected to be sensitive to DPDM-induced oscillation signals for couplings $g_{\mathrm{B-L}} \lesssim 10^{-24}$ near $f = 1 \mathrm{Hz}$.
The effective gravitational wave signal $h_{\mathrm{eff}}(t)$ due to $\delta \mathbf{x}(t)$ is given by~\cite{2021PRDImproved}:
\begin{align}
h_{\mathrm{eff}}(t)&=h_1(t)+h_2(t) \,, \\
h_1(t) &=   \frac{e g_{\mathrm{D}}}{\pi f L} \frac{q_{\mathrm{D}}}{M}  \sin ^2\big(\pi fL/c\big) \times \left(\mathbf{n}-\mathbf{m}\right) \cdot \mathbf{A}_0 \sin \phi(t) \,, \\[6pt]
h_2(t) &= -\frac{e g_{\mathrm{D}}}{c^2} \frac{q_{\mathrm{D}}}{M} \times \Big[\left(\mathbf{n} \cdot \mathbf{A}_0\right) \left(\mathbf{n} \cdot \mathbf{v}_{\mathrm{DM}}\right) -\left(\mathbf{m} \cdot \mathbf{A}_0\right)\left(\mathbf{m} \cdot \mathbf{v}_{\mathrm{DM}}\right) \Big]\cos \phi(t) \,,
\end{align}
where $\mathbf{n}, \mathbf{m}$ are unit vectors along the interferometer arms, and $\phi(t) = \omega \left(t - L / c\right) + \phi_A$. The term $h_1(t)$ accounts for the light travel time delay, while $h_2(t)$ results from the phase difference experienced by the test masses. The phase difference is estimated to be $\mathbf{k} \cdot \Delta \mathbf{x} \simeq 6.14 \times 10^{-8} f/\mathrm{Hz}$, which is negligible for LIGO but significant for longer interferometers like LISA (see Fig.~\textcolor{magenta}{2b} of the main text). The total effective strain, considering no interference between $h_1(t)$ and $h_2(t)$ after time averaging, is:
\begin{align}\label{Seqn:heff}
    \langle h_{\mathrm{eff}}^2 \rangle &=\langle h_{1}^2 \rangle +\langle h_{2}^2 \rangle \,, \\
   \langle h_{1}^2 \rangle &= \frac{e^2 g_{\mathrm{D}}^2 \rho_{\DM}}{6\pi^4 f^4 L^2 \epsilon_0} \left(\frac{q_{\mathrm{D}}}{M}\right)^2  \sin ^4\big(\pi fL/c\big) \times \left(1-\mathbf{n} \cdot \mathbf{m}\right) \,, \\
   \langle h_{2}^2 \rangle &= \frac{e^2 g_{\mathrm{D}}^2 \rho_{\DM}}{18\pi^2 f^2 c^4 \epsilon_0} \left(1-(\mathbf{n}\cdot\mathbf{m})^2\right)\,.
\end{align}
In this context, the notation $\langle ... \rangle$ represents an average over time, as well as averaging across the propagation direction and polarization states of the dark photon. 
The geometry factor $\mathbf{n}\cdot\mathbf{m}$ is given by $\mathbf{n}\cdot\mathbf{m} = 0$ for LIGO (as well as our sensor) and $\mathbf{n}\cdot\mathbf{m} = 1/2$ for LISA.
\subsection{ 1\texorpdfstring{$\sigma$}{} Upper Limits of \texorpdfstring{$g_{\mathrm{D}}$}{}}\label{Ssubsec:upper limit}
\begin{figure}[htp]
    \centering
    \subfloat[\textbf{a}][]{
        \includegraphics[width=0.47\textwidth]{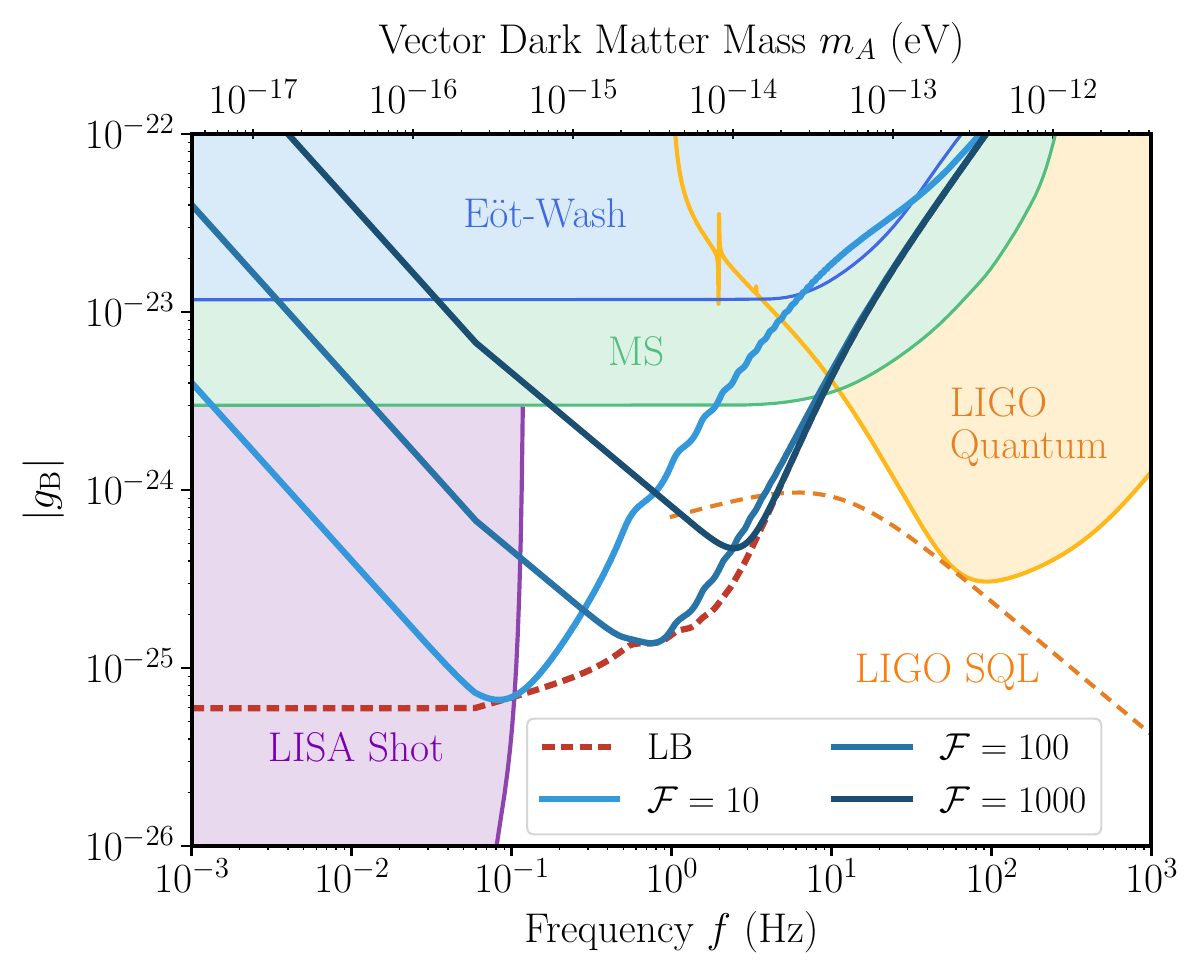}
    }
    \subfloat[\textbf{b}][]{
        \includegraphics[width=0.47\textwidth]{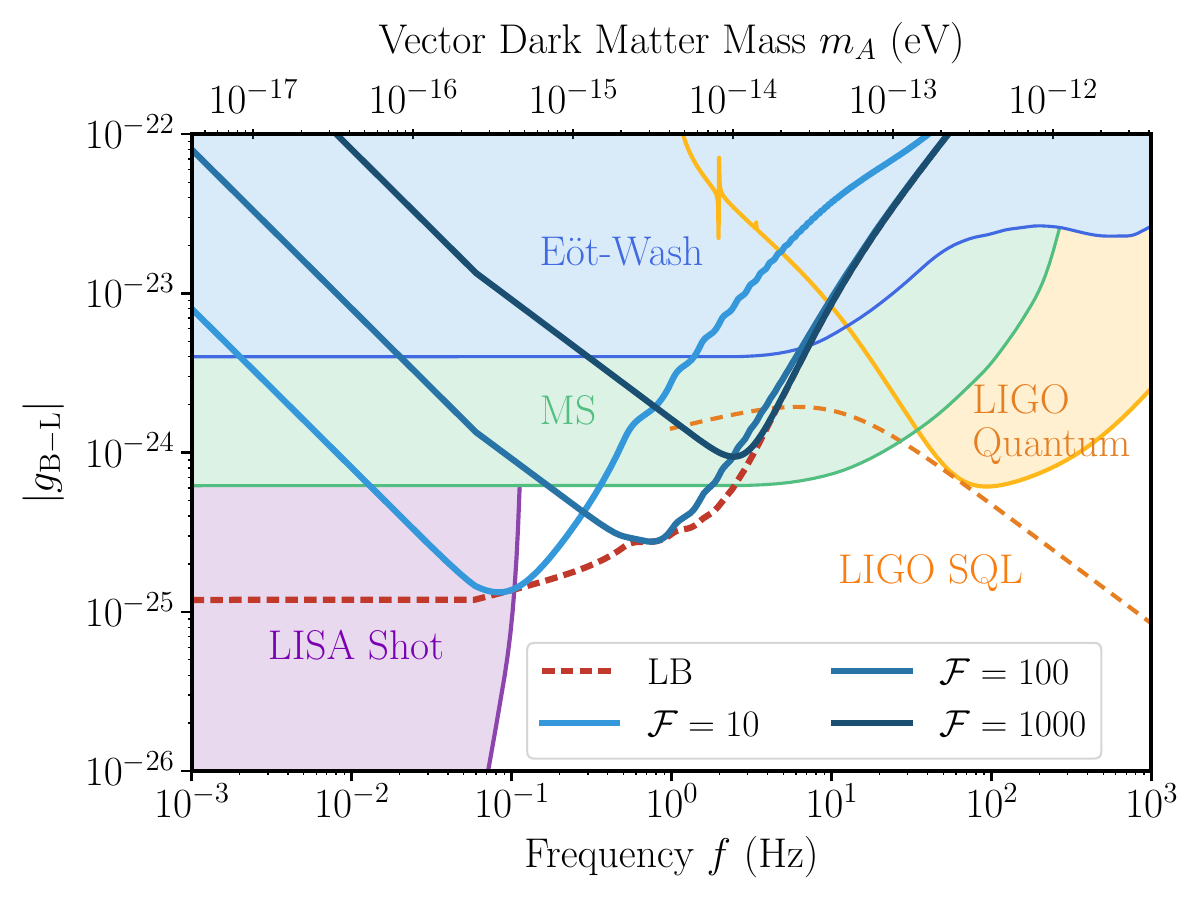}
    }
    \caption{\fontsize{12pt}{14pt}\selectfont 1$\sigma$ detection upper limit on dark matter detection by the orbital optomechanical sensor. \textbf{a}, Upper limit on the dark photon/baryon coupling $|g_{\mathrm{B}}|$; \textbf{b}, Upper limit on the dark photon/baryon coupling $|g_{\mathrm{B-L}}|$. For comparison, we show the limits of LIGO~\cite{ligo2022PRD}, LISA~\cite{2019LISA}, Eöt-Wash~\cite{EotWash1,EotWash2}, and MICROSCOPE (MS) experiments~\cite{MICROSCOPE1,MICROSCOPE2,MICROSCOPE3,MICROSCOPE4}.}
    \label{Sfig:g}
\end{figure}

To determine the detection threshold for DPDM, we employ a semi-coherent search method, where data is integrated over segments comparable to the signal's coherence time, $T_{\mathrm{coh}}$~\cite{2021PRDImproved,2021PRD:lightDM}. The $1\sigma$ detection upper limit for $g_{\mathrm{D}}$ is derived as:
\begin{equation}\label{Seqn:upper_limit}
    \langle h^2_{\mathrm{eff}} \rangle (g_{\mathrm{D}}) = \frac{S_n(f)}{T_{\mathrm{eff}}(f)} \,,
\end{equation}
where $\langle h^2_{\mathrm{eff}} \rangle$ represents the effective strain of dark photons, $S_n(f)$ is the full strain sensitivity of our atomic sensor, and $T_{\mathrm{eff}}(f)$ is the effective observation time, defined by:
\begin{equation}
    T_{\mathrm{eff}}(f) = 
    \begin{cases}
        T_{\mathrm{obs}} & \mathrm{if } T_{\mathrm{obs}} > T_{\mathrm{coh}} \\
        \sqrt{T_{\mathrm{obs}} T_{\mathrm{coh}}(f)} & \mathrm{if } T_{\mathrm{obs}} \leq T_{\mathrm{coh}} \\
    \end{cases}
    \,,
\end{equation}
Here, $T_{\mathrm{obs}}$ denotes the total observation time and $T_{\mathrm{coh}}$ the coherence time of dark photon dark matter given by Eq.~\eqref{Seqn:Tcoh}. The 1$\sigma$ detection upper limits for $g_{\mathrm{B}}$ and $g_{\mathrm{B-L}}$ of our orbital optomechanical sensor are shown in Fig.~\ref{Sfig:g}.

For generating Fig.~\textcolor{magenta}{2b} in the main text, we apply Eq.~\eqref{Seqn:upper_limit} to translate the strain sensitivity of gravitational waves from LISA~\cite{2019LISA}, LIGO~\cite{ligo2015cqg}, and our atomic sensor into the corresponding detection upper limit for $g_{\mathrm{D}}$. We assume a total observation time of $T_{\mathrm{obs}} = 2 \ \mathrm{y.r.}$ for the calculations. The parameters of our orbital optomechanical sensor are summarized in Table~\ref{tab:parameters}. 

\newpage
\section{ Parameters of the Orbital Optomechanical Sensor}
The following tables (\ref{tab:parameters},\ref{tab:noise_parameters},\ref{tab:requirements_summary}) summarize the key design parameters, noise model assumptions, and derived technical requirements for the proposed orbital optomechanical (OOM) sensor. These values form the basis for the sensitivity calculations and figures presented in this work.

\extrarowheight=3pt
\rowcolors[]{1}{RoyalBlue!20}{RoyalBlue!5}
    \begin{table}[!htp]
    \setlength\tabcolsep{0pt}
    \begin{center}
    \begin{tabular}{|L{60mm}|C{20mm}|C{30mm}|}
    \hline
    \rowcolor{Black!20} \textbf{\ \ Parameter} & \textbf{Units} & \textbf{Value}  \\\hline
    \rowcolor{Gray!20} \multicolumn{3}{|c|} {\textbf{Interferometer parameters}}\\\hline 
    \ \ Arms length $L$ & km  & $4$ \\\hline
    \ \ Mirror mass $M$& kg  & $40$  \\\hline
    \ \ FP cavity finesse $\mathcal{F}$ &  & $1\sim10^{4}$ \\\hline
    \ \ Primary laser power $P_{\mathrm{arm},{\lambda}}$ & W  & $1.77$ \\\hline
    \ \ Signal laser power $P_{\mathrm{arm},2{\lambda}}$ & mW  & $32.2$ \\\hline

    \rowcolor{Gray!20} \multicolumn{3}{|c|} {\textbf{Atomic parameters}} \\\hline
    \rowcolor{RoyalBlue!5}
    \ \ Atom numbers $N$ &  & $8\times 10^{5}$ \\\hline
    \rowcolor{RoyalBlue!20}
    \ \ Atom species  &  & $^7\mathrm{Li}$ \\\hline
    \rowcolor{RoyalBlue!5}
    \ \ $p$ orbital states lifetime $T_{\mathrm{life}}$ & s & $1$ \\\hline
    
    \rowcolor{Gray!20} \multicolumn{3}{|c|} {\textbf{Optical Lattice parameters}} \\\hline
  \ \ Primary laser wavelength $\lambda$  & nm & $532$ \\\hline
  \ \ Primary lattice depth $V_0$  & $E_{R}$ & $6$ \\\hline
  \ \ Signal laser wavelength $2\lambda$  & nm & $1064$ \\\hline
  \ \ Signal lattice depth $U_0$  & $E_{R}$ & $0.1$ \\\hline
  \ \ Efficiency $\eta$  &  & $0.6$ \\\hline
    \end{tabular}
    \caption{Core design parameters of the Orbital Optomechanical (OOM) Sensor.}
    \label{tab:parameters}
    \end{center}
    \end{table}

\extrarowheight=3pt
\rowcolors[]{1}{RoyalBlue!20}{RoyalBlue!5}
    \begin{table}[!htp]
    \setlength\tabcolsep{0pt}
    \begin{center}
    \begin{tabular}{|L{78mm}|C{20mm}|C{25mm}|}
    \hline
    \rowcolor{Black!20} \textbf{\ \ Parameter} & \textbf{Units} & \textbf{Value}  \\\hline

    \rowcolor{Gray!20} \multicolumn{3}{|c|} {\textbf{Signal Arm Cavity Imbalance Model}} \\\hline
    \ \ Average arm cavity reflectivity $ r_a$ &  &  0.9887 \\\hline
    \ \ Differential arm cavity reflectivity $\delta r_a$ &  & 0.0029  \\\hline
    \ \ Differential cavity pole $\delta\omega_c/\omega_c$ &   & 0.0057  \\\hline
    \ \ Differential reduced test mass $\delta\mu/\mu$&   & -259 ppm  \\\hline
    \ \ Differential power gain $\delta g_{\mathrm{cav}}/g_{\mathrm{cav}}$ &   &  -45 ppm\\\hline

    \rowcolor{Gray!20} \multicolumn{3}{|c|} {\textbf{Primary Mirror Mechanical Model}} \\\hline
        \rowcolor{RoyalBlue!5}
    \ \  Retroreflector effective mass $m_{\mathrm{eff}}$ & kg & 0.01 \\\hline
        \rowcolor{RoyalBlue!20}
    \ \ Mechanical resonance freq. $\omega_m/2\pi$ & kHz & 30 \\\hline
        \rowcolor{RoyalBlue!5}
    \ \ Mechanical quality factor $Q$ & & $10^3$ \\\hline
        \rowcolor{RoyalBlue!20}
    \ \ Differentialmirror reflectivity $\delta r_m$ &  & 50 ppm \\\hline

    \rowcolor{Gray!20} \multicolumn{3}{|c|} {\textbf{Atomic Readout Model (FMI Method)}} \\\hline
    \ \ On-resonance absorption cross section $\sigma$ & $\mu\mathrm{m}^2$ & $0.215$ \\\hline
    \ \ Effective saturation intensity $I_{\mathrm{sat}}$ & $\mathrm{W}/\mathrm{m}^2$ & $4.04$ \\\hline
    \ \ Photodetector responsivity $R_0$ & A/W & $0.40$ \\\hline
    \ \ TOF expansion time $t_{\mathrm{TOF}}$ & ms & 30\\\hline
    
    \end{tabular}
    \caption{Assumed parameters for the classical and technical noise budget calculations (Sec.~\ref{Ssec:classical_noise}).}
    \label{tab:noise_parameters}
    \end{center}
    \end{table}

\extrarowheight=3pt
\rowcolors[]{1}{RoyalBlue!20}{RoyalBlue!5}
    \begin{table}[!htp]
    \setlength\tabcolsep{0pt}
    \begin{center}
    \begin{tabular}{|L{68mm}|L{35mm}|L{45mm}|}
    \hline
    \rowcolor{Black!20} \textbf{\ \ Requirement Category} & \ \textbf{Parameter} & \ \textbf{Required Value}  \\\hline

    \rowcolor{Gray!20} \multicolumn{3}{|c|} {\textbf{Environmental Noise Suppression}} \\\hline
    \ \ Ground-Based Displacement Noise & $ \ \ S^{1/2}_{\Delta x, \text{req}}$ at 2 Hz & $ \ \ < 5\times 10^{-20} \ \mathrm{m}/\sqrt{\mathrm{Hz}}$ \\\hline
    \ \ Space-Based Acceleration Noise & $ \ \ S^{1/2}_{\Delta a, \text{req}}$ at 2 Hz & $\ \ < 10^{-12} \ \mathrm{m \cdot s^{-2}}/\sqrt{\mathrm{Hz}}$ \\\hline

    \rowcolor{Gray!20} \multicolumn{3}{|c|} {\textbf{Signal Laser Stability}} \\\hline
        \rowcolor{RoyalBlue!20}
    \ \ Frequency Noise ASD & $\ \ S^{1/2}_{\delta \nu, 2\lambda}$ at 2 Hz & $\ \ < 5 \times 10^{-7} \ \mathrm{Hz}/\sqrt{\mathrm{Hz}}$ \\\hline
        \rowcolor{RoyalBlue!5}
    \ \ Relative Intensity Noise ASD & $\ \ S^{1/2}_{\mathrm{RIN}, 2\lambda}$ at 2 Hz & $\ \  < 10^{-6}  /\sqrt{\mathrm{Hz}}$ \\\hline
    
    \rowcolor{Gray!20} \multicolumn{3}{|c|} {\textbf{Primary Laser Stability}} \\\hline
    \ \ Frequency Noise ASD & $ \ \ S^{1/2}_{\delta \nu, \lambda}$  & \ \ Standard BEC Exp. \\\hline
    \ \ Relative Intensity Noise ASD & $\ \ S^{1/2}_{\mathrm{RIN}, \lambda}$ at $\omega_m$ & $\ \ <3\times 10^{-4} /\sqrt{\mathrm{Hz}}$ \\\hline

    \rowcolor{Gray!20} \multicolumn{3}{|c|} {\textbf{Atomic Readout Performance}} \\\hline
        \rowcolor{RoyalBlue!20}
    \ \ Imaging Method & \multicolumn{2}{l|}{\ \ Frequency Modulation Imaging (FMI)} \\\hline
        \rowcolor{RoyalBlue!5}
    \ \ Critical Atomic Temperature & $ \ \ T_c$  & $ \ \ < 117.7 \ \mathrm{nK}$ \\\hline

    \end{tabular}
    \caption{Summary of key technical requirements for the orbital optomechanical sensor. The values are derived using the detection of a 2 Hz continuous gravitational wave as a consistent benchmark scenario. The complete, target-frequency-dependent requirements are detailed in Sec.~\ref{Ssec:classical_noise}.}
    \label{tab:requirements_summary}
    \end{center}
    \end{table}

\clearpage
\bibliographystyle{apsrev4-2}
\bibliography{supp}